\documentclass[a4paper,fleqn,usenatbib]{mnras}

\usepackage{newtxtext,newtxmath}

\usepackage[T1]{fontenc}
\usepackage{ae,aecompl}

\usepackage{graphicx}	
\usepackage{amsmath}	
\usepackage{amssymb}	
\usepackage{subcaption}
\usepackage{float}
\usepackage{lipsum}

\usepackage{lscape}
\usepackage{multirow}
\usepackage{xcolor}
\usepackage{afterpage}
\usepackage{alphalph}
\usepackage{adjustbox}

\title[K-band PNe Spectroscopy]{The excitation mechanisms and evolutionary stages of UWISH2 planetary nebula candidates}

\author[A. M. Jones et al.]{
A. M. Jones$^{1}$\thanks{E-mail: aj16acd@herts.ac.uk}, T. M. Gledhill$^{1}$, D. Froebrich$^{2}$ and M. D. Smith$^{2}$
\\
\\
$^{1}$Centre for Astrophysics Research, University of Hertfordshire, College Lane, Hatfield AL10 9AB, UK \\
$^{2}$Centre for Astrophysics \& Planetary Science, The University of Kent, Canterbury, Kent CT2 7NH, UK
}

\date{Accepted XXX. Received YYY; in original form ZZZ}

\pubyear{2018}

\begin{document}
\label{firstpage}
\pagerange{\pageref{firstpage}--\pageref{lastpage}}
\maketitle

\begin{abstract}
We present medium-resolution K-band long-slit spectroscopy of 29 true, likely, possible and candidate Galactic Plane planetary nebulae (PNe) from the UWISH2 survey - many of which have only been recently discovered. These objects are bright in molecular hydrogen (H$_2$) emission, and many have bipolar morphologies. Through the detection of the Br$\gamma$ emission line, which traces ionized hydrogen, we find that the majority of the candidate PNe are indeed likely to be PNe, while 2 of the targets are more likely young stellar objects (YSOs) or pre-planetary nebulae (pPNe). We detect Br$\gamma$ in 13 objects which have no detection in IPHAS or SHS H$\alpha$ surveys. This implies they are potential members of the little-known optically-obscured PN population, hidden from wide-field optical surveys. We use the spatial extent of the H$_2$ 1-0 S(1) and Br$\gamma$ lines to estimate the evolutionary stage of our targets, and find that W-BPNe (bipolar PNe with pinched waist morphologies) are likely to be younger objects, while R-BPNe (bipolar PNe with large ring structures) are more evolved. We use line ratios to trace the excitation mechanism of the H$_2$, and find the 1-0 S(1) / 2-1 S(1) and 1-0 S(1) / Br$\gamma$ ratios are higher for R-BPNe, implying the H$_2$ is thermally excited. However, in W-BPNe, these ratios are lower, and so UV-fluorescence may be contributing to the excitation of H$_2$.
\end{abstract}




\section{Introduction}


\begin{table*} 		
\centering
\caption{Observation log and additional information for the 29 targets. Morphologies and sizes in H$_2$ and H$\alpha$ are taken from table A1 of \citet{2018MNRAS.tmp.1499G}. Morphologies make use of the `ERBIAS' classification system \citep{2006MNRAS.373...79P}, and sizes are the major and minor axis dimensions in arcsec. Where no H$\alpha$ morphology is given, the object is not detected in H$\alpha$ emission. PN status comes from the HASH PN database \citep{2016JPhCS.728c2008P}, where `T', `L', `P' and `C' refer to true, likely, possible and candidate PNe - see text for details.}
\resizebox{1.0\linewidth}{!}{
\begin{tabular}{lllllllllll} 
\hline
PN G & Other names & RA & Dec & Integration (s) & Airmass & \multicolumn{2}{c}{H$_2$} & \multicolumn{2}{c}{H$\alpha$} & PN status \\
          &			 &	   &       &                        &               & Morph. & Size & Morph. & Size & \\
\hline
004.7-00.8 & --- & 269.95542 & -25.26527 & 1200 & 1.71 & Bams & 25x12 & --- & --- & C \\

009.7-00.9 & SSTGLMC G009.7612-00.9575 & 272.71231 & -20.96230 & 1200 & 1.56 & Bs & 18x6 & --- & --- & C  \\ 

020.7-00.1 & --- & 277.37398 & -10.94099 & 1200 & 1.46 & Bs & 14x9 & A & 15x9 & C \\

020.8+00.4 & SSTGLMC G020.8543+00.4857 & 276.84922 & -10.50625 & 1200 & 1.34 & B & 2.8x1.2 & --- & --- & C \\

024.8+00.4 & SSTGLMC G024.8959+00.4586 & 278.76581 & -6.93617 & 1200 & 1.23 & Bs & 14x6 & --- & --- & C \\

025.9-00.5 & --- & 280.21143 & -6.44546 & 1200 & 1.46 & Ers & 8x5 & --- & --- & C \\

032.6-01.2 & MPA J1855-0048 & 283.85720 & -0.80638 & 1200 & 1.31 & Bps & 12x5 & E & 10x5 & T  \\

034.8+01.3 & --- & 282.55901 & +2.30305 & 1200 & 1.17 & Bs & 14x7 & --- & --- & C \\

035.7-01.2 & NVSS 190102+015727 & 285.26293 & +1.95644 & 1200 & 1.17 & Bs & 26x12 & --- & --- & C \\

036.4+00.1 & GPSR5 36.481+0.155 & 284.34075 & +3.23021 & 1200 & 1.25 & Bps & 12x5 & --- & --- & C  \\

037.4-00.1 & --- & 285.07885 & +3.90133 & 1080 & 1.28 & Er & 6x5 & --- & --- & C \\

040.4+01.1 & --- & 285.32726 & +7.20967 & 1200 & 1.11 & Bs & 12x4 & --- & --- & C \\

040.5-00.7 & --- & 287.02679 & +6.41554 & 1200 & 1.24 & Bs & 12x7 & --- & --- & C \\

042.1+00.4 & --- & 286.67062 & +8.38580 & 1200 & 1.55 & Bs & 9x9 & --- & --- & C \\

047.1+00.4 & --- & 289.05872 & +12.86745 & 1200 & 1.27 & Brs & 25x14 & S? & --- & C \\

047.5-00.3 & --- & 289.95370 & +12.76896 & 1200 & 1.24 & Es & 12x8 & --- & --- & C \\

048.2-00.4 & --- & 290.40474 & +13.37742 & 1200 & 1.47 & Bs & 10x8 & --- & --- & C \\

050.0-00.7 & --- & 291.57894 & +14.80423 & 1320 & 1.41 & Bs & 12x12 & --- & --- & C \\

050.5+00.0 & NVSS J192414+153909 & 291.06046 & +15.65315 & 1320 & 1.23 & Bs & 16x5 & S & --- & L \\

057.9-00.7 & Kn 7 & 295.60822 & +21.75634 & 1080 & 1.33 & Brs & 26x20 & B & 27x30 & T \\

058.1-00.8 & IPHASX J194301.3+215424 & 295.75604 & +21.90672 & 1320 & 1.61 & Bs & 20x12 & B & 15x12 & L \\

059.7-00.8 & IPHASX J194633.0+231659 & 296.63714 & +23.28371 & 1440 & 1.17 & Bs & 12x9 & Ea & 13x9 & P \\

060.5-00.3 & K 3-45 & 296.56513 & +24.18437 & 1200 & 1.93 & Bps & 11x6 & Bs & 11x6 & T \\

061.8+00.8 & --- & 296.14467 & +25.92826 & 1200 & 1.40 & Bs & 12x13 & B? & 12x12 & C \\

062.1+00.1 & --- & 297.01712 & +25.81331 & 1200 & 1.71 & Ear & 12x10 & E? & 10x8 & C \\

062.2+01.1 & --- & 296.15164 & +26.44203 & 1200 & 1.00 & Bs & 16x10 & E? & 11 & C \\

062.7+00.0 & IPHASX J194940.9+261521 & 297.42017 & +26.25520 & 1200 & 2.20 & Bs & 17x8 & Bps & 13x7 & T \\

064.1+00.7 & --- & 297.57807 & +27.89812 & 1080 & 1.05 & Bs & 11x4.5 & E? & 2.3x4 & C \\

064.9+00.7 & --- & 298.03161 & +28.54425 & 1200 & 1.96 & Brs & 10x8 & B? & 10x8.5 & C
\\
\hline
\end{tabular}}
\label{tab:observations}
\end{table*}


A planetary nebula (PN, or PNe plural) is an expanding gaseous shell, formed from the material released during periods of heavy mass loss on the asymptotic giant branch (AGB) phase of stellar evolution \citep{1970AcA....20...47P}. It is the penultimate stage of evolution for a low to intermediate mass star (0.8 $M_{\odot}$ to 8 $M_{\odot}$). The star responsible for the formation of the PN (central star, or CSPN), will move to the left of the Hertzsprung-Russell diagram as its surface temperature increases, until eventually becoming a white dwarf. UV radiation from the CSPN excites and ionizes the gaseous material, creating strong emission lines in the optical and infrared.

There are currently over 3000 PNe known in the Milky Way \citep{2016JPhCS.728c2008P}. Many of these were discovered by H$\alpha$ emission surveys, including the SuperCOSMOS H$\alpha$ Survey (SHS; \citeauthor{2005MNRAS.362..689P} \citeyear{2005MNRAS.362..689P}) and INT Photometric H$\alpha$ Survey of the Northern Galactic Plane (IPHAS; \citeauthor{2005MNRAS.362..753D} \citeyear{2005MNRAS.362..753D}). However it is likely that many PNe are yet to be discovered, obscured by dust in the Galactic Plane. It is therefore necessary to switch to infrared observations, which can penetrate the dust and reveal a variety of new objects. Studies such as those of \citet{2011MNRAS.413..514C} and \citet{2012MNRAS.427.3016P} have shown how certain infrared diagnostics can be used in order to successfully separate PNe from possible imposters.

The near-IR region is ideal for studying molecular hydrogen (H$_2$) in PNe - something known to us since its first detection by \cite{1976ApJ...209..793T}. Many PNe contain large reservoirs of H$_2$ in and outside the photodissociation region (PDR) \citep{1993IAUS..155..155T}, and it has also been proposed that H$_2$ emission can originate from the ionized region \citep{2011A&A...528A..74A}. However, extended H$_2$ emission can also be detected at large distances from the main nebula site, as shown recently by \citet{2018ApJ...859...92F}. H$_2$ is therefore an excellent tracer of the morphology of PNe, and many studies have been conducted to investigate relationships between morphology and H$_2$ emission. For example, there is strong observational evidence that bipolar PNe (BPNe) have brighter H$_2$ emission - known as Gatley's Rule (\citeauthor{1988ApJ...324..501Z} \citeyear{1988ApJ...324..501Z}; \citeauthor{1994ApJ...421..600K} \citeyear{1994ApJ...421..600K}). However, H$_2$ can also be detected in ellipsoidal or barrel-like PNe \citep{2013MNRAS.429..973M}. Bipolar PNe can be further divided into those showing broad, ring-like features (R-BPNe) and those with narrow waists or compact centres (W-BPNe) \citep{1996iacm.book.....M}. The nature of this divide is uncertain - they may form an evolutionary sequence, or originate from different progenitor populations \citep{2000ApJS..127..125G}. 

The K-band (2 - 2.4 $\muup$m) is home to a wide range of ro-vibrational lines of H$_2$, including the v = 1-0 S(1) (2.1218~$\muup$m) and 2-1 S(1) (2.2477~$\muup$m) lines. Ratios of these lines can indicate the mechanisms responsible for the excitation of the H$_2$ \citep{1976ApJ...203..132B}. Also in this waveband lie recombination lines of hydrogen and helium. The detection of the Br$\gamma$ H recombination line at 2.1661~$\muup$m signifies that the CSPN has begun to photoionize its environment, however this line can also be produced in shocks. Imaging of this line is a powerful tool for studying the evolution of PNe, especially in the transition from proto-planetary nebulae (pPNe) to PNe \citep{2015MNRAS.447.1080G}. The presence of these emission lines makes the K-band particularly advantageous to study both the molecular and ionized components of PNe.

In this work, we present K-band long-slit spectra of a sample of PNe and candidate PNe taken from the recent UWISH2 imaging survey (UKIRT Widefield Infrared Survey for H$_2$) \citep{2011MNRAS.413..480F}. We focus on the mechanisms governing the excitation of H$_2$, and how these relate to the evolutionary stages and morphologies of the targets. Along with spectra, we make use of near-IR H$_2$ and optical H$\alpha$ imaging, and mid-IR colours where available. In Sect.~\ref{sec:sample}, we describe the sample and the available data. Sect.~\ref{sec:observations} outlines our observations, and methods used to reduce the data. In Sect.~\ref{sec:results} we describe the spectra and images of the objects, and discuss the links between morphology, line ratios and evolution in Sect.~\ref{sec:discussion}. Finally, we make our conclusions in Sect.~\ref{sec:conclusions}.


\section{Sample}
\label{sec:sample}

UWISH2 was an unbiased Galactic Plane survey, focussed on the H$_2$ 1-0 S(1) line at 2.1218~$\muup$m. A total of 284 candidate PNe were found to lie in its survey area, including those already known to us, and many new discoveries with no known optical counterparts \citep{2015MNRAS.454.2586F}. \citet{2018MNRAS.tmp.1499G}, hereafter G18, have since updated this number to 291 candidates. Our sample for spectroscopic follow-up consisted of 29 targets, selected mainly due to their bright H$_2$ fluxes. The objects lie at low Galactic latitudes ($-1.5^{\circ} < b < 1.5^{\circ}$) and have longitudes in the range ($0^{\circ} < l < 65^{\circ}$). Table~\ref{tab:observations} lists all the targets observed.

Four of our targets have a PN status of `true', and have been confirmed as PNe through available multi-wavelength data. Two have the status `likely', for objects that are likely to be PNe but do not have completely conclusive morphology/spectroscopy data. One is labelled `possible', indicating a possible PN with currently insufficient data. The remaining 22 targets have the status `new candidate' - these have been classified as candidate PNe by the UWISH2 survey on the basis of morphology and lack of association with known star-forming regions. These statuses come from the HASH PN database\footnote{\url{http://hashpn.space}} \citep{2016JPhCS.728c2008P}. H$\alpha$ images are available for all 29 targets, however only 14 show evidence of H$\alpha$ emission, with 7 of these having been previously reported.

It is also apparent that some of our targets are bright in the mid-IR, with Spitzer/IRAC imaging available at 3.6, 4.5, 5.8, and 8.0~$\muup$m from the GLIMPSE survey\footnote{\url{http://irsa.ipac.caltech.edu/data/SPITZER/GLIMPSE/ index_cutouts.html}}. Mid-IR colour-colour plots can be used as an additional diagnostic to imaging and spectroscopy, to distinguish PNe from possible mimics, as their various infrared properties cause them to populate different regions of the diagrams (e.g. \citeauthor{2008AJ....136.2413R} \citeyear{2008AJ....136.2413R}; \citeauthor{2011MNRAS.413..514C} \citeyear{2011MNRAS.413..514C}). G18 have measured IRAC colours for targets in the UWISH2 survey with detectable mid-IR emission, which includes the targets from this work. Fluxes in all four bands are available for 18 of the targets.


\section{Observations and data reduction}
\label{sec:observations}

Observations were obtained on the 22nd - 24th July, 2016, using the Long-slit Intermediate Resolution Infrared Spectrograph (LIRIS: \citeauthor{2003RMxAC..16...43M} \citeyear{2003RMxAC..16...43M}; \citeauthor{2003INGN....7...15A} \citeyear{2003INGN....7...15A}) on the William Herschel Telescope (WHT), located at the Observatorio del Roque de los Muchachos on La Palma, Canary Islands. LIRIS uses a 1024 $\times$ 1024 HAWAII detector, with a pixel scale of 0.25~arcsec~pixel$^{-1}$, yielding a field of view of 4.27~arcmin x 4.27~arcmin.

We used a slit width of 1~arcsec for the brighter targets, and increased it to 2.5~arcsec for the fainter ones. The slit was generally oriented along the major axis of the nebula, as defined by UWISH2 images, or to contain the brightest H$_2$ features. However, this sometimes led to nearby stars falling within the slit, rendering certain sections of the data unusable. The extraction sections are marked on the UWISH2 H$_2$ - K images in Fig.~\ref{fig:h2images}, along with Fig.~\ref{fig:0505_h2} for PN\,G050.5+00.0, which includes the K band continuum, and Fig.~\ref{fig:0597_h2} for PN\,G059.7-00.8. Table~\ref{tab:observations} gives observational details for each target. 

Standard stars with spectral types B8 to A3 were observed throughout the 3 nights, as their spectra in the K-band are relatively featureless, apart from absorption at Br$\gamma$, which needed to be removed. Both the targets and standards were observed in a A-B-B-A nodding sequence, in order to effectively subtract the sky contribution. After flatfield correction, the images were aligned and stacked to form two-dimensional spectra with an increased signal to noise, which were then extracted at various positions along the slit to form multiple one-dimensional spectra. Arc images were used to wavelength calibrate the spectra, giving a spectral resolution of $\approx$~3.6~\AA~$\!$~pixel$^{-1}$. This was followed by telluric correction and flux calibration using the standard star spectra. Standard \textsc{iraf} routines \citep{1993ASPC...52..173T} were used for flatfield correction, wavelength calibration, telluric correction, flux calibration and image combination. \textsc{python} with the \textsc{astropy} module \citep{2013A&A...558A..33A} was used to extract the spectra. In order to calculate emission line fluxes, Gaussians were fitted to the spectra using the \textsc{curve fit} function from the \textsc{python scipy} module \citep{cite...scipy}, which uses non-linear least squares in order to fit a function to the data.

We also obtained narrow-band imaging of PN\,G050.5+00.0, using LIRIS. The observations were carried out on the 26th September, 2017, using the Br$\gamma$ and K-band continuum filters, which included flatfield frames. The average airmass of the target frames was 1.33, with a seeing of $\approx$~1~arcsec. Images were taken at 5 noddings - stacking the unaligned images together created a sky frame, which was subtracted from each target frame to remove the sky contribution. These were then aligned and stacked to form the final image. Standard \textsc{iraf} routines were used in flatfielding, image alignment and stacking. In order to highlight any structures within the bright central region of PN\,G050.5+00.0, we deconvolved the image using the Richardson-Lucy algorithm (\citeauthor{Richardson:72} \citeyear{Richardson:72}; \citeauthor{1974AJ.....79..745L} \citeyear{1974AJ.....79..745L}). This was achieved using a star within the same field to act as a point spread function, in order to improve the resolution. The procedure was carried out using the \textsc{lucy} routine in the \textsc{starlink} package \textsc{kappa} \citep{2014ASPC..485..391C}.


\section{Results}
\label{sec:results}

\begin{figure*}
\centering
	\begin{subfigure}{0.27\textwidth}
	\includegraphics[width=\textwidth]{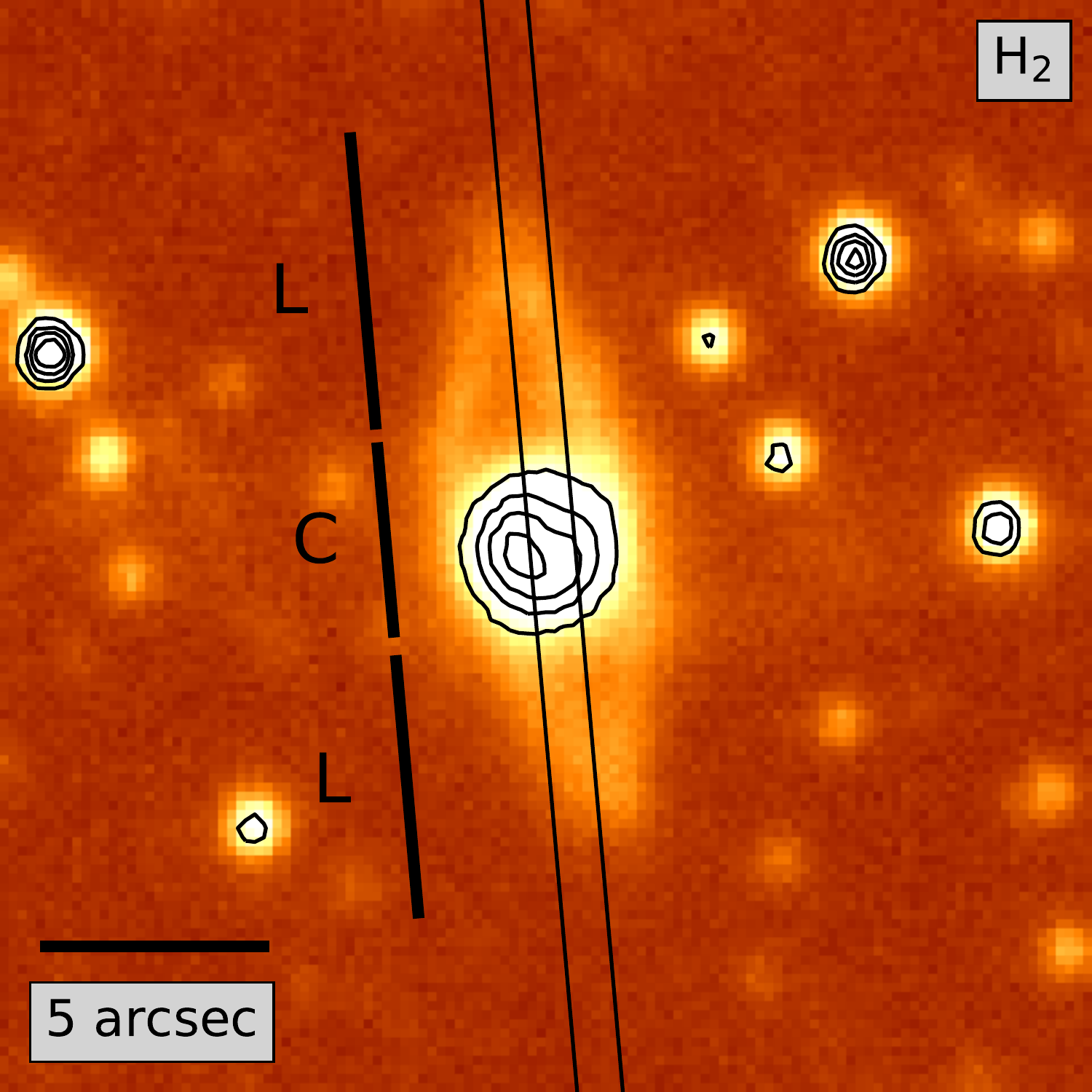}
	\caption{}
	\label{fig:0505_h2}
	\end{subfigure}
	\hspace{3mm}
	\begin{subfigure}{0.27\textwidth}
	\includegraphics[trim = 1.8cm 1.8cm 1.8cm 1.8cm, clip=true, width=\textwidth]{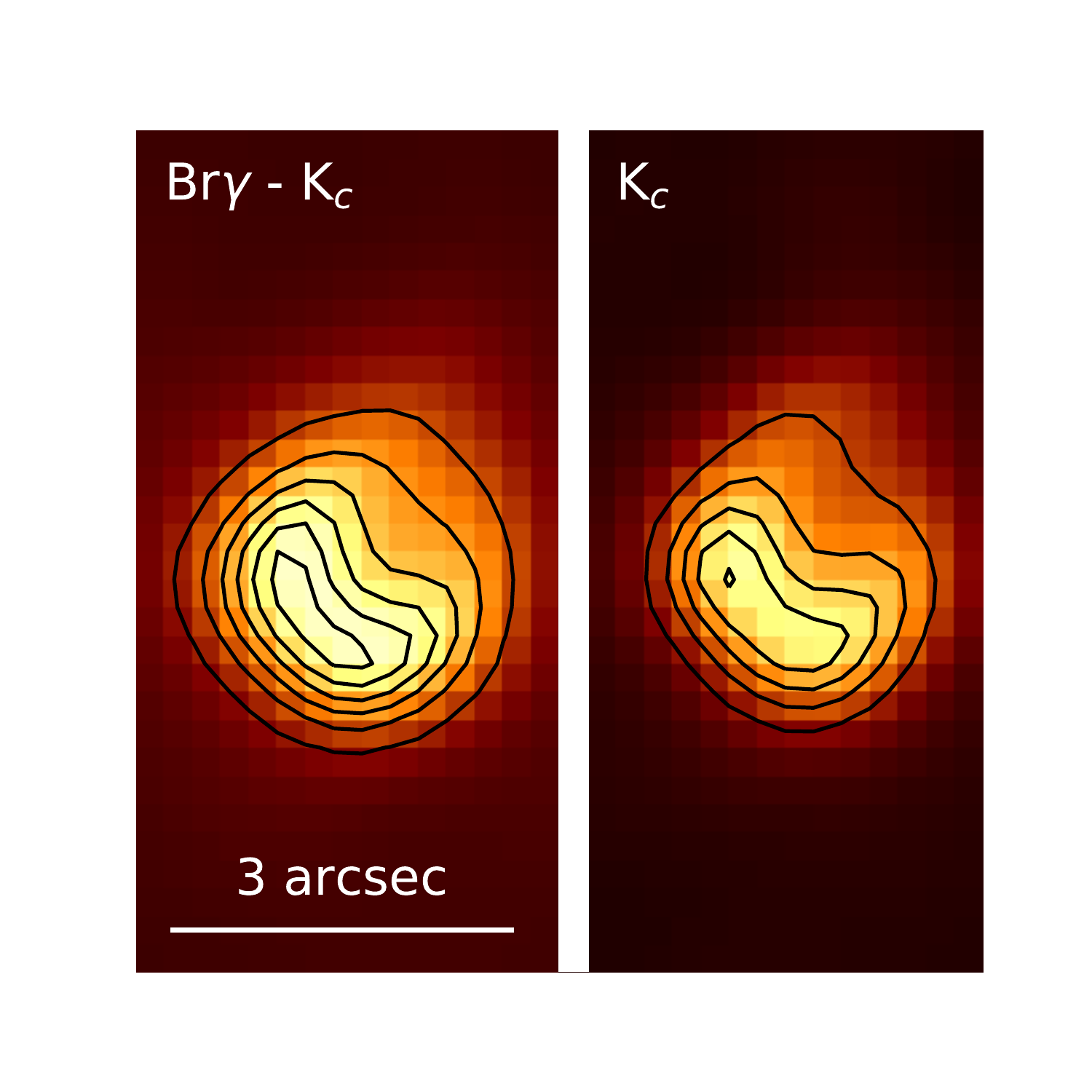}
	\caption{}
	\label{fig:0505_brg}
	\end{subfigure}
	\hspace{3mm}
	\begin{subfigure}{0.27\textwidth}
	\includegraphics[trim = 2.4cm 2.4cm 2.4cm 2.4cm, clip=true, width=\textwidth]{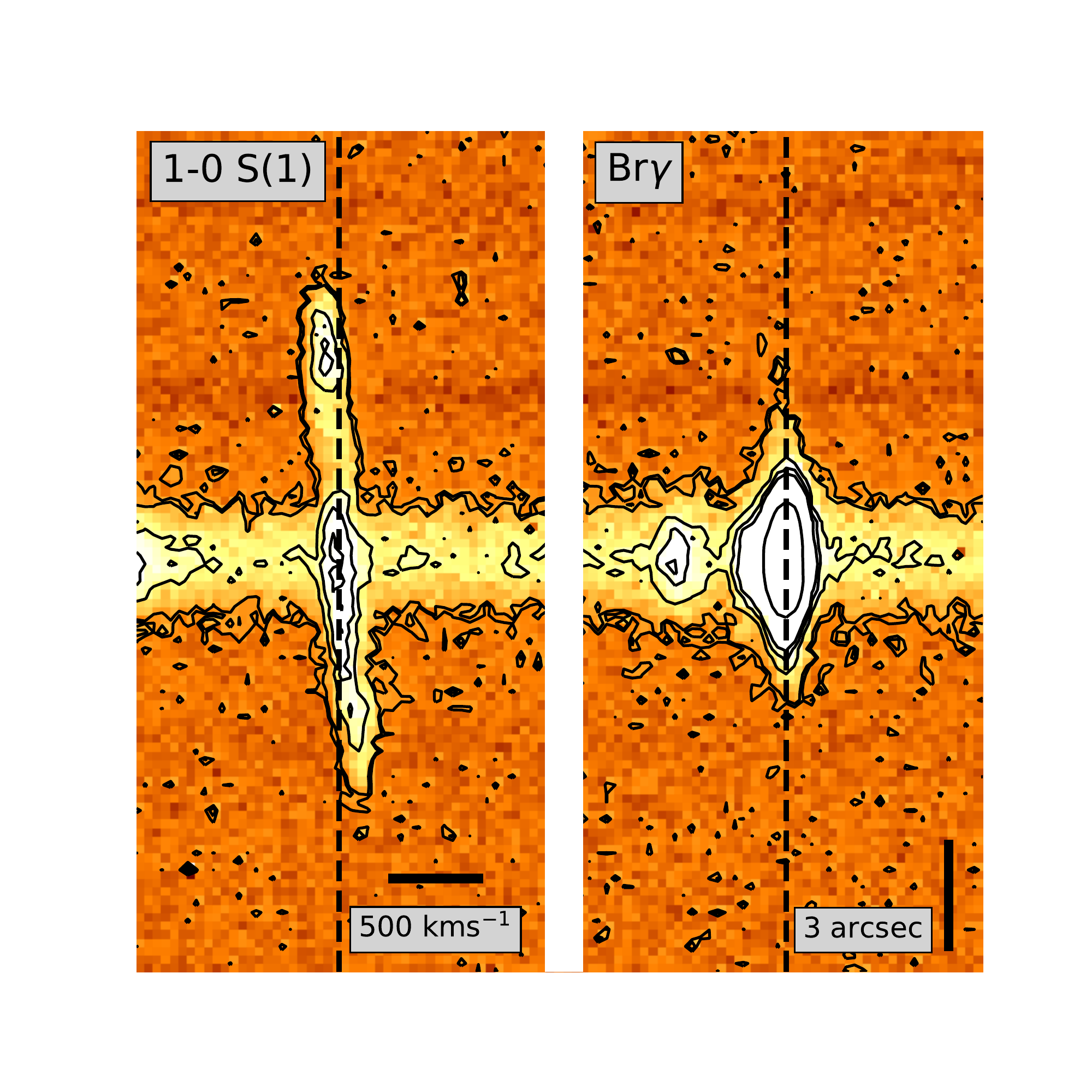}
	\caption{}
	\label{fig:0505_pv}
	\end{subfigure}
	\begin{subfigure}{0.4\textwidth}
	\includegraphics[width=\textwidth]{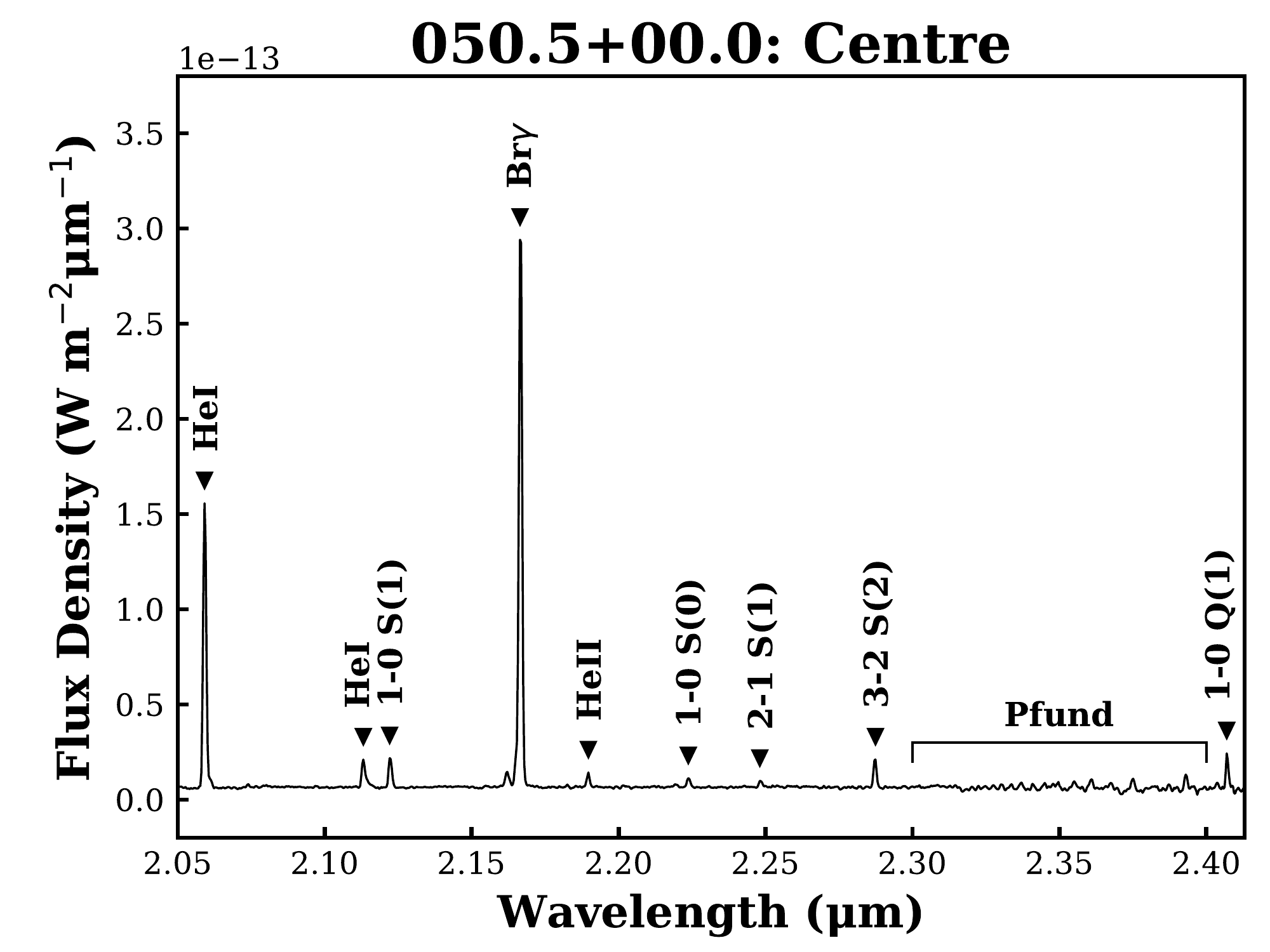}
	\caption{}
	\label{fig:0505_spec_centre}
	\end{subfigure}
	\begin{subfigure}{0.4\textwidth}
	\includegraphics[width=\textwidth]{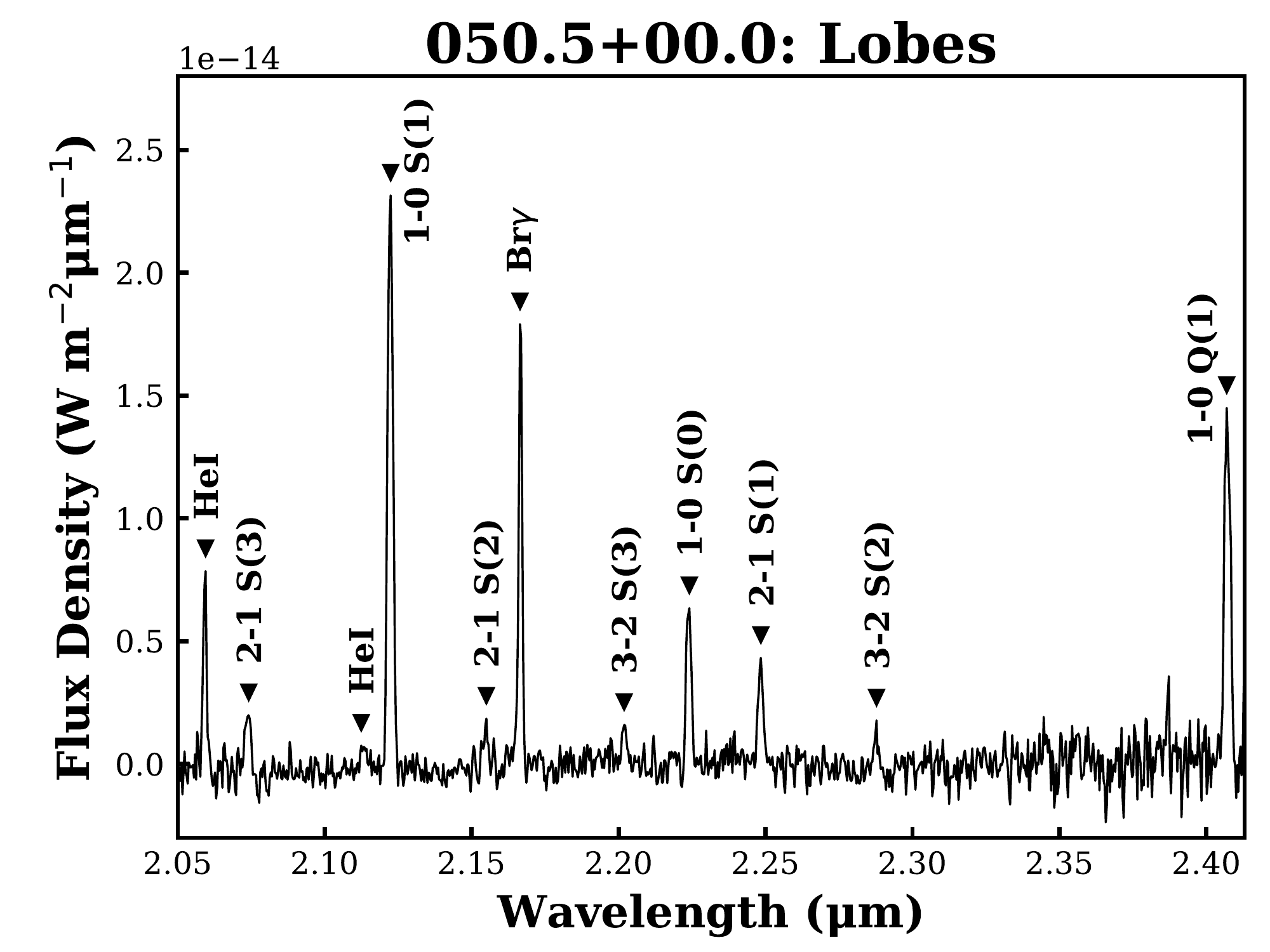}
	\caption{}
	\label{fig:0505_spec_lobes}
	\end{subfigure}
	\caption{K-band images and spectra for PN\,G050.5+00.0. a) H$_2$ image from UWISH2, showing position and width of slit, and extraction sections. Contours range from 2400 to 3600 counts, in steps of 400. b) Deconvolved narrow-band Br$\gamma$ - K$_c$ (left) and K-band continuum (right) images of the central region. Contours range from 500 to 3000 counts, in steps of 500. c) Two-dimensional spectra for the 1-0 S(1) and Br$\gamma$ lines, with dotted lines drawn through their centres. d) K-band spectrum of central region e) K-band spectrum of the sum of the lobes.}
	\label{fig:0505}
\end{figure*}

\begin{figure}
\centering
	\hspace{5mm}
	\begin{subfigure}{0.3\textwidth}
	\includegraphics[width=\textwidth]{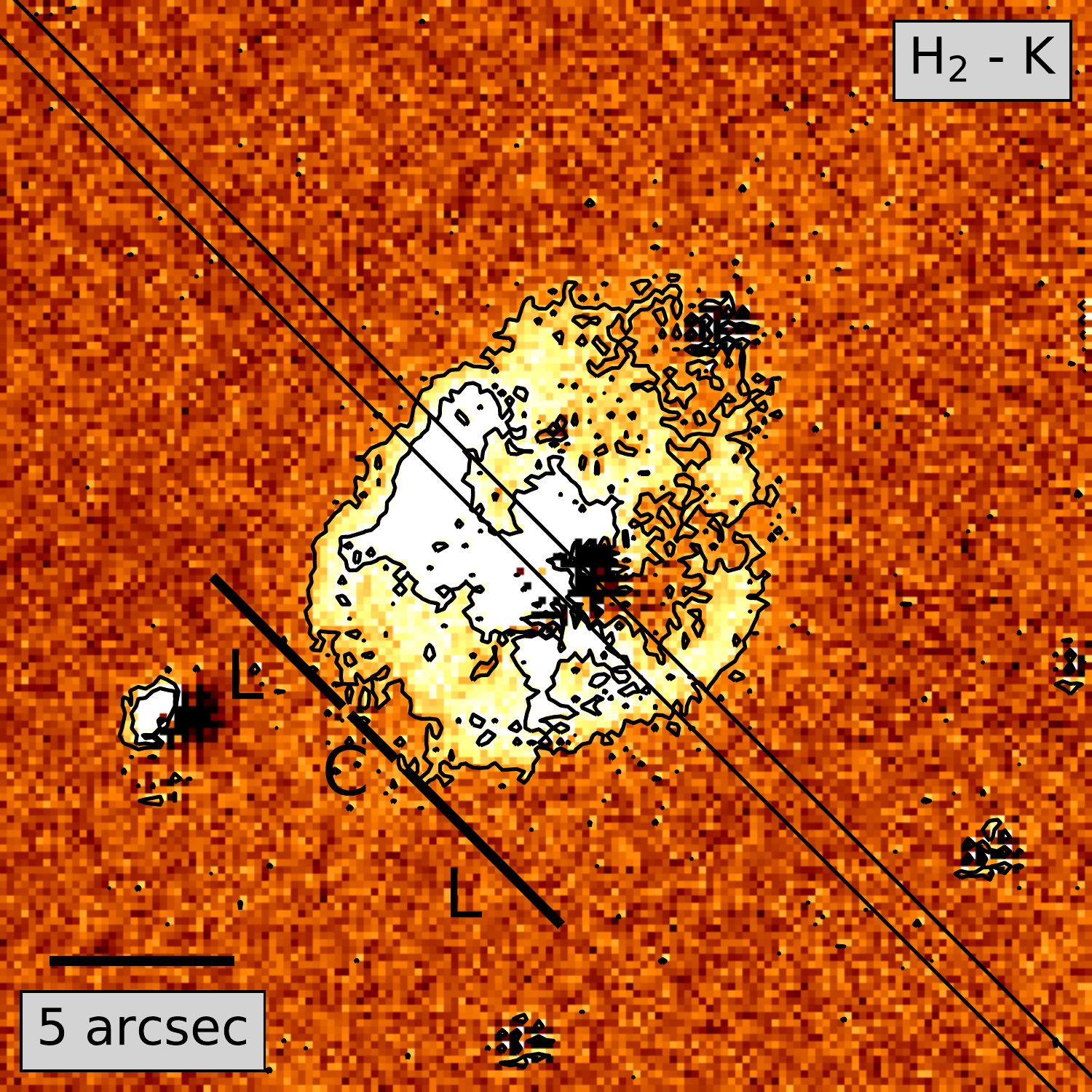}
	\caption{}
	\label{fig:0597_h2}
	\end{subfigure}
	\hspace{3mm}
	\begin{subfigure}{0.4\textwidth}
	\includegraphics[width=\textwidth]{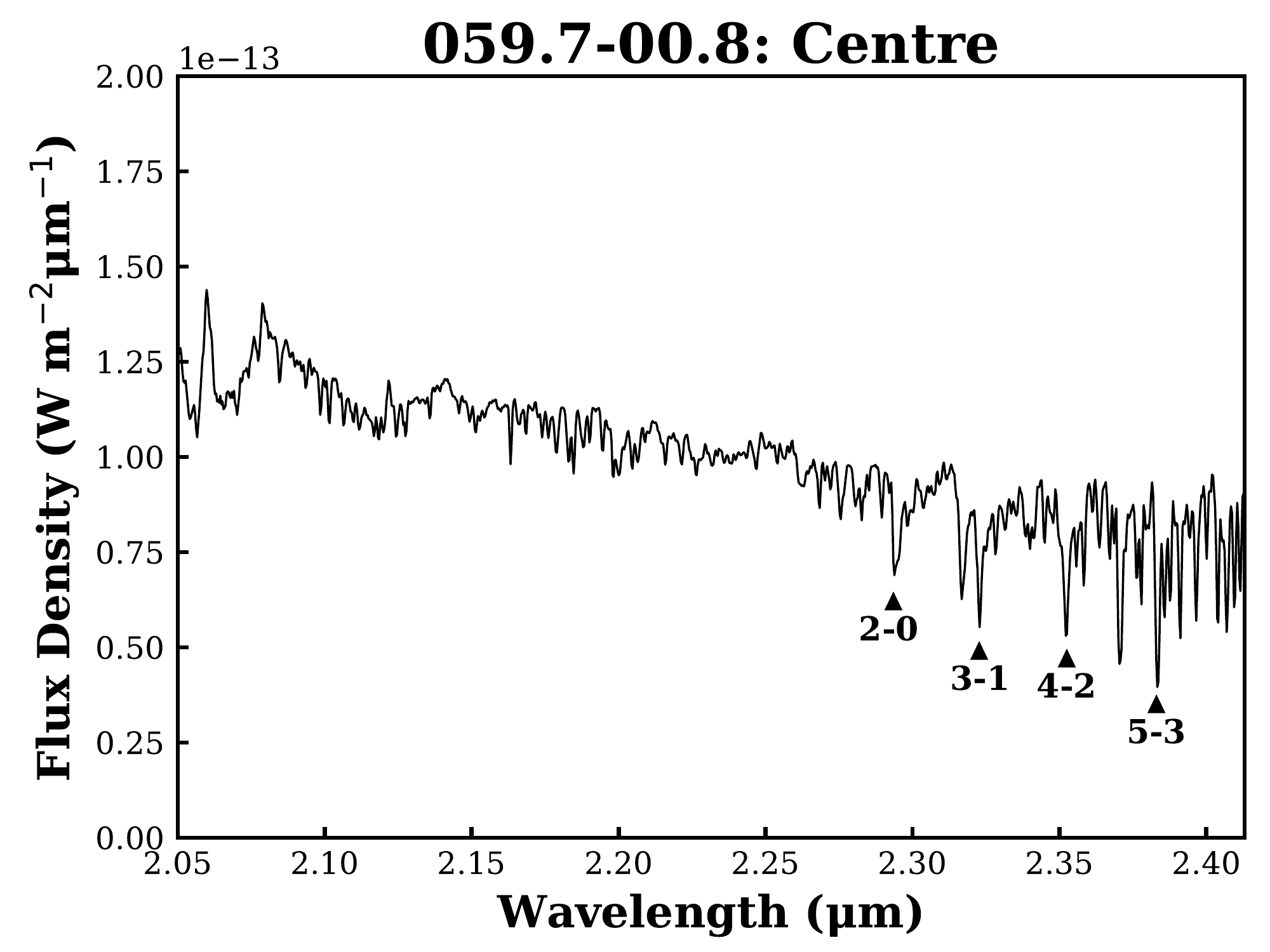}
	\caption{}
	\label{fig:0597_spec_centre}
	\end{subfigure}
	\begin{subfigure}{0.4\textwidth}
	\includegraphics[width=\textwidth]{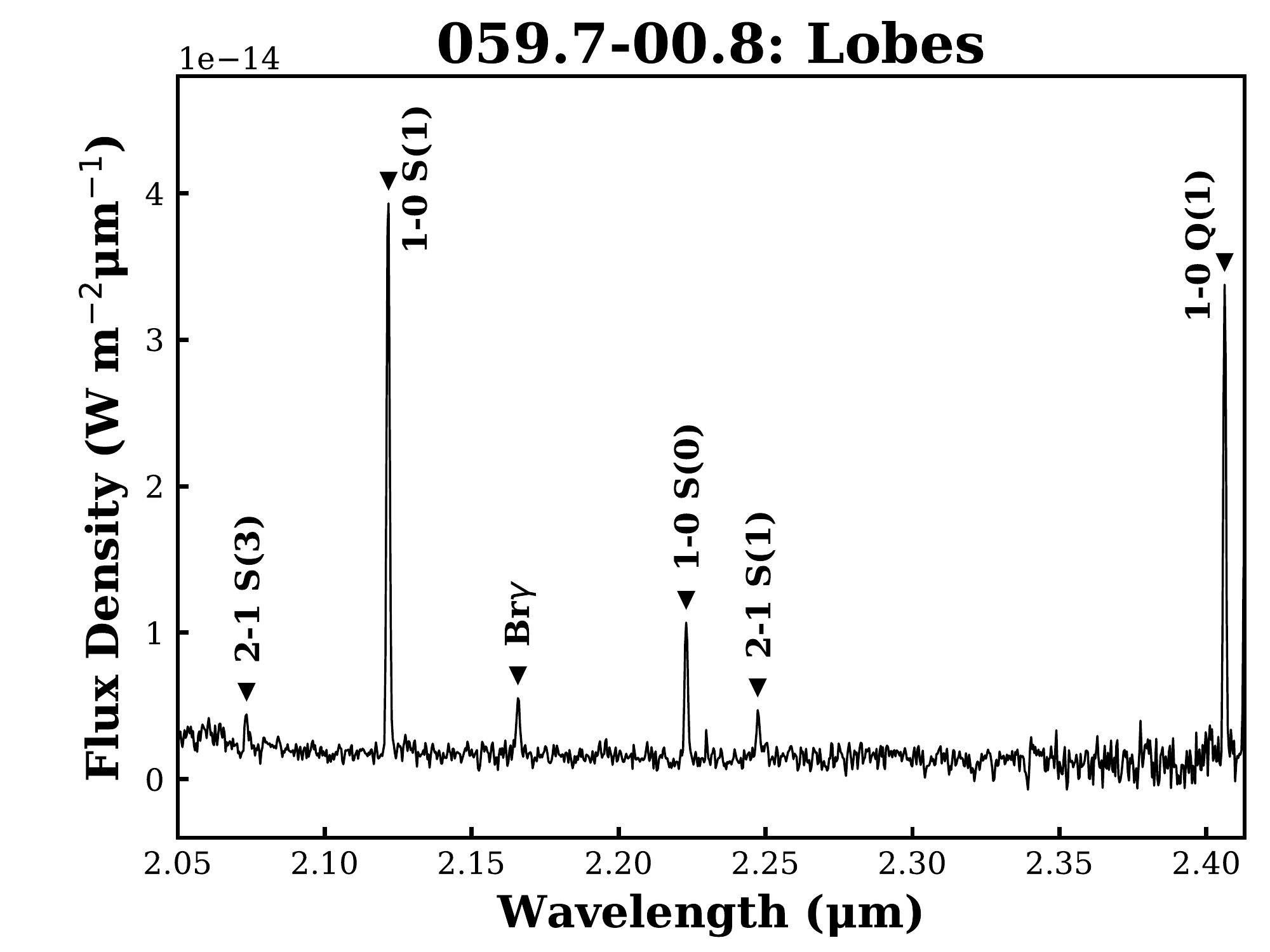}
	\caption{}
	\label{fig:0597_spec_lobes}
	\end{subfigure}
	\caption{K-band image and spectra for PN\,G059.7-00.8. a) H$_2$ - K image from UWISH2, showing position and width of slit, and extraction sections. Contours are located at 50 and 150 counts. b) K-band spectrum of central region, with CO bandheads labelled c) K-band spectrum of the sum of the lobes.}
	\label{fig:0597}
\end{figure}

Figures \ref{fig:0505_spec_centre} and \ref{fig:0505_spec_lobes} for PN\,G050.5+00.0, \ref{fig:0597_spec_centre} and \ref{fig:0597_spec_lobes} for PN\,G059.7-00.8, and Fig.~\ref{fig:spectra} for the remaining targets, show the fully calibrated emission spectra, with identified features labelled. All targets have the v = 1-0 S(1), 1-0 S(0) and 1-0 Q(1) H$_2$ emission lines, albeit the latter line is often weak due to poor atmospheric transmission. Higher vibrational level H$
_2$ lines are identified in some objects, including the 3-2 S(2) and 3-2 S(3) lines. Br$\gamma$ (2.1661~$\muup$m) is seen in all but 4 objects\footnote{We do not detect Br$\gamma$ in PN\,G062.1+00.1 or 064.9+00.7, however these objects have clear H$\alpha$ emission in their IPHAS images. We attribute this lack of detection to a combination of high airmass during observations, and nearby stars preventing extraction of the central regions, where the H$^{+}$ is thought to be concentrated.}, signifying the presence of ionized hydrogen. We detect the \ion{He}{i} (2.0587~$\muup$m) line in 9 targets, which means their central stars have become hot enough that their UV photons are capable of ionizing He. We also see the weaker \ion{He}{i} (2.1127~$\muup$m) line in 3 targets, and the \ion{He}{ii} (2.1891~$\muup$m) line in 4 targets. These, and a range of other H$_2$ emission lines seen in our targets are listed in Table~\ref{tab:linefluxes}, along with their fluxes.

\subsection{PN\,G050.5+00.0}
\label{sec:0505}

PN\,G050.5+00.0 is a particularly interesting object, with a range of radio data available. It has been classified as a `likely PN', (e.g. the CORNISH 5.0~GHz survey; \citeauthor{2012PASP..124..939H} \citeyear{2012PASP..124..939H}; and the RMS survey; \citeauthor{2009A&A...501..539U} \citeyear{2009A&A...501..539U}), while the Green Bank Telescope (GBT) \ion{H}{ii} Region Discovery Survey \citep{2011ApJS..194...32A} classifies PN\,G050.5+00.0 as a \ion{H}{ii} region. The UWISH2 image is shown in Fig.~\ref{fig:0505_h2}, without K-band subtraction\footnote{We include the H$_2$ image to give a better view of the central region, as the H$_2$ - K image of the centre is dominated by negative flux, due to a strong Br$\gamma$ flux in the K-band leading to an over-subtraction.}. PN\,G050.5+00.0 has a cigar-shaped H$_2$ morphology, surrounded by a rim of stronger emission. The object has a very bright central region, which is likely a combination of H$_2$ and K-band continuum emission (K$_c$). We also find strong Br$\gamma$ emission here, which extends slightly into the lobes where there is a void of H$_2$; more noticeable in the northern lobe (Fig.~\ref{fig:0505_pv}). 

We overlay contours on Fig.~\ref{fig:0505_h2} to highlight structure in the centre. We find the emission here is not uniform, but instead peaks at a point slightly to the southeast. This encouraged us to attempt narrow band Br$\gamma$ and K$_c$ imaging, to see if it would be possible to further resolve the central region, and to investigate the structure of the ionized and continuum emission. We show the deconvolved Br$\gamma$~-~K$_c$, and K$_c$ images in (Fig.~\ref{fig:0505_brg}). Both have elongated structures, slightly offset from the centre of the object, which are similar to the structure seen in Fig.~\ref{fig:0505_h2}. This leads us to believe that the Br$\gamma$ and continuum emission both originate from the unresolved central region, and are being scattered by the same spatially extended dusty structure.

Higher vibrational level H$_2$ lines, and recombination lines of helium (including those of \ion{He}{i} and \ion{He}{ii}) are seen in the spectra of PN\,G050.5+00.0 (Figures~\ref{fig:0505_spec_centre} and \ref{fig:0505_spec_lobes}). The Pfund series is also visible in the spectrum of the central region, from 2.3~$\muup$m onwards. In the two-dimensional spectrum of PN\,G050.5+00.0, we find the H$_2$ emission lines are slanted, indicating that the northern lobe is blueshifted, while the southern lobe is redshifted. We suggest two possible scenarios - either the object has curved outflows (where the northern lobe curves towards us, while the southern lobe curves away), or the outflows are accelerating. We show a region of the two-dimensional spectrum, centred on the 1-0 S(1) line, in Fig.~\ref{fig:0505_pv}, which is essentially a position-velocity diagram. We measure the wavelength shift from the centre to the end of each lobe to be $7.2 \pm 3.6$~\AA , which gives a line-of-sight velocity of $100 \pm 50$~km\,s$^{-1}$, relative to the systemic velocity. This is a lower limit on the expansion velocity, as any inclination to the line-of-sight will act to raise this value. Mid-IR colours measured by G18 suggest that PN\,G050.5+00.0 is more likely to be a PN rather than a \ion{H}{ii} region.

\subsection{PN\,G059.7-00.8}

PN\,G059.7-00.8 is a `possible' PN, with an optical spectrum and imaging available\footnote{These can be found on the HASH PN database.}. The object has a steady rising continuum in the optical (4000 to 8000~\AA), with strong emission lines of [\ion{O}{iii}], [\ion{N}{ii}] and H$\alpha$. It has an elliptical morphology in H$\alpha$ emission, however its H$_2$ structure is spider-like, with stronger shells of emission to the northeast and southwest (Fig~\ref{fig:0597_h2}). In the centre of PN\,G059.7-00.8, we find a very bright source, and investigate to see if it is related to the object. The spectrum of this source, with a bright continuum, is shown in Fig.~\ref{fig:0597_spec_centre}. The continuum falls at longer wavelengths, and we find multiple CO bandheads in absorption, including the v = 2-0 (2.2935~$\muup$m), 3-1 (2.3227~$\muup$m), 4-2 (2.3535~$\muup$m) and 5-3 (2.3829~$\muup$m). These are characteristics common among YSOs (\citeauthor{1996A&A...306..427C} \citeyear{1996A&A...306..427C}; \citeauthor{1997AJ....114.2700R} \citeyear{1997AJ....114.2700R}) but are also present in main sequence dwarf stars of spectral type K and later. The nebula spectrum at optical wavelengths shows emission lines typical of a PN, plus G18 measure mid-IR colours consistent with a PN. We suspect, therefore, that the bright central source is an unrelated field star projected onto the PN.

\subsection{Candidate PNe}

22 targets in our sample are candidate PNe selected by the UWISH2 survey, with the remaining 7 targets either true, likely or possible PNe. 7 of the candidates have H$\alpha$ emission in either IPHAS or SHS images (see Table~\ref{tab:observations}). 13 candidate PNe have the Br$\gamma$ line, but show no signs of H$\alpha$ emission in optical surveys. These are likely to be members of the optically-obscured PN population. Two of the candidates, PN\,G004.7-00.8 and PN\,G020.8+00.4, show no signs of Br$\gamma$ or H$\alpha$ emission - a lack of H$^{+}$ suggests these may not be PNe. Both objects have bipolar morphologies, however the former has multiple extended structures, including a large sweeping tail of H$_2$ extending eastward before turning south, while the latter consists of two small blobs of H$_2$ emission. G18 find mid-IR colours which place PN\,G004.7-00.8 in the YSO region, while the mid-IR colours of PN\,G020.8+00.4 are not consistent with PNe, YSOs or \ion{H}{ii} regions; instead it seems more likely to be a proto-planetary nebula (pPN).

We classified 4 of the candidates as W-BPNe, including PN\,G009.7-00.9, G024.8+00.4, G036.4+00.1 and G040.4+01.1. These objects have either a compact core or a narrow, pinched waist in H$_2$ emission. All of these have strong Br$\gamma$ and \ion{He}{i} (2.0587~$\muup$m) lines in their central regions, while the first two have v = 3-2 H$_2$ lines and additional helium lines (\ion{He}{i} at 2.1127~$\muup$m and \ion{He}{ii} at 2.1891~$\muup$m) in their spectra. None show evidence of H$\alpha$ emission, implying strong line-of-sight extinction. The cores of PN\,G009.7-00.9 and G036.4+00.1 can be resolved into two knots of H$_2$, which could be signs of molecular tori being viewed side-on \citep{1997A&A...318..561K}. PN\,G024.8+00.4 has well-separated H$_2$ lobes, with a smaller knot of emission just south of the centre. Its spectrum has a continuum gradually rising with wavelength, which is often a sign of dust. PN\,G040.4+01.1 also has two separated lobes, with two smaller knots to the northwest and southeast of the centre, which again could be due to a molecular torus.

Most of the remaining candidates we consider to be R-BPNe, which are bipolars with equatorial ring structures\footnote{It is sometimes difficult to see a clear ring feature, however we label any bipolar objects as R-BPNe if their morphology is obviously not of the W-BPN or elliptical type.}, with the exception of PN\,G025.9-00.5, G037.4-00.1 and G047.5-00.34, whose H$_2$ morphologies are more elliptical. The R-BPNe in our sample have fairly similar spectra which lack the v = 3-2 H$_2$ lines. The \ion{He}{i} (2.0587~$\muup$m) line is observed in 3 of these objects, and one of the elliptical PNe (PN\,G037.4-00.1). The most impressive of the R-BPNe is PN\,G035.7-01.2, with the largest angular size of the sample. The object has arcs of H$_2$, which are likely to be cavity walls. It is invisible in H$\alpha$, but Br$\gamma$ lines are present in our spectra. PN\,G047.1+00.4 also has an interesting morphology, which is point-symmetric resembling an eye rotated by 90$^{\circ}$. It has a ring of H$_2$ emission, and two curved arms extending in the north and south directions. Spectra and H$\alpha$ images reveal an ionized region localised to the centre, where we also observe the \ion{He}{ii} (2.1891~$\muup$m) line - this is the only R-BPN in our sample which shows this feature. However, this could be an orientation effect, where we are looking down the major axis of the object, and from another viewing angle, the object could look more like a W-BPN.

\begin{figure*}
\centering
	\begin{subfigure}{0.45\textwidth}
	\includegraphics[width=\textwidth]{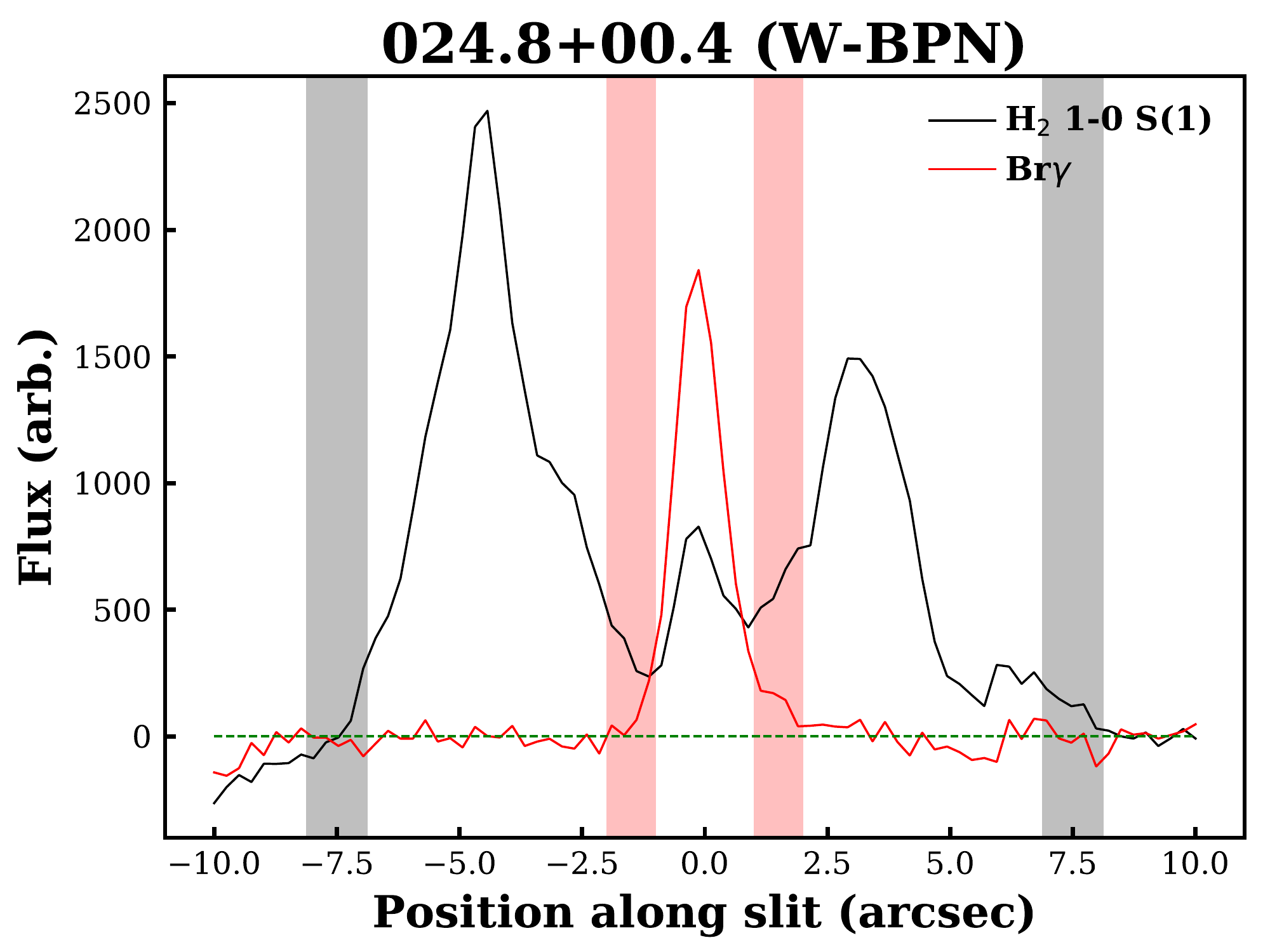}
	\caption{}
	\label{fig:slices_0248}
	\end{subfigure}
	\begin{subfigure}{0.45\textwidth}
	\includegraphics[width=\textwidth]{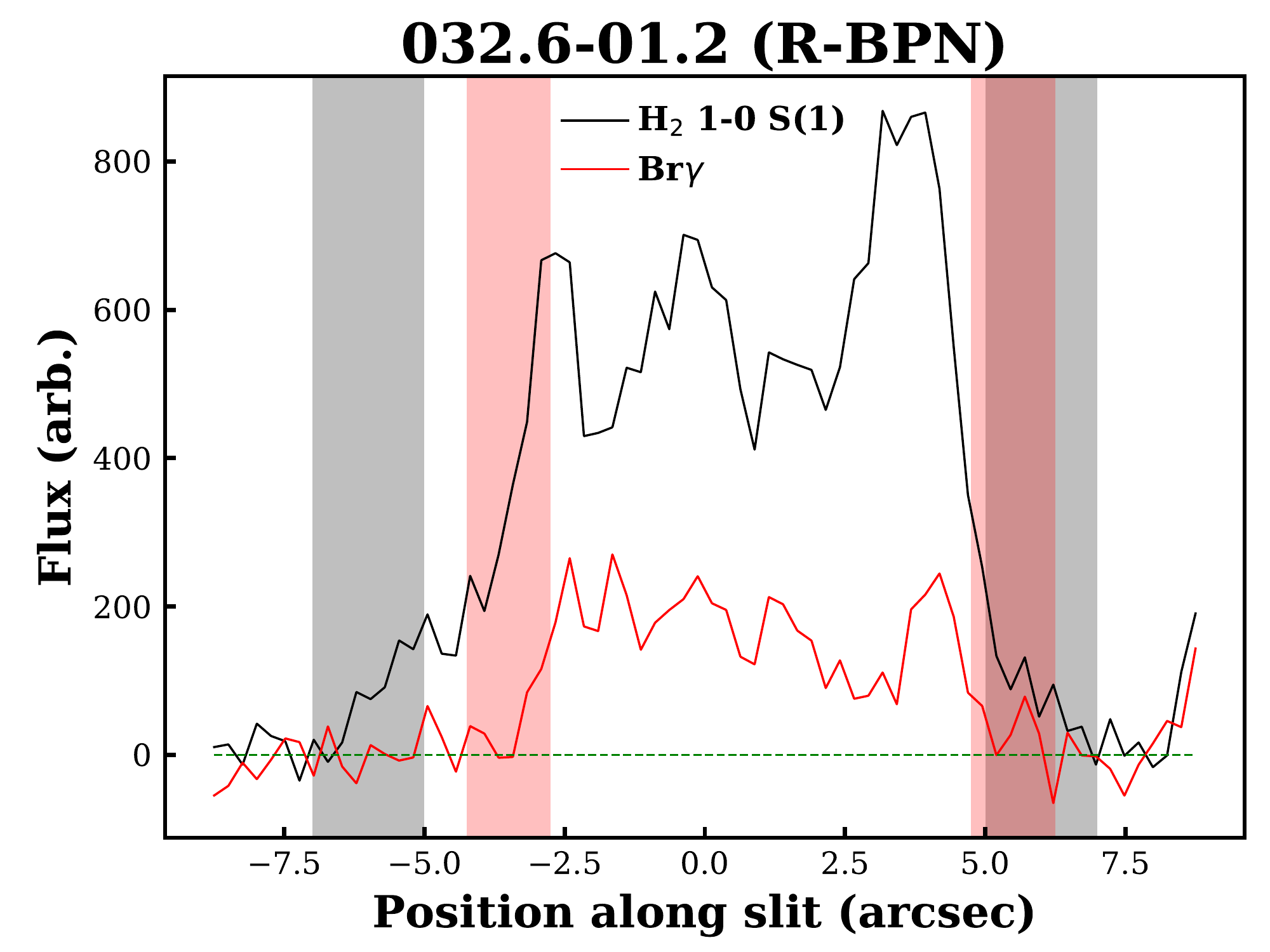}
	\caption{}
	\label{fig:slices_0326}
	\end{subfigure}
\caption{Slices showing the variation of the flux along the slit for PN\,G024.8+00.4 (a) and PN\,G032.6-01.2 (b), where the black and red lines are for the 1-0 S(1) and Br$\gamma$ hydrogen emission lines respectively. The shaded regions represent the regions where the radii are estimated to lie. The green dotted line represents zero flux.}
\label{fig:slices}
\end{figure*}


\section{Discussion}
\label{sec:discussion}

\begin{table*}
\centering
\caption{Key line flux ratios and errors, uncorrected for extinction. The regions of the objects these fluxes have been extracted from are shown in Appendix~\ref{fig:h2images}.}
\fontsize{8.0}{9.6}\selectfont
\begin{tabular}{llcccc}
\hline
ID & Region & 1-0 S(1) / 2-1 S(1) & 1-0 S(1) / 3-2 S(3) & 1-0 S(1) / Br$\gamma$ & \ion{He}{i} (2.0587 $\muup$m) / Br$\gamma$ \\ 
\hline

004.7-00.8 & Centre & 12 $\pm$ 2 & --- & --- & --- \\

009.7-00.9 & Centre & 7.3 $\pm$ 0.6 & 54 $\pm$ 19 & 2.4 $\pm$ 0.1 & 0.58 $\pm$ 0.04 \\ 
 & Lobes & 7.2 $\pm$ 0.7 & 25 $\pm$ 7 & --- & --- \\

020.7-00.1 & Centre & --- & --- & 6.4 $\pm$ 1.5 & --- \\
 & Lobes & 12 $\pm$ 2 & --- & 19 $\pm$ 6 & --- \\

020.8+00.4 & Centre & 7.1 $\pm$ 1.3 & --- & --- & --- \\
 & Lobes & 6.3 $\pm$ 1.0 & --- & --- & --- \\

024.8+00.4 & Centre & 5.7 $\pm$ 2.0 & --- & 0.50 $\pm$ 0.07 & 0.67 $\pm$ 0.03 \\
 & Lobes & 8.6 $\pm$ 1.0 & 43 $\pm$ 9 & --- & --- \\
 
025.9-00.5 & North lobe & ---                 & ---                 & 11 $\pm$ 4      & ---             \\

032.6-01.2 & Centre     & 12 $\pm$ 2          & ---                 & 2.9 $\pm$ 0.2   & 0.30 $\pm$ 0.05 \\
                   & Lobes      & 14 $\pm$ 2          & ---                 & 7.5 $\pm$ 0.8   & ---             \\

034.8+01.3 & Centre     & ---                 & ---                 & 3.4 $\pm$ 0.4   & 0.25 $\pm$ 0.08 \\
                            & Lobes      & 11 $\pm$ 2          & ---                 & 11 $\pm$ 1      & 0.48 $\pm$ 0.14 \\

035.7-01.2 & Centre     & ---                 & ---                 & 6.4 $\pm$ 1.3   & ---             \\
                            & Lobes      & 12 $\pm$ 2          & ---                 & 9.1 $\pm$ 1.1   & ---             \\

036.4+00.1 & Centre     & 7.0 $\pm$ 1.5       & ---                 & 1.6 $\pm$ 0.1   & 0.34 $\pm$ 0.06 \\
                            & South lobe & 8.4 $\pm$ 1.7       & ---                 & 2.2 $\pm$ 0.2   & 0.31 $\pm$ 0.04 \\

037.4-00.1 & Centre     & 6.3 $\pm$ 1.7       & ---                 & 0.44 $\pm$ 0.07 & 0.28 $\pm$ 0.07 \\
                            & Lobes      & 9.3 $\pm$ 1.7       & ---                 & 3.7 $\pm$ 0.4   & 0.28 $\pm$ 0.0
\\  

040.4+01.1 & Centre & 7.4 $\pm$ 2.5       & ---                 & 1.4 $\pm$ 0.1   & 0.50 $\pm$ 0.10 \\
                            & Lobes  & 7.8 $\pm$ 0.7       & ---                 & ---             & ---             \\

040.5-00.7                  & All    & ---                 & ---                 & 9.8 $\pm$ 3.7   & ---             \\

042.1+00.4                  & All    & ---                 & ---                 & 12 $\pm$ 2      & ---             \\

047.1+00.4 & Centre & ---                 & ---                 & 0.59 $\pm$ 0.03 & 0.46 $\pm$ 0.06 \\
                            & Lobes  & ---                 & ---                 & 9.2 $\pm$ 2.6   & ---             \\

047.5-00.3                  & All    & ---                 & ---                 & 5.5 $\pm$ 0.7   & ---             \\

048.2-00.4                  & All    & ---                 & ---                 & 12 $\pm$ 3    & ---             \\

050.0-00.7 & Centre & ---                 & ---                 & 6.0 $\pm$ 1.5   & ---             \\
                            & Lobes  & ---                 & ---                 & ---             & ---             \\

050.5+00.0 & Centre & 4.5 $\pm$ 0.4       & ---                 & 0.05 $\pm$ 0.01 & 0.49 $\pm$ 0.01 \\
                            & Lobes  & 5.8 $\pm$ 0.4       & 16 $\pm$ 3          & 2.0 $\pm$ 0.1   & 0.43 $\pm$ 0.04 \\

057.9-00.7 & Centre & 14 $\pm$ 4          & ---                 & 6.4 $\pm$ 0.7   & ---             \\
                            & Lobes  & 13 $\pm$ 1          & ---                 & 23 $\pm$ 7      & ---             \\

058.1-00.8 & Centre & ---                 & ---                 & 7.9 $\pm$ 3.2   & ---             \\
                            & Lobes  & 14 $\pm$ 4          & ---                 & 10 $\pm$ 3      & ---             
\\

059.7-00.8    & Lobes  & 13 $\pm$ 2          & ---                 & 8.5 $\pm$ 1.0   & --- \\

060.5-00.3 & Centre     & 12 $\pm$ 2          & ---                 & 4.6 $\pm$ 0.3  & ---       \\
                            & Lobes      & 14 $\pm$ 5          & ---                 & 12 $\pm$ 3     & ---       \\

061.8+00.8 & Centre     & ---                 & ---                 & 7.0 $\pm$ 1.2  & ---       \\
                            & Lobes      & 13 $\pm$ 2          & ---                 & 11 $\pm$ 3     & ---       \\

062.1+00.1                  & North lobe & ---                 & ---                 & ---            & ---       \\

062.2+01.1 & Centre     & ---                 & ---                 & 12 $\pm$ 3     & ---       \\
                            & Lobes      & ---                 & ---                 & ---            & ---       \\

062.7+00.0 & Centre     & 12 $\pm$ 3          & ---                 & 2.7 $\pm$ 0.3  & ---       \\
                            & North lobe & ---                 & ---                 & 6.3 $\pm$ 1.0  & ---       \\

064.1+00.7 & Centre     & 9.3 $\pm$ 1.8       & ---                 & 9.1 $\pm$ 1.6  & ---       \\
                            & Lobes      & 13.3 $\pm$ 3.7      & ---                 & 20 $\pm$ 5     & ---       \\

064.9+00.7                  & Lobes      & ---                 & ---                 & ---            & --- \\

\hline
\end{tabular}
\label{tab:lineratios}
\end{table*}

\begin{figure*}
\centering
	\begin{subfigure}{0.4\textwidth}
	\includegraphics[width=\textwidth]{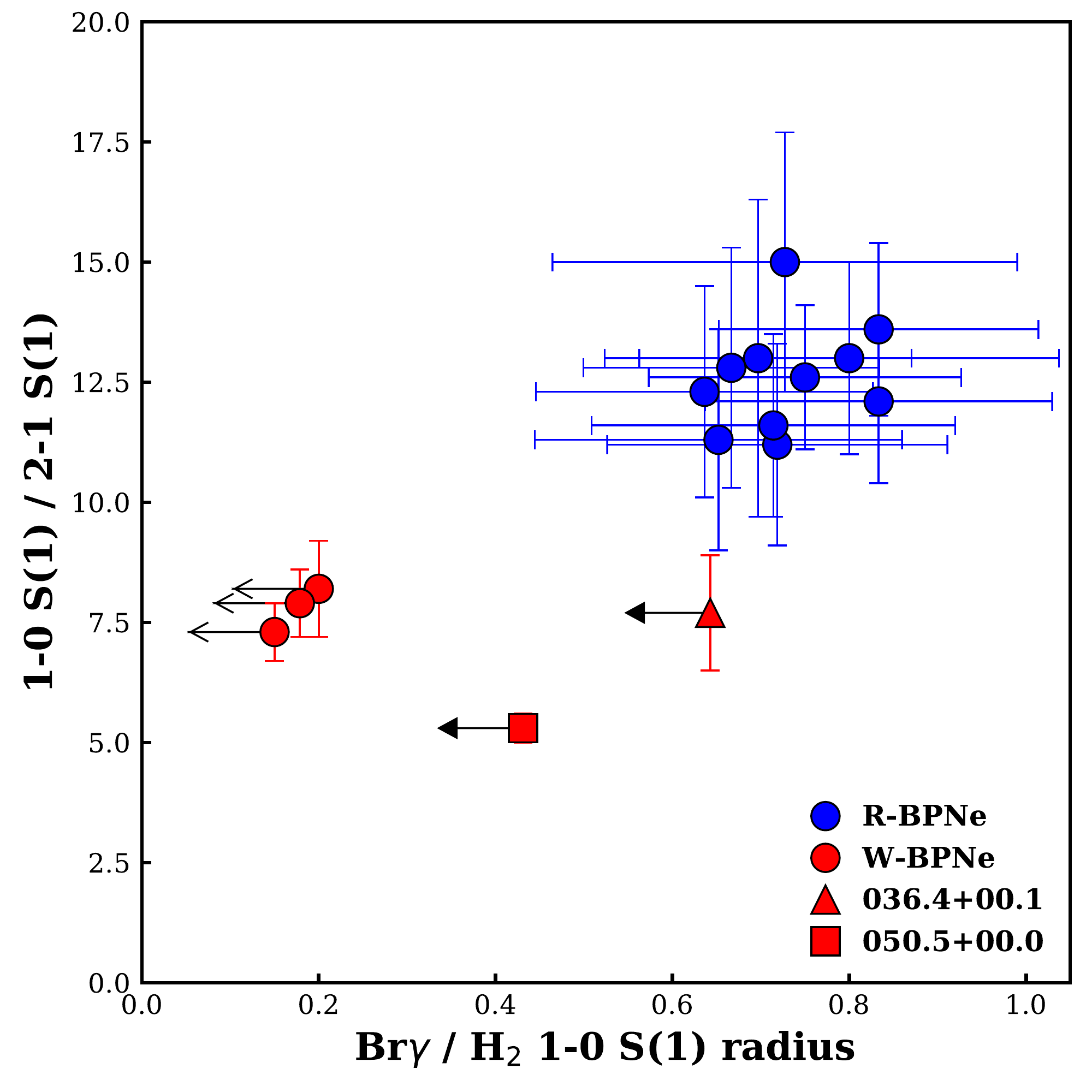}
	\caption{}
	\label{fig:radiusvratios}
	\end{subfigure}
	\begin{subfigure}{0.4\textwidth}
	\includegraphics[width=\textwidth]{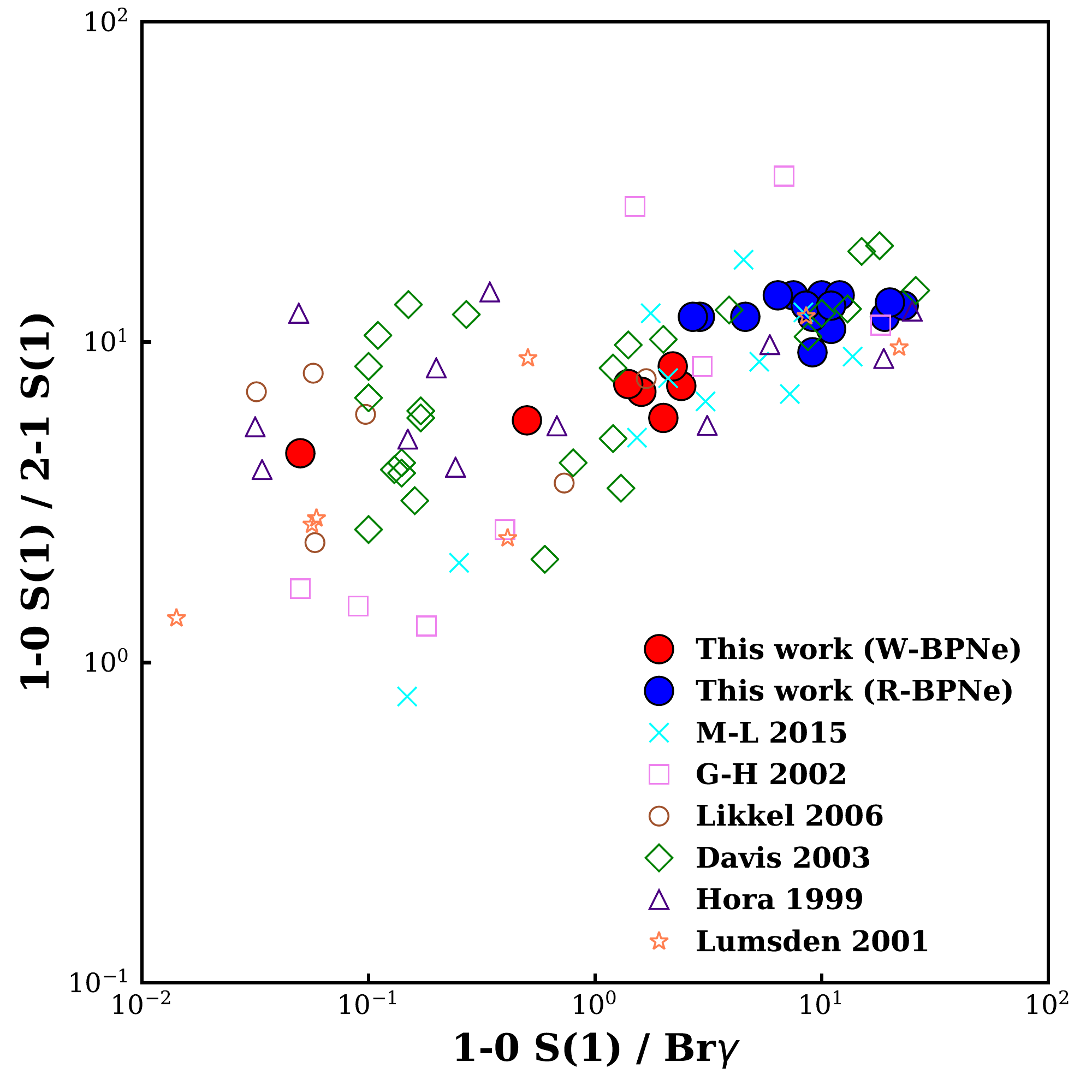}
	\caption{}
	\label{fig:lineratios}
	\end{subfigure}
\caption{a) Ionized to molecular radius ratio versus 1-0 S(1) / 2-1 S(1) ratio for bipolar PNe. Error bars for radii are estimated from two-dimensional spectra. We mark solid arrows (not to scale) on two objects as we believe these should move to the left of the diagram, and hollow arrows on objects thought to represent upper limits (see text for details) b) Line ratio plot with data from this work overlaid onto previous results. This includes the individual ratios extracted at different positions along the slit.}
\label{fig:ratiosandradii}
\end{figure*}

A simple means of determining the evolutionary stage of a PN is to observe the spatial extent of ionized material, traced by Br$\gamma$ emission, in relation to the extent of H$_2$ emission. It follows that in pPNe, Br$\gamma$ emission should be absent, as their central stars are not yet hot enough to ionize the surrounding envelope. When the temperature reaches $\approx$~25000~K, ionization produces Br$\gamma$ emission localised to the central region. As the PN evolves, the ionization front moves outwards, gradually replacing the molecular material. This process can be clearly seen in young PNe \citep{2015MNRAS.447.1080G}. As not all of our targets are visible at H$\alpha$, we can use the two-dimensional spectra to measure the radius of the ionized region along the slit using the Br$\gamma$ emission, which can then be compared to the radius of H$_2$ 1-0 S(1) emission. In Fig.~\ref{fig:slices}, we demonstrate how the ionized and molecular radii are estimated using this technique for two objects, PN\,G024.8+00.4 (left) and PN\,G032.6-01.2 (right). The position along the slit is given on the x-axis, and the flux (summed over the slit width) is on the y-axis. The flux of the molecular and ionized hydrogen emission lines along the slit are given by the black and red lines respectively. The point at which the flux reaches the noise level is the maximum extent of the emission line, and therefore the radius, where the shaded areas show the regions where the radii are estimated to lie. For the two objects in Fig.~\ref{fig:slices}, we estimate the ratio of the diameters (or equivalently the radii) of the Br$\gamma$ and 1-0~S(1) lines to be $0.2\pm0.07$ and $0.75\pm0.18$ for PN\,G024.8+00.4 and PN\,G032.6-01.2 respectively. We plot the ratio of the ionized to molecular radius on the x-axis of Fig.~\ref{fig:radiusvratios} for the bipolar objects in our sample. Error bars on these values reflect the regions in which the radii are estimated to lie, as the exact point at which the emission reaches the background level when using plots such as Fig.~\ref{fig:slices} is not always clear. This is certainly the case when the signal to noise ratio is low, or when nearby stars contaminate the two-dimensional spectra. In cases where it was deemed too difficult to estimate a radius using this technique alone, H$_2$ and H$\alpha$ imaging were also inspected when available. 

We separate the objects in Fig.~\ref{fig:radiusvratios} according to their bipolar type, with W-BPNe (compact core or narrow waist) in red and the R-BPNe (broad ring structures) in blue. We classify five of our targets as W-BPNe, including PN\,G009.7-00.9, G024.8+00.4, G036.4+00.1, G040.4+01.1 and G050.5+00.0. Most of the remaining sample we classify as R-BPNe. Three of the W-BPNe have ionized regions localised to their centres, so they are positioned to the very left of the figure. These values are upper limits, indicated by hollow arrows, as their Br$\gamma$ extents match the average seeing of the observations, given by the width of stellar continua in the same field of view. Fig.~\ref{fig:slices_0248}, for PN\,G024.8+00.4, clearly shows the small angular extent of the Br$\gamma$ emission when compared to the much more extended 1-0 S(1) emission. PN\,G050.5+00.0 lies to the right of this group, however we showed in Sec~\ref{sec:0505} that this object's Br$\gamma$ emission is likely to be generated in an unresolved central region, and then scattered, possibly by a dusty torus, which increases its apparent Br$\gamma$ extent. If this scattering process were not occurring, PN\,G050.5+00.0 would move to the left, which we indicate by a solid arrow in Fig.~\ref{fig:radiusvratios}. PN\,G036.4+00.1 lies to the right of PN\,G050.5+00.0, however we note the slit was not positioned along the major axis of the nebula, which we believe to run northeast to southwest. Therefore, we have no information as to what extent the ionization front has travelled into the lobes. We believe if the slit were positioned along the major axis, the Br$\gamma$ to H$_2$ radius would decrease, and so we mark this object with a solid arrow to show it could also move to the left. 

Once these considerations are taken into account, Fig.~\ref{fig:radiusvratios} shows a clear divide between the two morphological types, where the W-BPNe have a smaller ratio of Br$\gamma$ to H$_2$ radius, and so are less evolved, while the R-BPNe have larger Br$\gamma$ to H$_2$ radii and are therefore more evolved. Fig.~\ref{fig:slices_0326} compares the extents of the ionized and molecular emission for an R-BPN (PN\,G032.6-01.2), and it can be seen these are comparable. This idea makes sense, considering that many pPNe, in the stage of evolution immediately before the PN phase, have morphologies closely resembling those of W-BPNe. It is also known that strong emission lines of \ion{He}{i} (2.0587~$\muup$m) and Br$\gamma$ are seen in young PNe \citep{2015MNRAS.447.1080G}, which is what we observe. On the other hand, the fact that the ionization front in the R-BPNe has moved further from the central region approaching the outer bound of the H$_2$ radius, means that R-BPNe have had more time to form broad, ring-like structures. \citet{2017MNRAS.470.3707R} measure physical sizes, kinematic ages and luminosities of a sample of bipolar PNe, and also find that R-BPNe are more evolved than W-BPNe.

Information about how the H$_2$ is being excited can be found by comparing the fluxes of emission lines. These line ratios have the advantage that the effect of differential extinction on their values is relatively small, due to small wavelength separations between lines. H$_2$ can be excited thermally (e.g. in shocks) or non-thermally (e.g. UV-fluorescence). In the thermal process, H$_2$ molecules are collisionally heated to a few thousand degrees, and radiate near-IR photons as they cool \citep{1987PhDT.......204B}. In the non-thermal process, a H$_2$ molecule is excited when it absorbs a UV photon, and quickly decays to to a vibrationally-excited level of the electronic ground state. Further decays result in the emission of infrared photons \citep{1976ApJ...203..132B}. A mixture of these processes contribute to the emission lines we observe in our spectra. However, thermal processes populate the H$_2$ levels from the lowest (v = 0) to the highest (v $\geq$ 2) states, whereas non-thermal processes populate from the top (v $\geq$ 3) down. Therefore, ratios such as the 1-0 S(1) / 2-1 S(1) and 1-0 S(1) / 3-2 S(3) will have typical values depending on the excitation environment. A purely UV pumped spectrum will have a 1-0 S(1) / 2-1 S(1) ratio equal to 1.8 (\citeauthor{1976ApJ...203..132B} \citeyear{1976ApJ...203..132B}, \citeauthor{1987ApJ...322..412B} \citeyear{1987ApJ...322..412B}), while in the thermal case, indicative of shocks, this ratio will be at least 10 (\citeauthor{1977ApJ...216..419H} \citeyear{1977ApJ...216..419H}, \citeauthor{1992ApJ...399..563B} \citeyear{1992ApJ...399..563B}). We present key line ratios in Table~\ref{tab:lineratios}.

On the y-axis of Fig.~\ref{fig:radiusvratios} we plot the 1-0 S(1) / 2-1 S(1) ratio, summed over the object, and again there is good evidence for a dichotomy between the two bipolar types. Using Fig.~\ref{fig:radiusvratios}, the mean 1-0 S(1) / 2-1 S(1) ratios for the W-BPNe and R-BPNe are $7.3\pm1.0$ and $12.6\pm1.1$ respectively. This leads us to believe that on average, thermal excitation is the main mechanism exciting H$_2$ in R-BPNe, while the lower ratios of W-BPNe, and the fact that we only observe v = 3-2 H$_2$ transitions in these objects, could mean that UV-fluorescence plays a more important role in their excitation. While a mean 1-0 S(1) / 2-1 S(1) ratio of 7.3 is higher than the theoretical value for a purely UV pumped spectrum, a fluoresced dense gas (n $\geq$ 10$^{5}$~cm$^{-3}$) subjected to intense UV radiation will have an increased 1-0 S(1) / 2-1 S(1) ratio due to collisional heating (\citeauthor{1989ApJ...338..197S} \citeyear{1989ApJ...338..197S}, \citeauthor{1995ApJ...455..133H} \citeyear{1995ApJ...455..133H}). Further observational evidence is provided by \citet{2015MNRAS.453.1888M}, who find UV excitation is likely occurring in the cores of W-BPNe, while R-BPNe are dominated by shock excitation. If R-BPNe are more evolved than W-BPNe, then it follows that UV-fluorescence is a process associated with younger objects, while thermal excitation prevails as the PN evolves; a trend found observationally by \citet{2003MNRAS.344..262D}. This broadly agrees with the theoretical work of \citet{1998A&A...337..517N}, however once high resolution integral field spectroscopy is obtained, it is evident that line ratios, and therefore excitation mechanisms, can vary across the surface of young PNe, and highlights the importance of the dependence of line ratios on density \citep{2015MNRAS.447.1080G}. This is to be explored further in subsequent modelling papers.

There are also indications that the 1-0 S(1) / Br$\gamma$ ratio is linked with the evolutionary stage, for example \citet{2000ApJS..127..125G} find this ratio is low for young PNe, while more evolved objects have larger ratios. This is supported by G18, who find that for a small sample of 23 PNe covering a range in physical radii of 0.03 to 0.6~pc, the average H$_2$ surface brightness is approximately independent of size, and therefore age. Combining this with the fact that the average H$\alpha$, and therefore Br$\gamma$, surface brightness decreases with size \citep{2016MNRAS.455.1459F}, it follows that the average H$_2$ 1-0 S(1) / Br$\gamma$ surface brightness ratio increases with size and age. The trend between the 1-0 S(1) / Br$\gamma$ and 1-0 S(1) / 2-1 S(1) ratios has been investigated in \citet{2015MNRAS.453.1888M}. We have taken fig. 7 from this work, included our data, and reproduced the graph here as Fig.~\ref{fig:lineratios}. Additional data comes from \citet{1999ApJS..124..195H}, \citet{2001MNRAS.328..419L}, \citet{2002A&A...387..955G}, \citet{2003MNRAS.344..262D}, \citet{2006AJ....131.1515L} and \citet{2015MNRAS.453.1888M}. This plot shows a loose positive correlation between the 1-0 S(1) / 2-1 S(1) and 1-0 S(1) / Br$\gamma$ ratios, and our data is no exception to this trend. Again, we separate our targets into W-BPNe and R-BPNe, and there is a clear separation between the two bipolar types, where W-BPNe have lower 1-0 S(1) / Br$\gamma$ ratios than the R-BPNe (mean values of $1.5\pm0.8$ and $10.6\pm6.0$ respectively). Fig.~\ref{fig:lineratios} may then suggest an evolutionary sequence, where bipolar PNe move to larger 1-0 S(1) / Br$\gamma$ and 1-0 S(1) / 2-1 S(1) ratios as they evolve.

PN\,G050.5+00.0 has the lowest 1-0 S(1) / 2-1 S(1) and 1-0 S(1) / Br$\gamma$ ratios in our sample, with values of 4.5 and 0.05 respectively in the central region. If PNe do generate higher line ratios as they evolve, this would mean that PN\,G050.5+00.0 is the youngest PN in our sample. It is likely that UV fluorescence is largely contributing to the excitation of H$_2$ here, in regions closer to the central star. Assuming a fluoresced gas, a 1-0 S(1) / 3-2 S(3) value of 16 for PN\,G050.5+00.0 could indicate a high FUV flux with a density $\geq 10^{5}$~cm$^{-3}$ \citep{1990ApJ...365..620B}. We find good evidence for fast outflows in the lobes of PN\,G050.5+00.0 (see Sec~\ref{sec:0505}), which increase the chance of the H$_2$ being shock-excited. This is likely the reason behind the small increase of the 1-0 S(1) / 2-1 S(1) ratio to 5.8 in the lobes.

A few of our objects have had their distances estimated in previous works, including PN\,G050.5+00.0 (5.4 kpc; \citeauthor{2015ApJ...799...29E} \citeyear{2015ApJ...799...29E}), PN\,G057.9-00.7 and PN\,G062.7+00.0 (6.8 kpc and 6.1 kpc respectively; \citeauthor{2017MNRAS.470.3707R} \citeyear{2017MNRAS.470.3707R}). The angular resolution of the UWISH2 survey is 0.2 arcsec~pixel$^{-1}$, which converts to roughly 1000 to 1400 AU~pixel$^{-1}$ for these three objects. Distances to the remaining objects could be estimated using the H$\alpha$ surface brightness - radius relation \citep{2016MNRAS.455.1459F} for objects with H$\alpha$ (or even Br$\gamma$) imaging available, so long as the extinction could be accurately determined. It is likely that many of the objects in this study suffer significant extinction as they lie in the Galactic Plane, especially those with detected Br$\gamma$ emission and no H$\alpha$ emission - large extinctions indicate they may be located at large distances. As \citet{2015ApJ...808..115M} have shown, the ability to resolve structures within PNe, such as knots and filaments, depends largely on the spatial resolution of the observations. The large distances, combined with generally small angular sizes, mean it is difficult to describe in detail the fine-scale structures of the objects in this study without access to higher resolution observations.


\section{Conclusions}
\label{sec:conclusions}

In this work, we have presented medium resolution, K-band spectra of a sample of 29 Galactic Plane objects, including 4 true, 2 likely, 1 possible and 22 candidate PNe, taken from the UWISH2 survey. The candidate PNe were selected on the basis of their morphologies, and lack of association with known star-forming regions. Evidence for ionized material, in the form of either Br$\gamma$ emission in our spectra or H$\alpha$ emission in narrow-band surveys, is found in all but 2 of the targets. One of these, PN\,G020.8+00.4, is likely to be a pPN - at an earlier stage of evolution with a central star not yet hot enough to ionize its surrounding environment. PN\,G004.7-00.8 has mid-IR colours indicative of a YSO. 13 of the candidate PNe show no clear H$\alpha$ emission, however we detect Br$\gamma$ emission in their spectra. These objects potentially contribute to the optically-obscured PN population, and their discovery clearly highlights the need for multi-wavelength studies of PNe, to accurately predict the number of PNe in the Milky Way.

We have used our spectra to calculate line ratios, which have been used to constrain the mechanisms dominating the excitation of H$_2$. Most of our targets that we believe to be PNe are either R-BPNe (large ring structures) or W-BPNe (pinched waist), while the remaining 3 are considered to be elliptical. In agreement with previous studies, we find the former are predominantly thermally excited, while in the latter, UV fluorescence may have more influence. The link between line ratios and the spatial extent of ionized emission could mean that W-BPNe are younger objects, while the R-BPNe are more evolved, and an evolutionary scheme in which one class evolves into the other is worth further investigation. While long-slit spectroscopy is a useful tool for measuring line ratios with the advantage of spatial information in one dimension, more detailed spatial information can be achieved using integral field spectroscopy (IFS). This would allow excitation mechanisms to be inferred over the entire target, while comparing to the two-dimensional structure of the ionized region.

\section*{Acknowledgements}

The authors would like to thank the referee A. Manchado for his report, along with J. E. Drew and P. W. Lucas for discussions which helped improve the manuscript. A. M. Jones acknowledges support from the UK's Science and Technology Facilities Council (STFC) [grant number ST/K502029/1]. The William Herschel Telescope and its service programme are operated on the island of La Palma by the Isaac Newton Group of Telescopes in the Spanish Observatorio del Roque de los Muchachos of the Instituto de Astrofísica de Canarias. This research has made use of the HASH PN database at \url{http://hashpn.space}, and the SIMBAD database, operated at CDS, Strasbourg, France.




\bibliographystyle{mnras}
\bibliography{Refs}



\appendix

\section{H$_2$ - K images}

\begin{figure*}
	\begin{subfigure}{0.22\textwidth}
	\includegraphics[width=\textwidth]{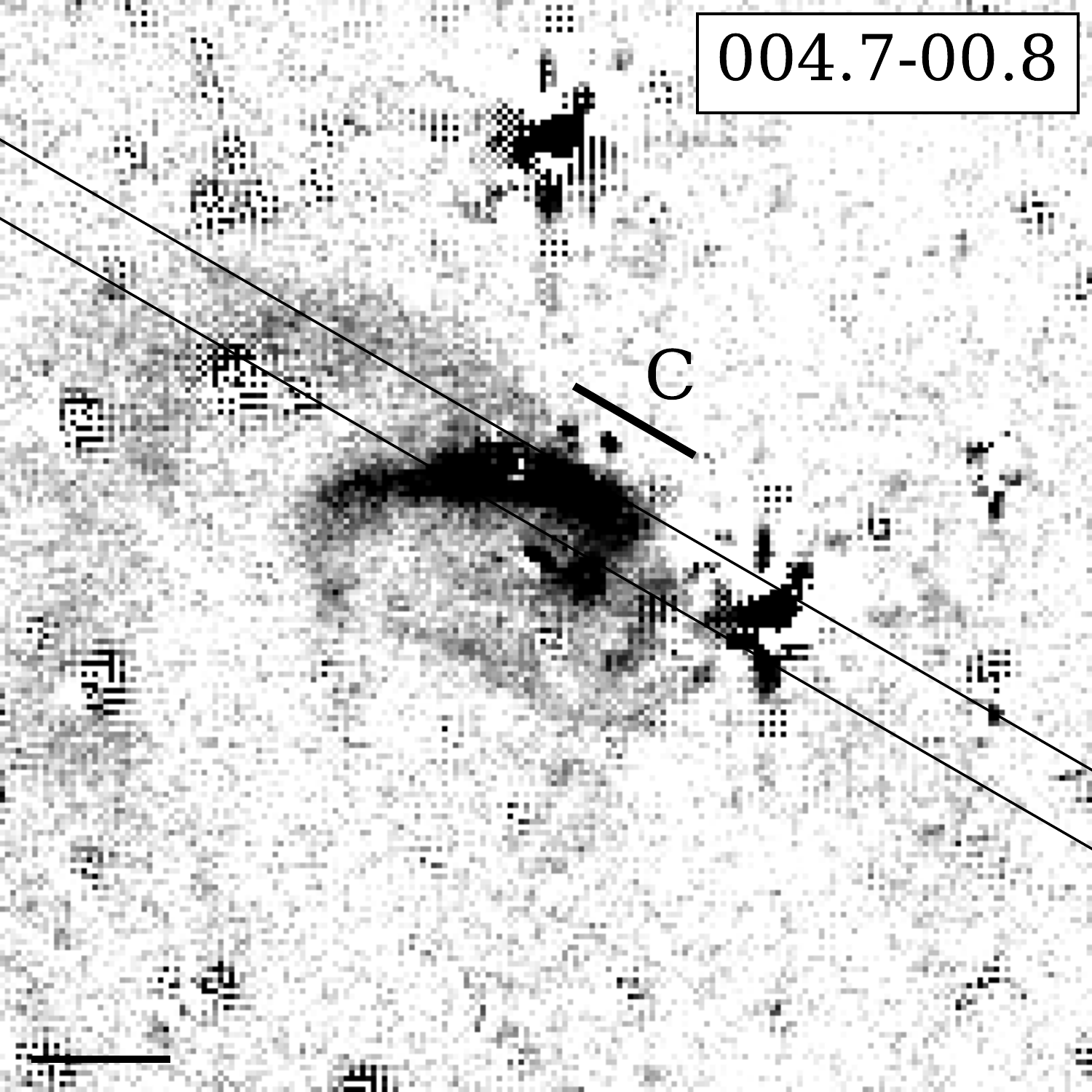}
	\end{subfigure}
	\vspace{4mm}
	\hspace{2mm}
	\begin{subfigure}{0.22\textwidth}
	\includegraphics[width=\textwidth]{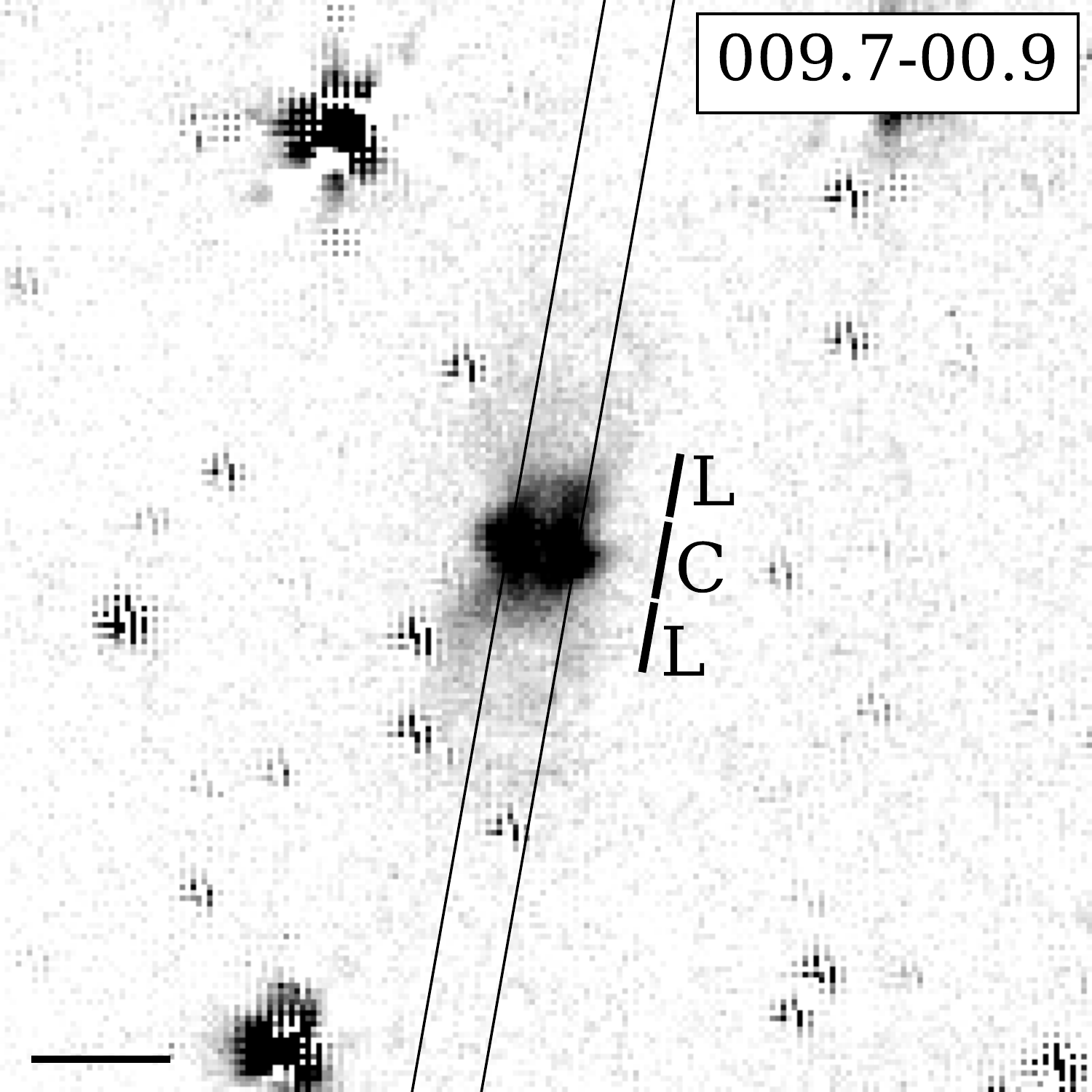}
	\end{subfigure}
	\hspace{2mm}
	\begin{subfigure}{0.22\textwidth}
	\includegraphics[width=\textwidth]{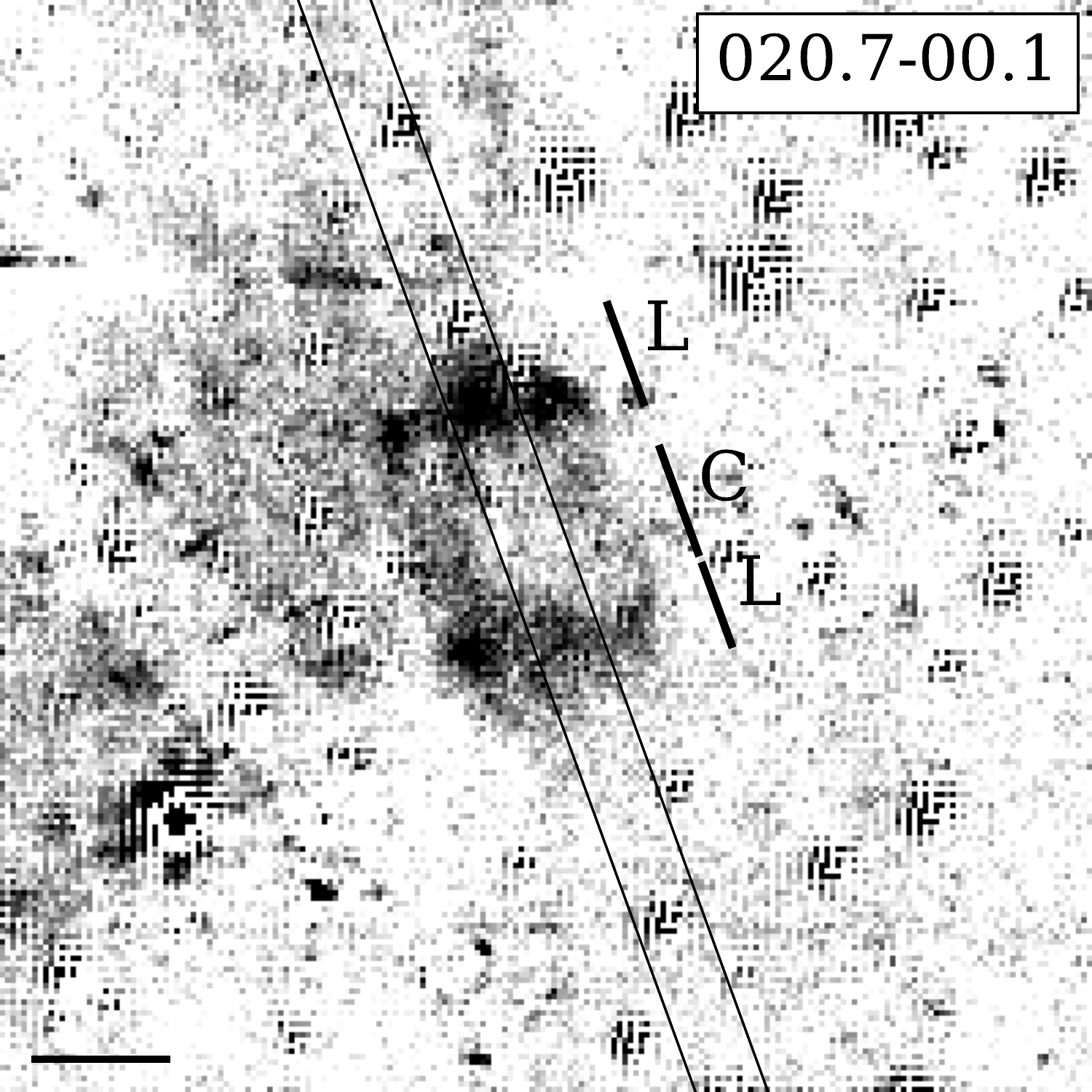}
	\end{subfigure}
	\hspace{2mm}
	\begin{subfigure}{0.22\textwidth}
	\includegraphics[width=\textwidth]{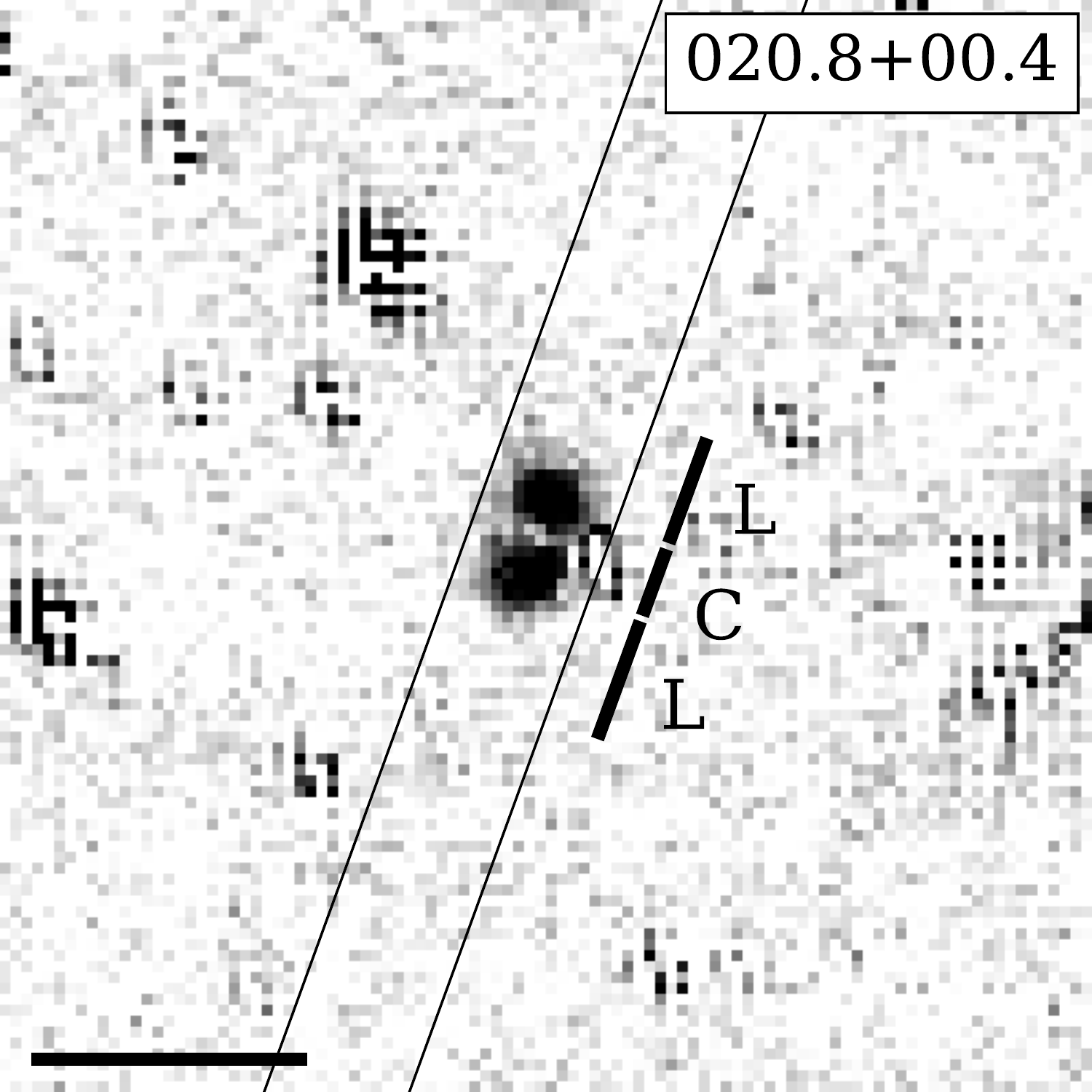}
	\end{subfigure}
	\begin{subfigure}{0.22\textwidth}
	\includegraphics[width=\textwidth]{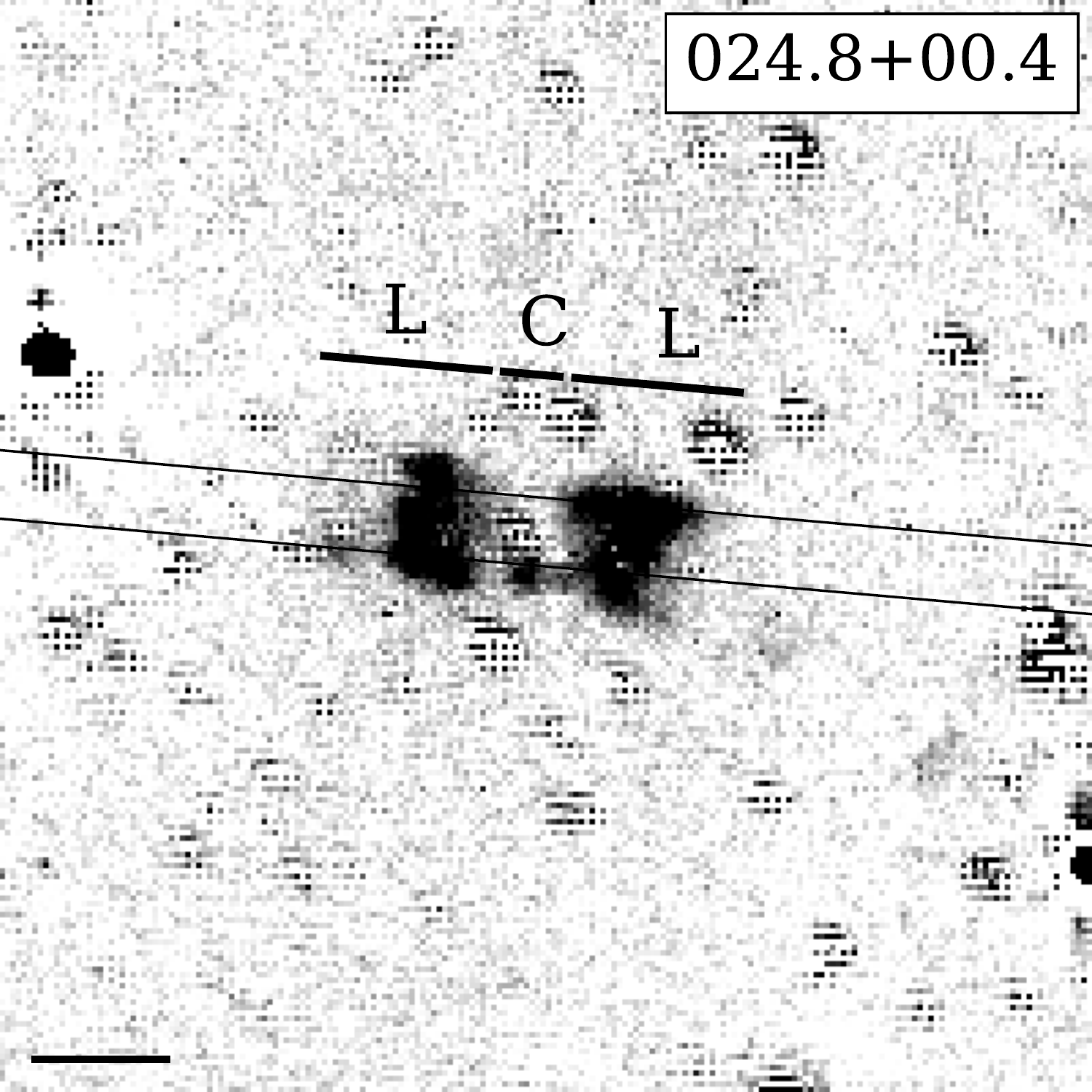}
	\end{subfigure}
	\vspace{4mm}
	\hspace{2mm}	
	\begin{subfigure}{0.22\textwidth}
	\includegraphics[width=\textwidth]{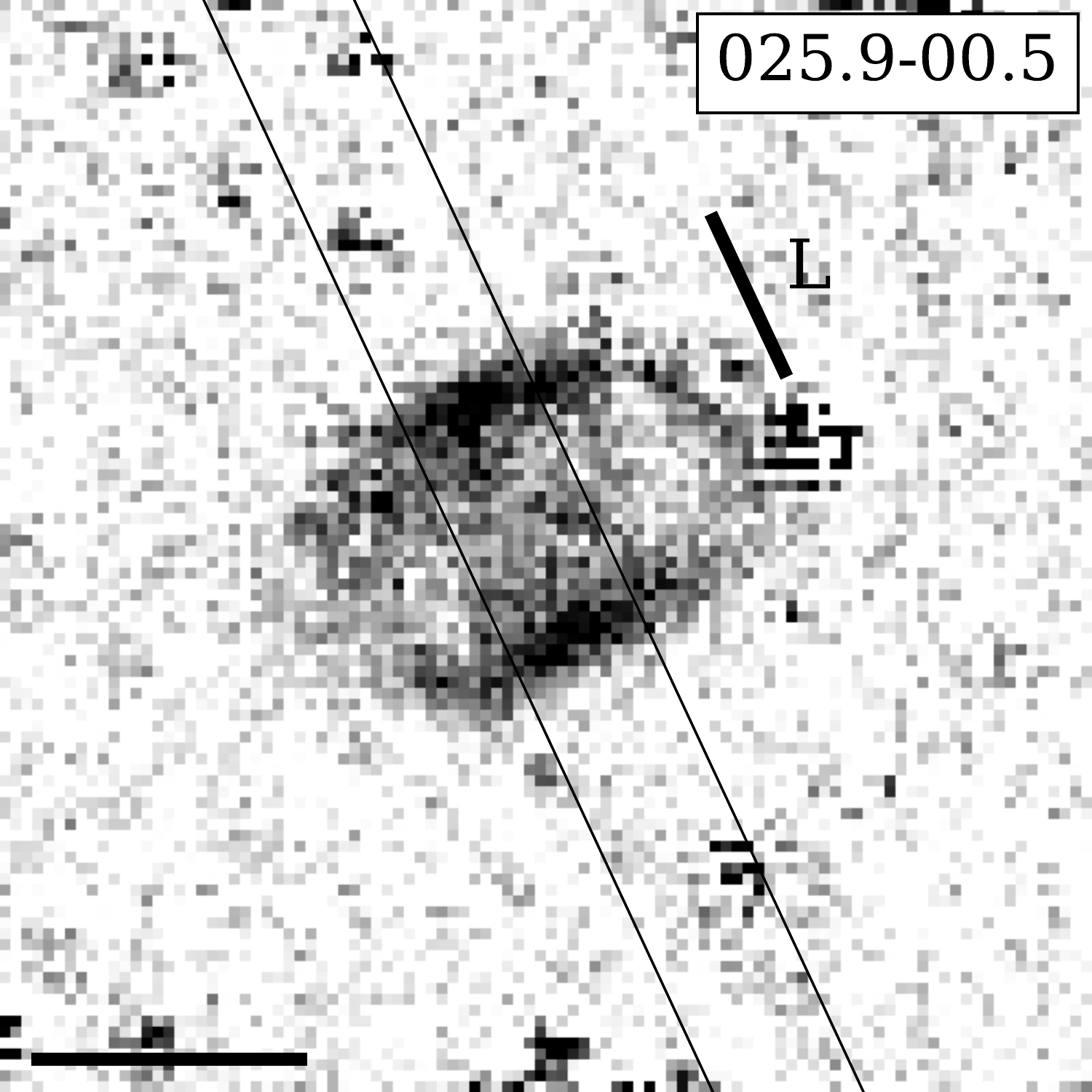}
	\end{subfigure}
	\hspace{2mm}	
	\begin{subfigure}{0.22\textwidth}
	\includegraphics[width=\textwidth]{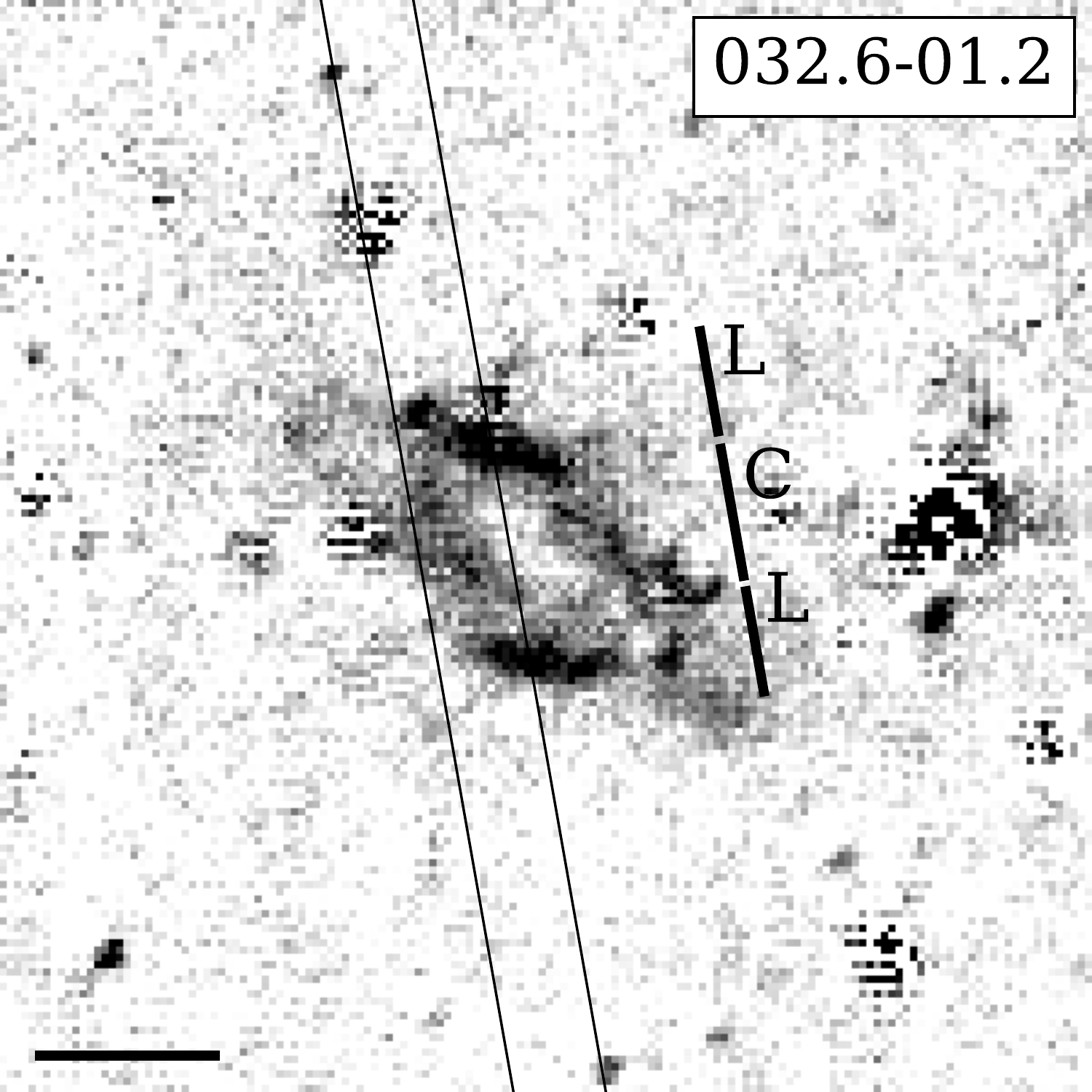}
	\end{subfigure}
	\hspace{2mm}
	\begin{subfigure}{0.22\textwidth}
	\includegraphics[width=\textwidth]{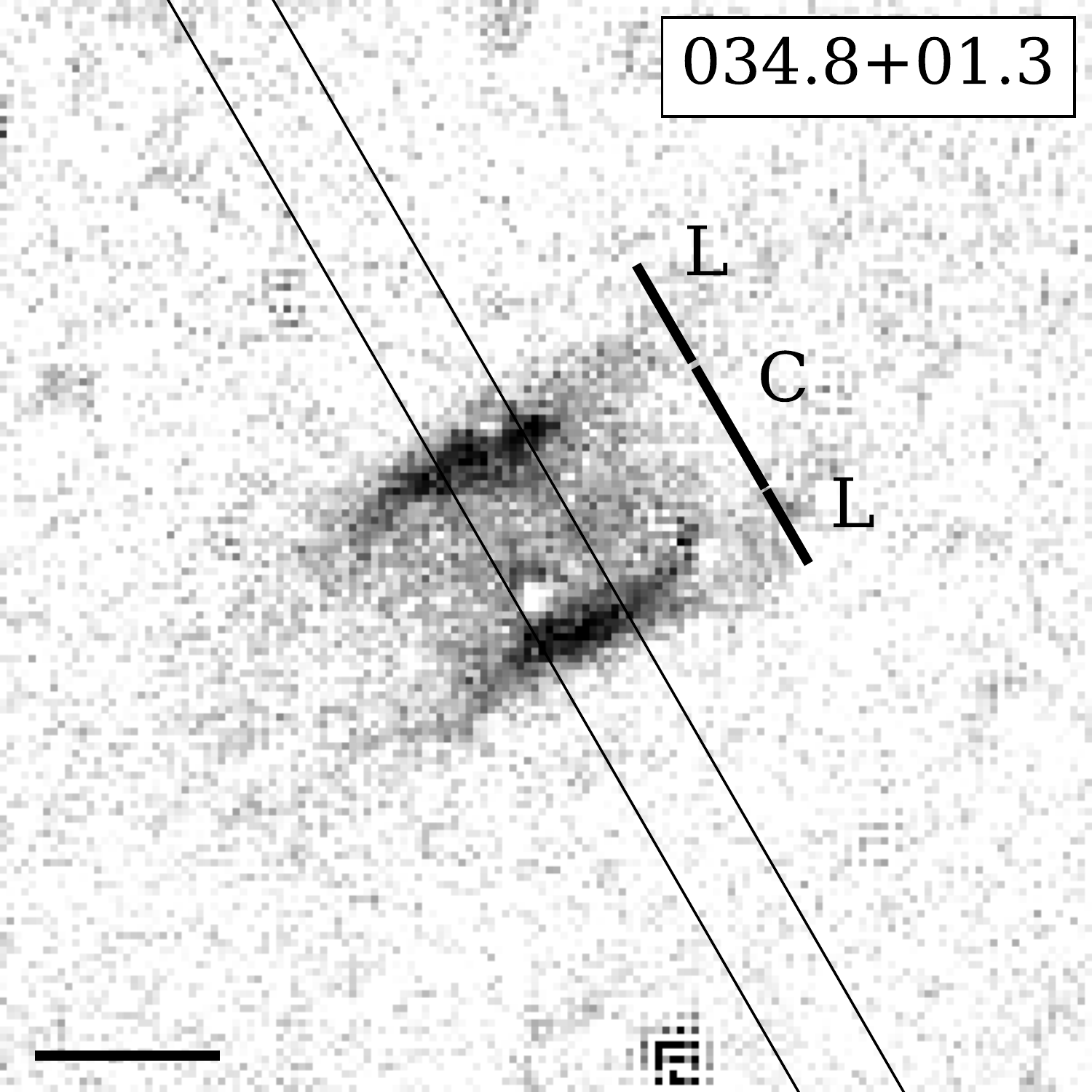}
	\end{subfigure}
	\begin{subfigure}{0.22\textwidth}
	\includegraphics[width=\textwidth]{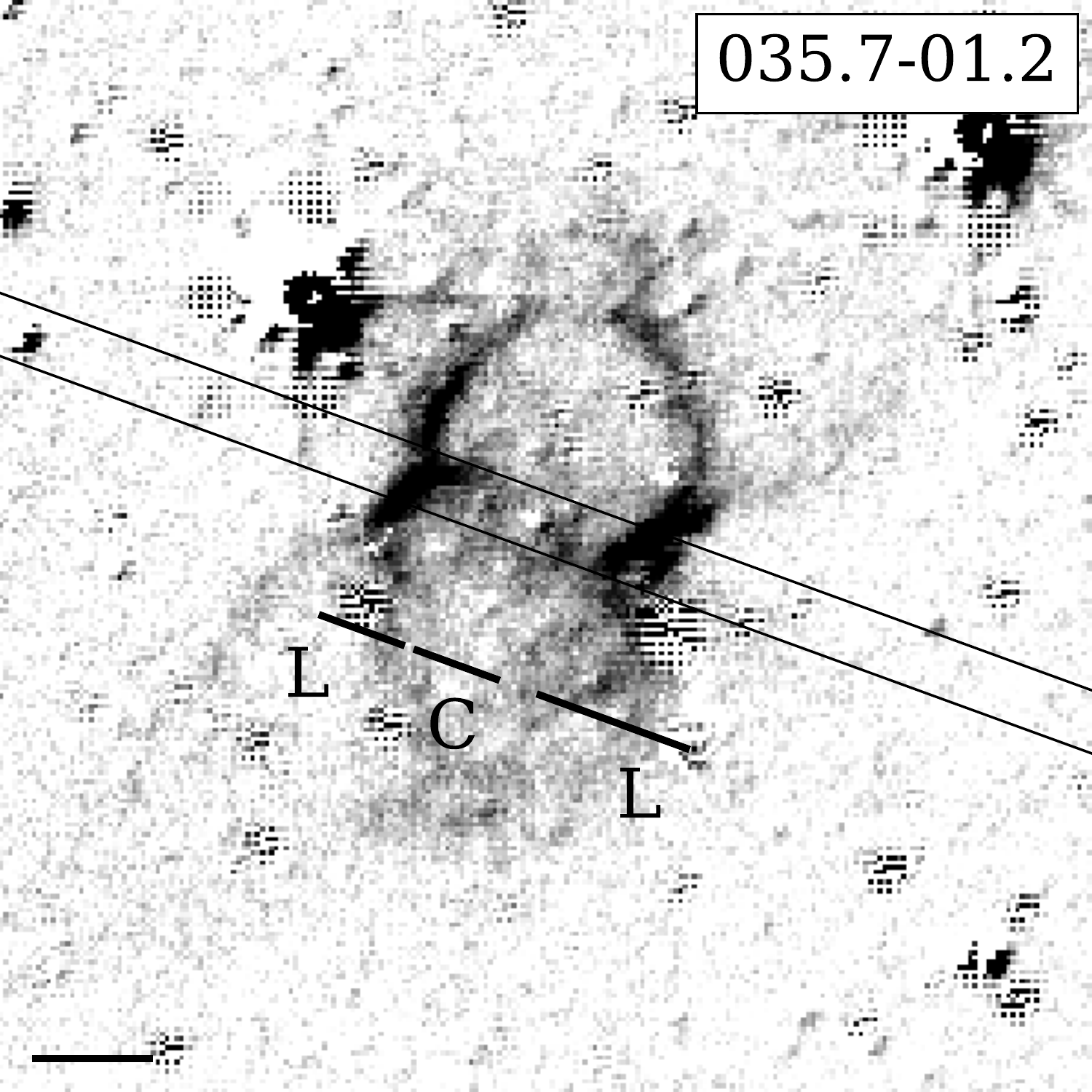}
	\end{subfigure}
	\vspace{4mm}
	\hspace{2mm}
	\begin{subfigure}{0.22\textwidth}
	\includegraphics[width=\textwidth]{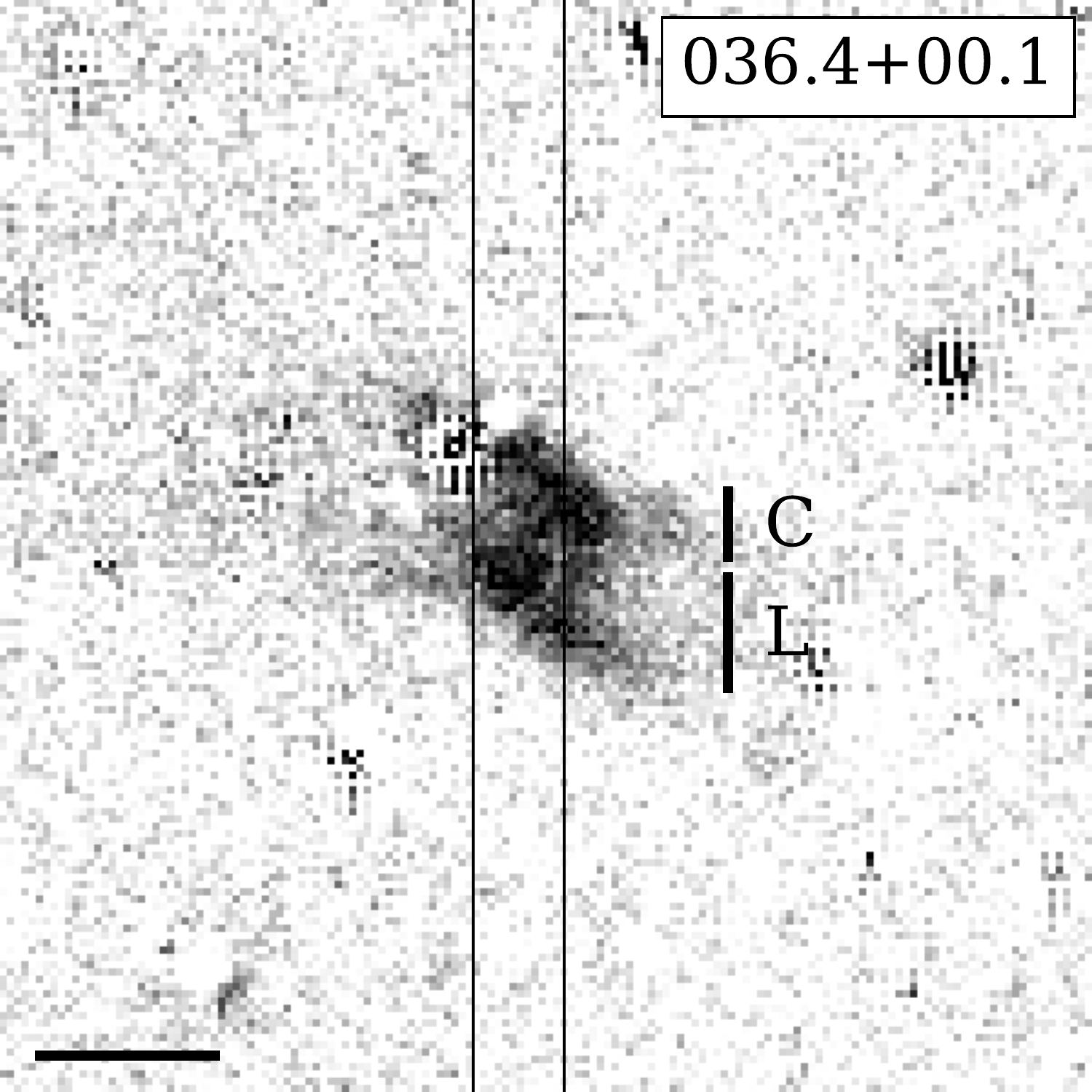}
	\end{subfigure}
	\hspace{2mm}
	\begin{subfigure}{0.22\textwidth}
	\includegraphics[width=\textwidth]{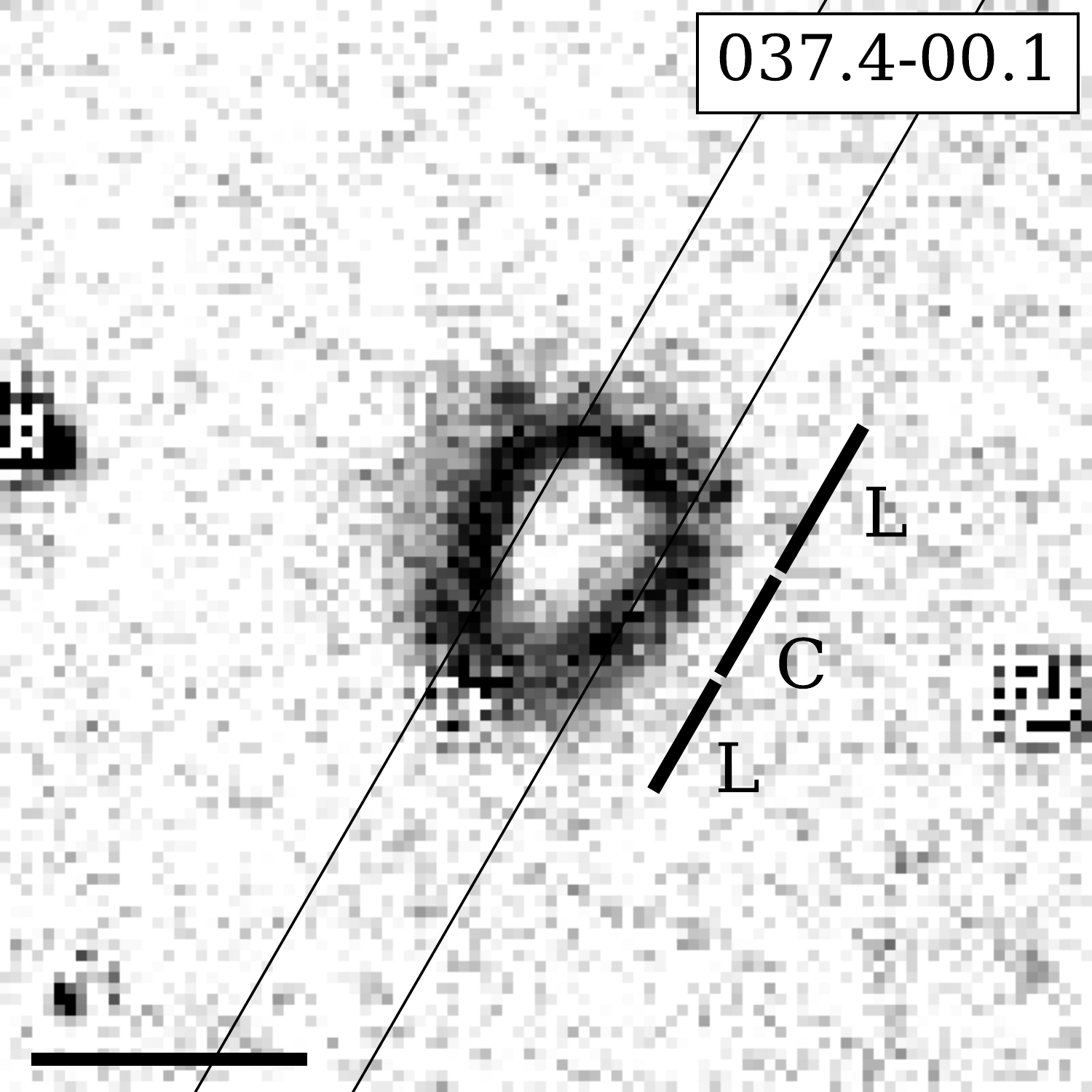}
	\end{subfigure}
	\hspace{2mm}
	\begin{subfigure}{0.22\textwidth}
	\includegraphics[width=\textwidth]{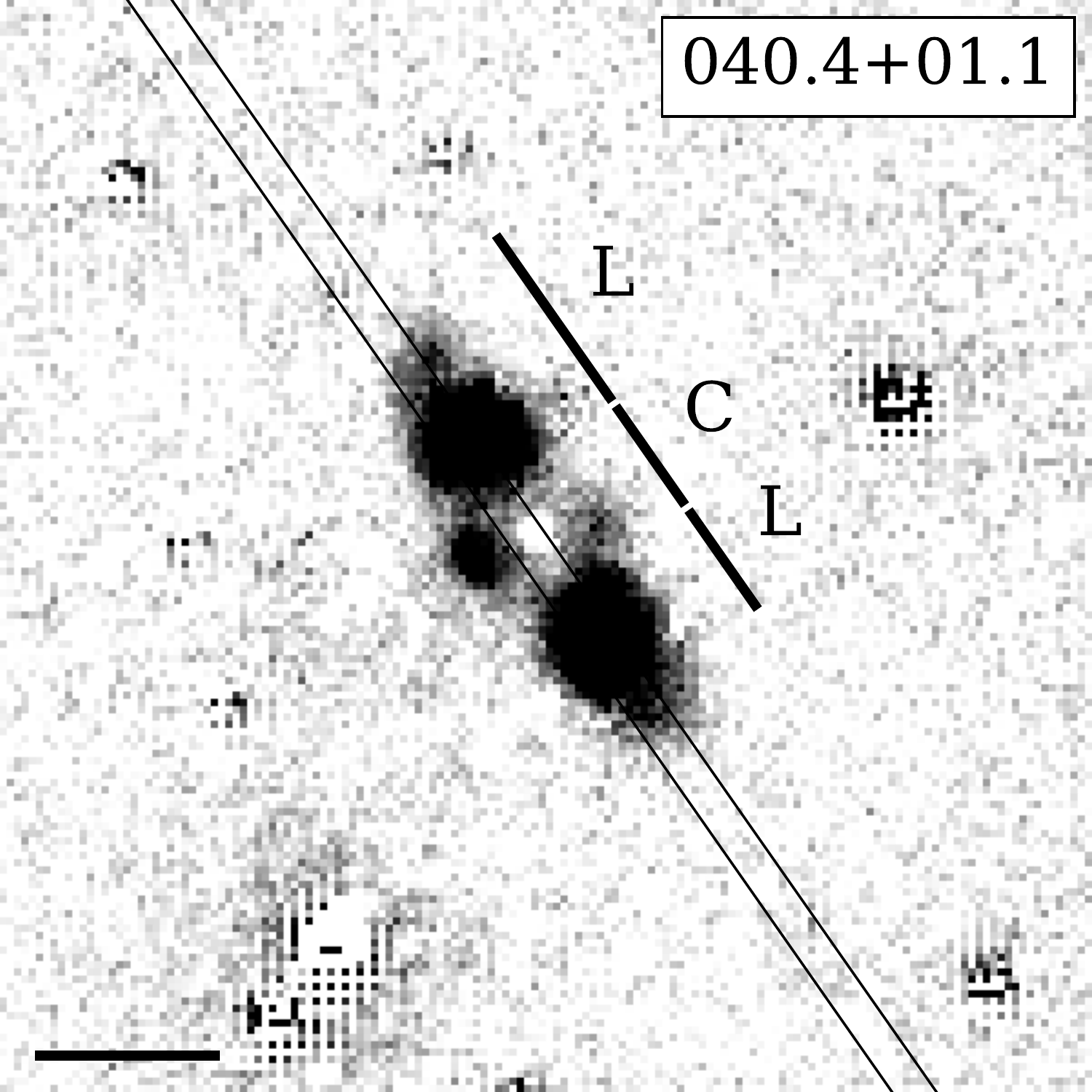}
	\end{subfigure}
	\begin{subfigure}{0.22\textwidth}
	\includegraphics[width=\textwidth]{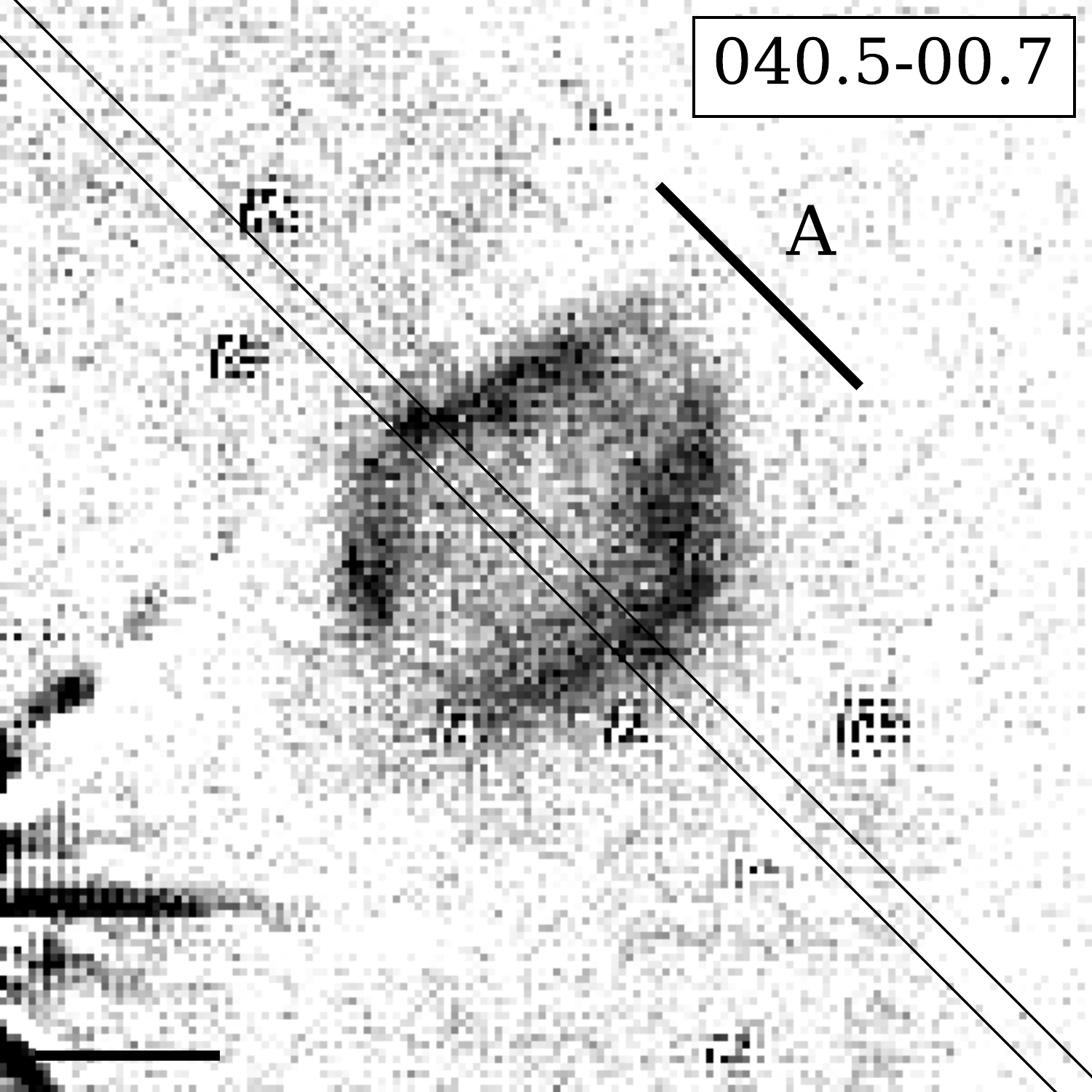}
	\end{subfigure}
	\vspace{4mm}
	\hspace{2mm}
	\begin{subfigure}{0.22\textwidth}
	\includegraphics[width=\textwidth]{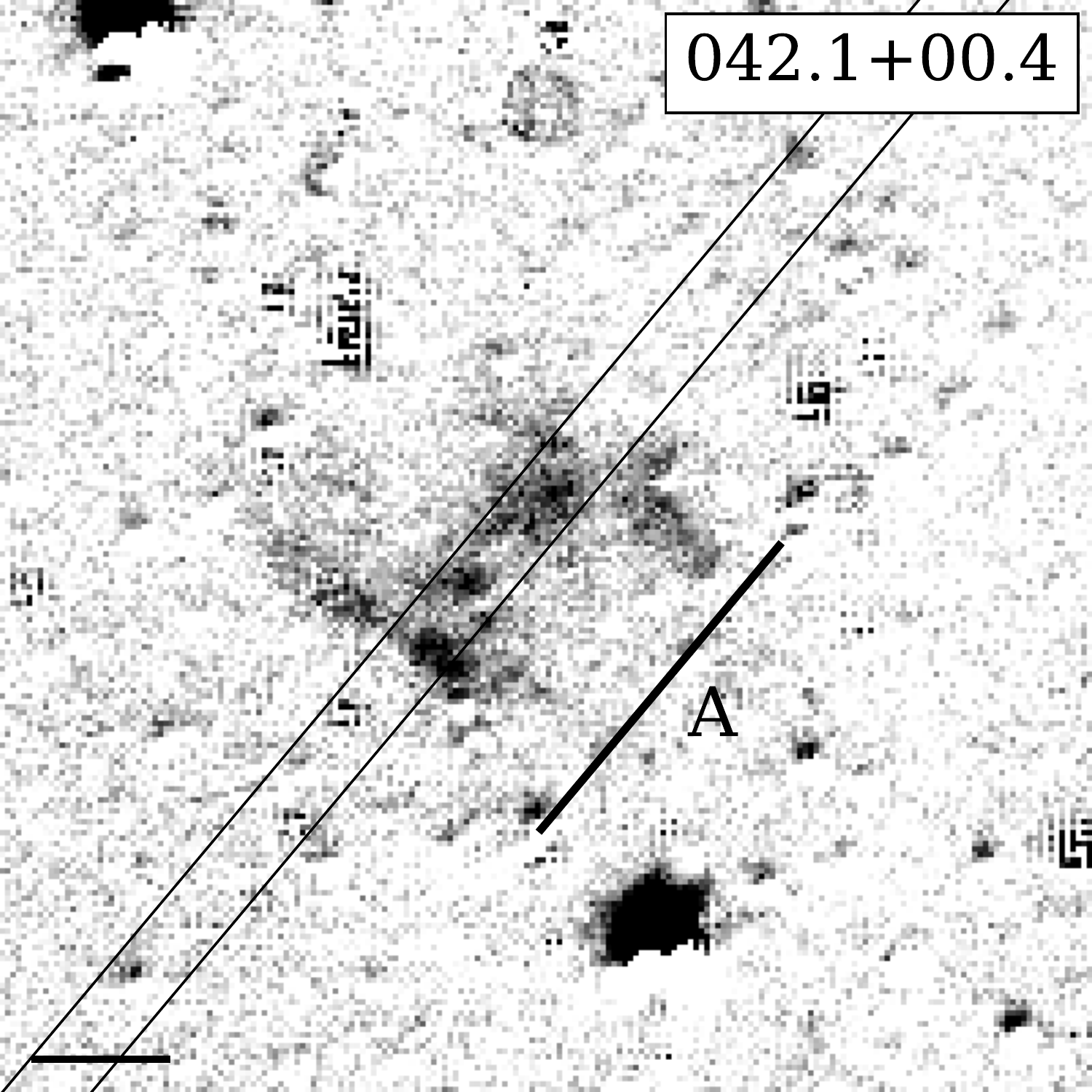}
	\end{subfigure}
	\hspace{2mm}
	\begin{subfigure}{0.22\textwidth}
	\includegraphics[width=\textwidth]{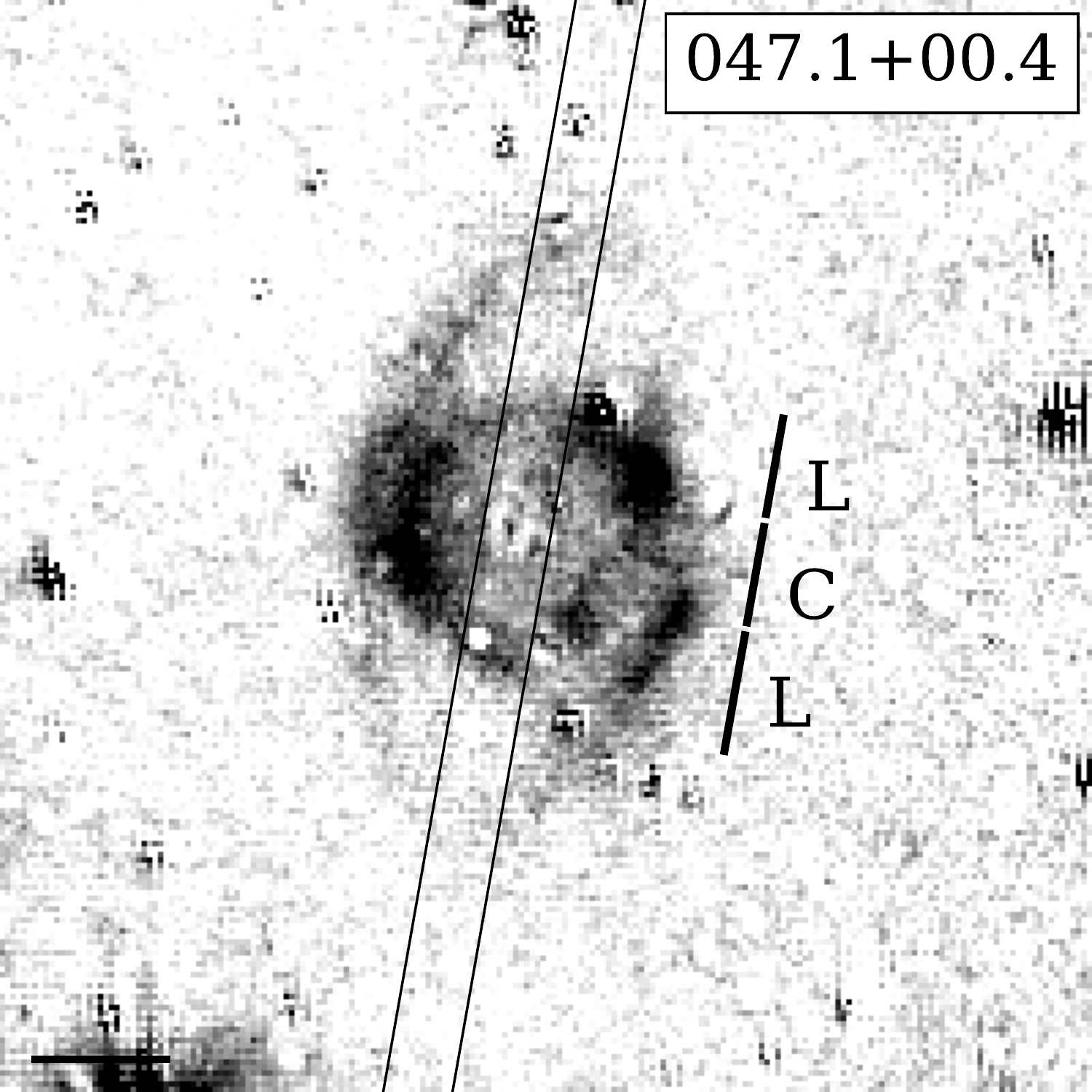}
	\end{subfigure}
	\hspace{2mm}
	\begin{subfigure}{0.22\textwidth}
	\includegraphics[width=\textwidth]{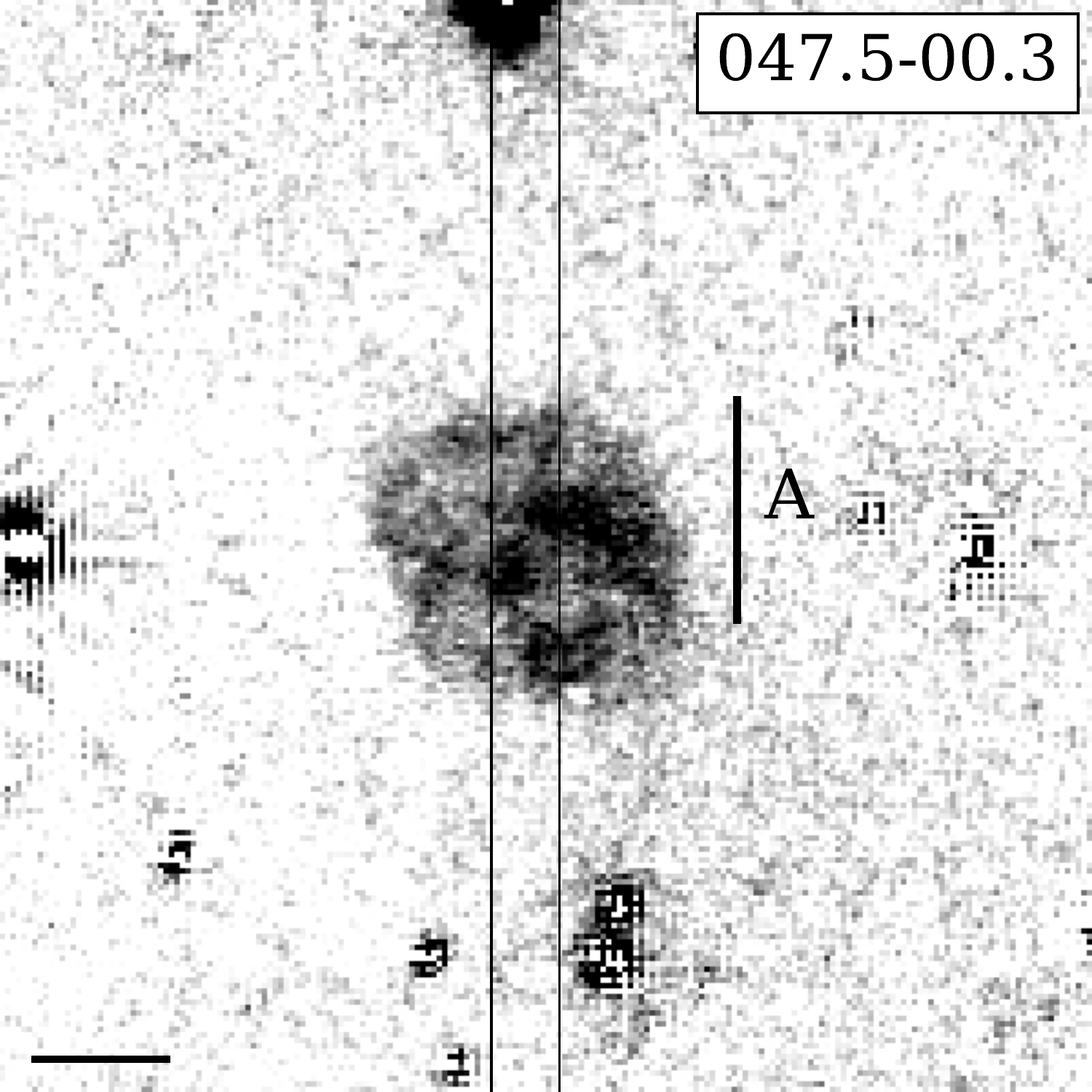}
	\end{subfigure}
\caption{H$_2$ - K images taken from the UWISH2 survey, for targets PN\,G004.7-00.8 to G047.5-00.3, in order of increasing Galactic longitude. We overlay the LIRIS slit size and position, and the sections used for extraction - these are denoted `C' for the centre, `L' for a lobe and `A' for all the target. The horizontal bar in the bottom left corner is a scale bar representing 5 arcsec. In all figures, north is up, east is left.}
\label{fig:h2images}
\end{figure*}

\begin{figure*}\ContinuedFloat
	\begin{subfigure}{0.22\textwidth}
	\includegraphics[width=\textwidth]{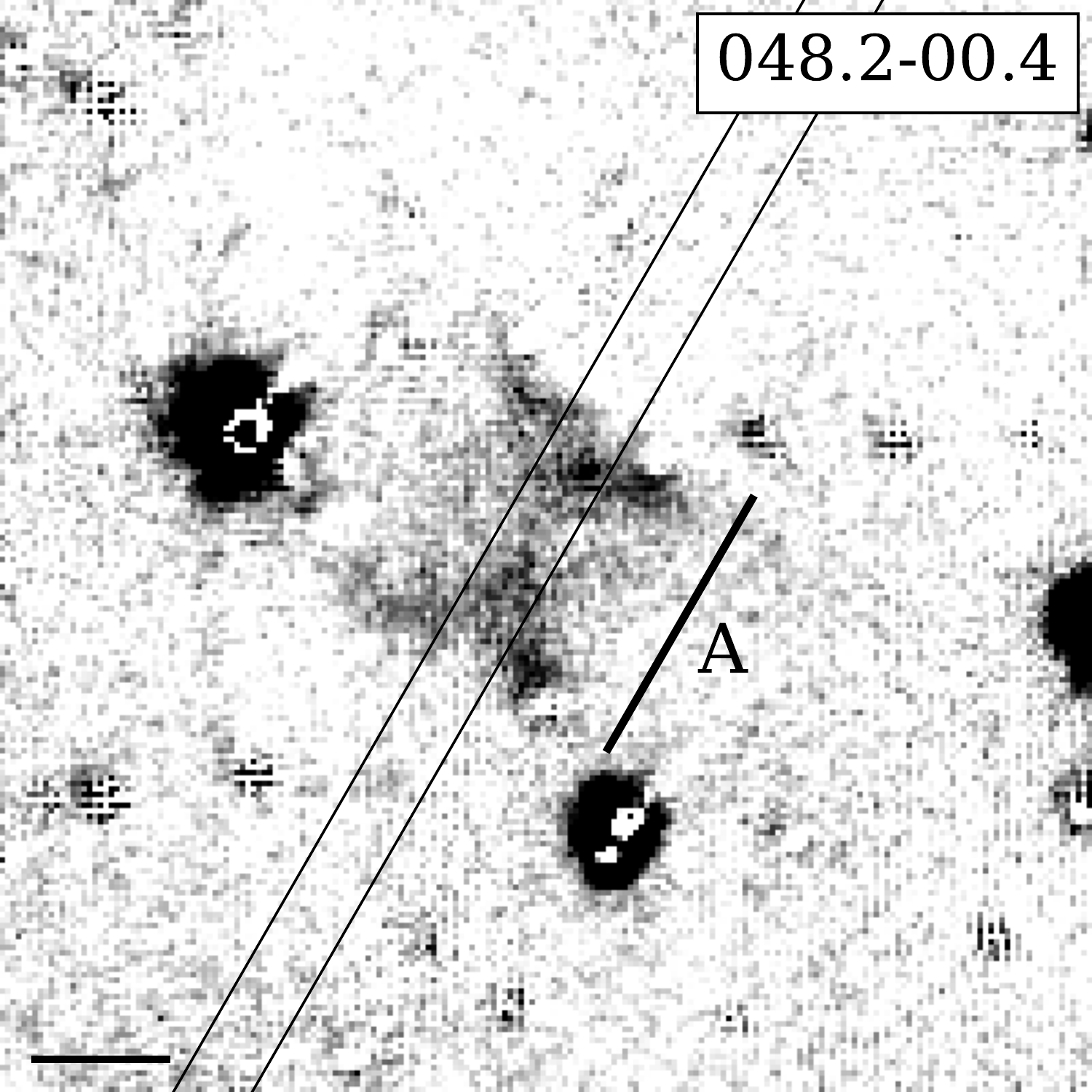}
	\end{subfigure}
	\vspace{4mm}
	\hspace{2mm}
	\begin{subfigure}{0.22\textwidth}
	\includegraphics[width=\textwidth]{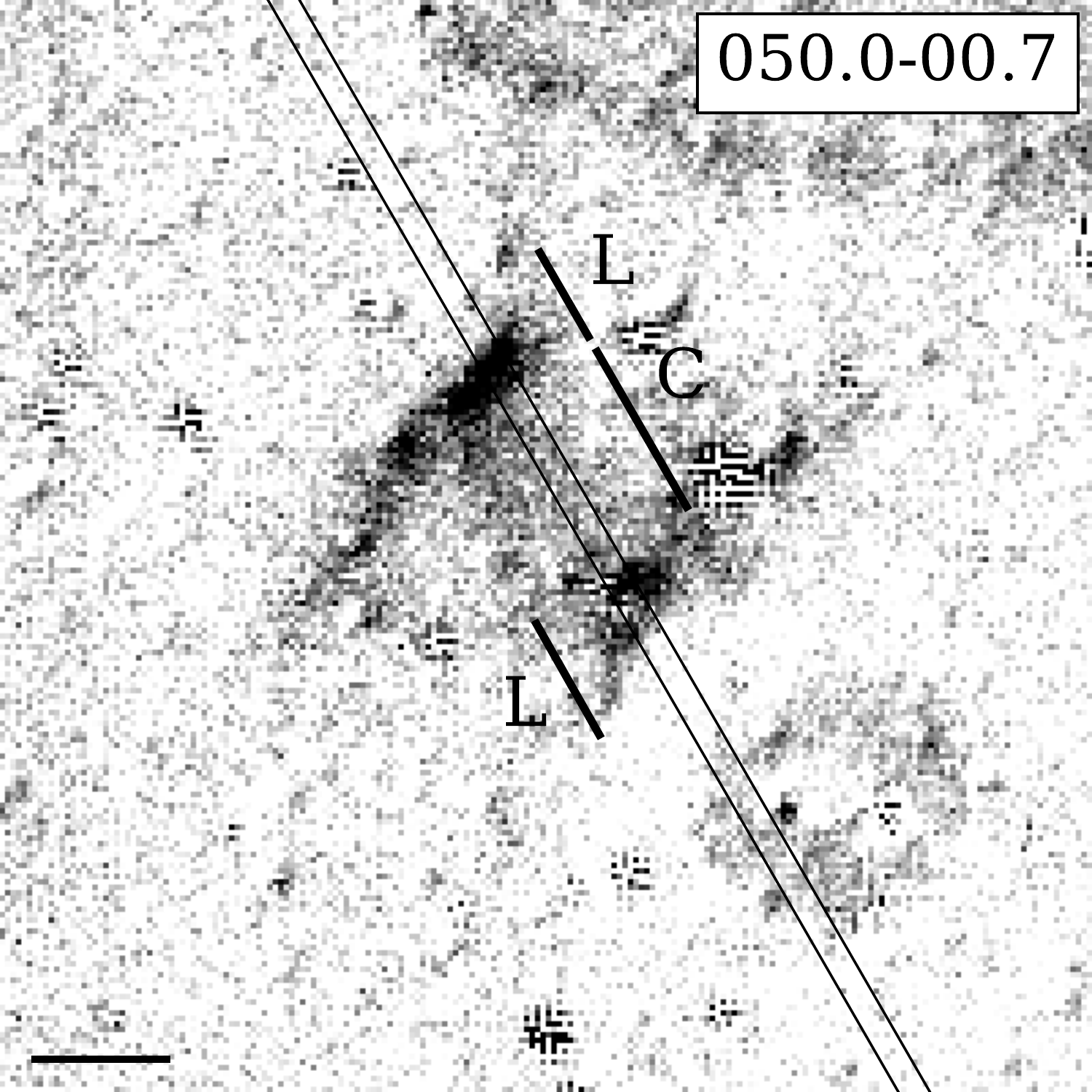}
	\end{subfigure}
	\hspace{2mm}
	\begin{subfigure}{0.22\textwidth}
	\includegraphics[width=\textwidth]{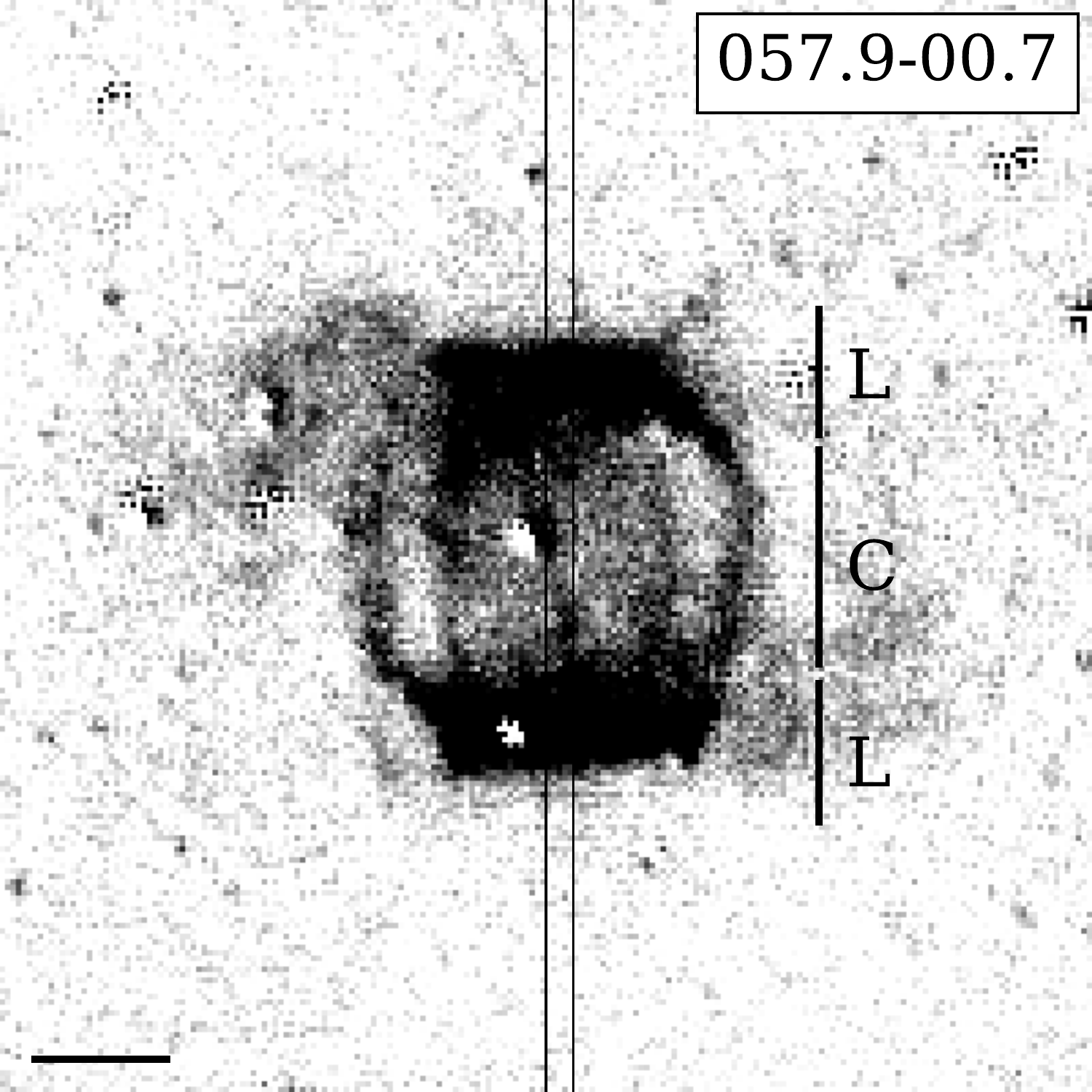}
	\end{subfigure}
	\hspace{2mm}
	\begin{subfigure}{0.22\textwidth}
	\includegraphics[width=\textwidth]{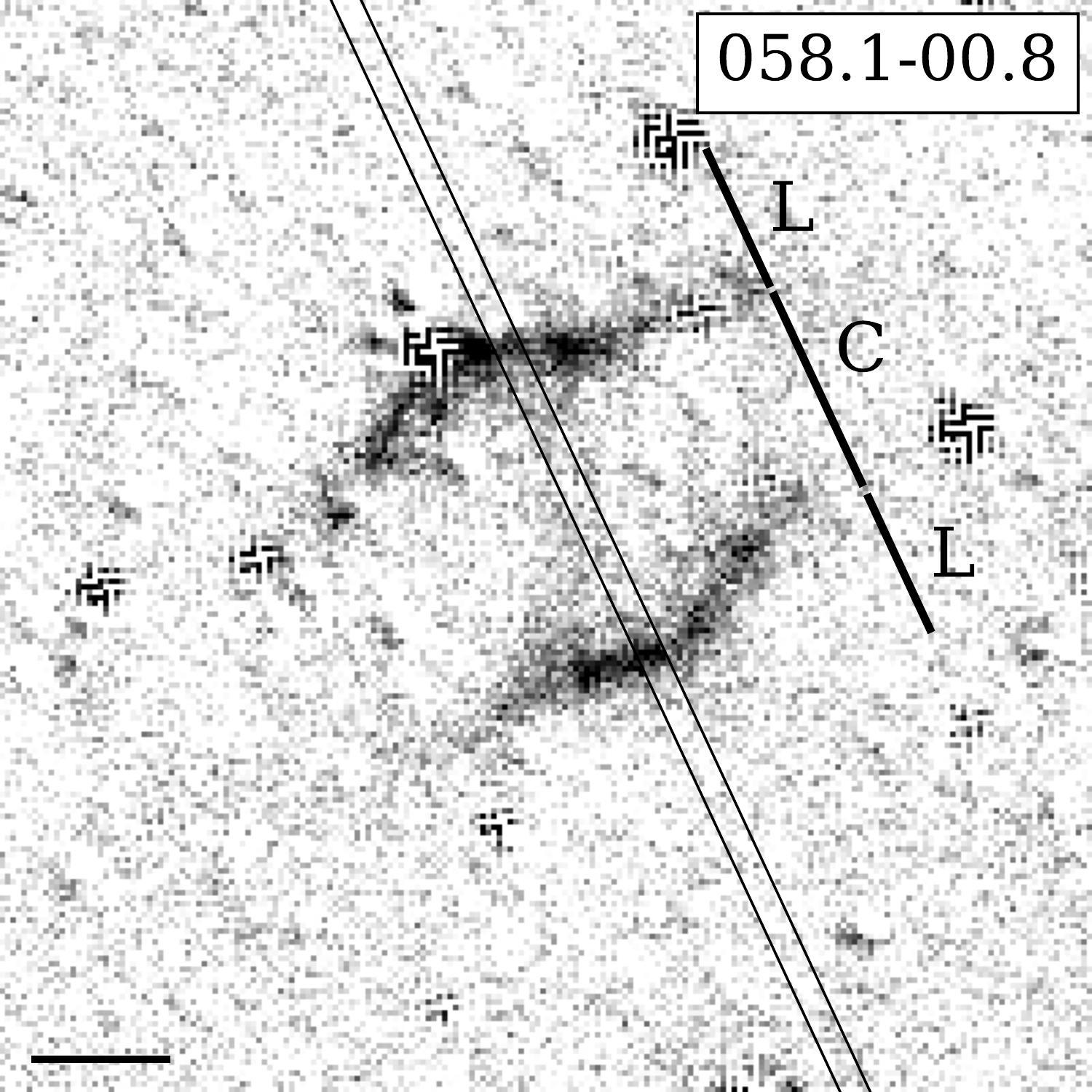}
	\end{subfigure}
	\begin{subfigure}{0.22\textwidth}
	\includegraphics[width=\textwidth]{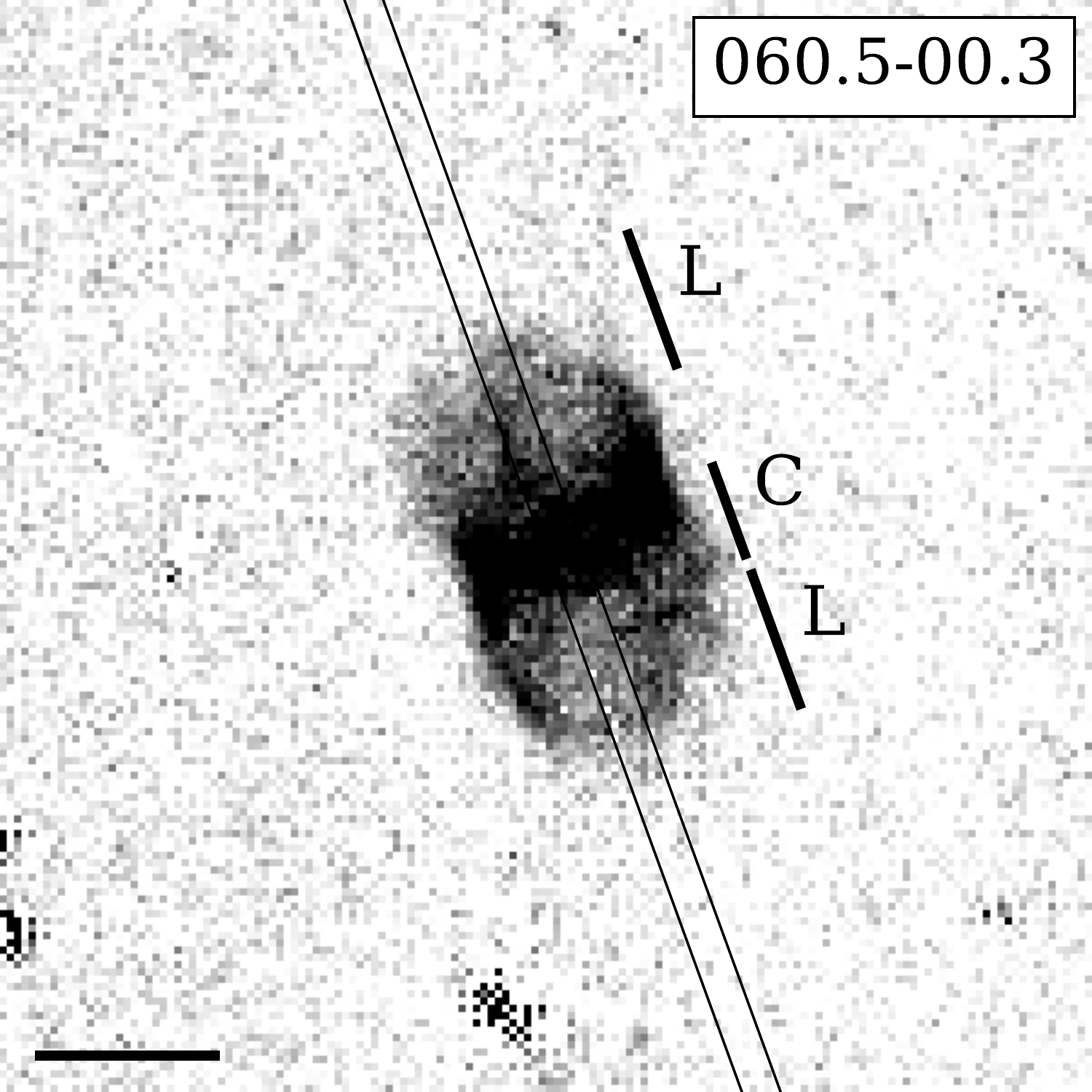}
	\end{subfigure}
	\vspace{4mm}
	\hspace{2mm}
	\begin{subfigure}{0.22\textwidth}
	\includegraphics[width=\textwidth]{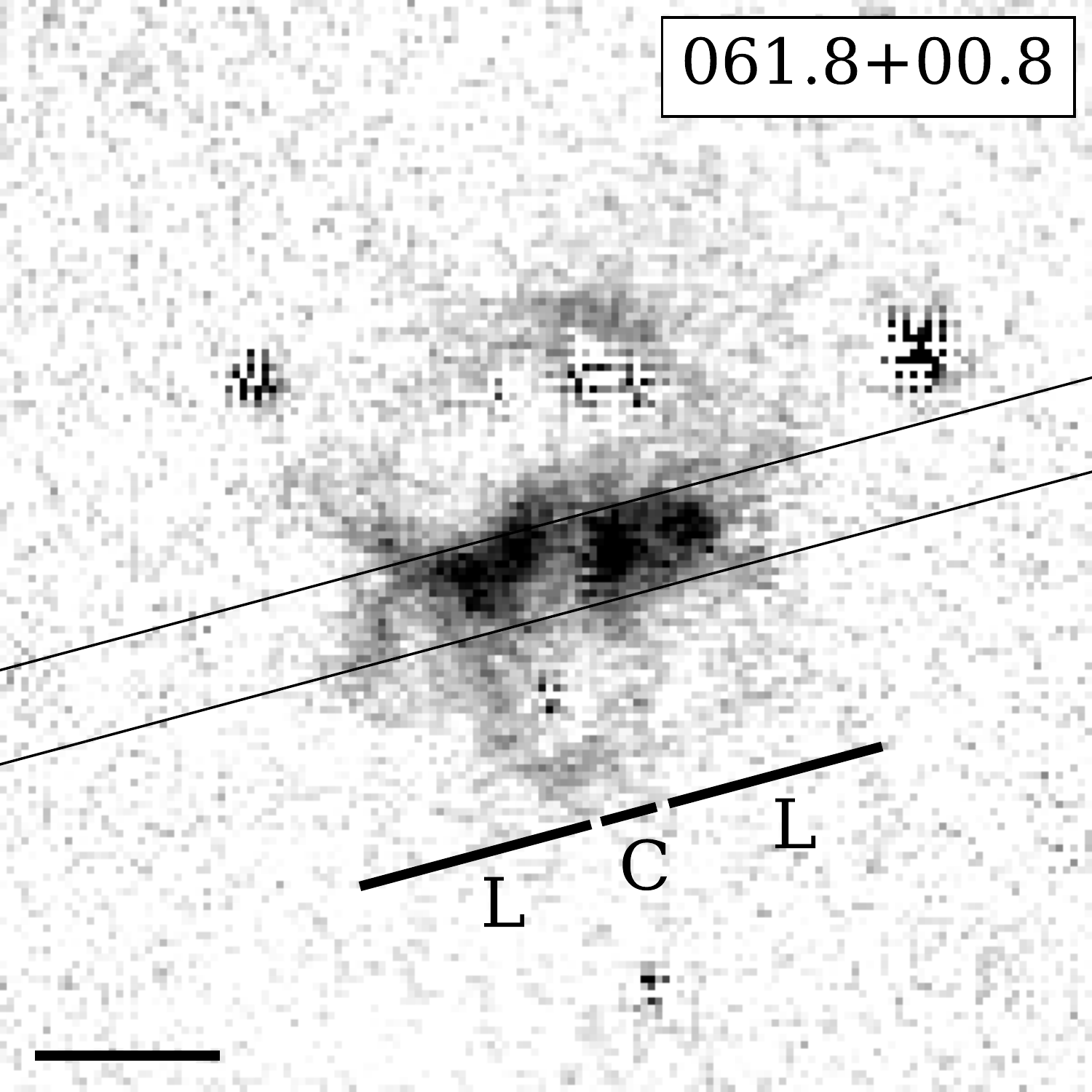}
	\end{subfigure}
	\hspace{2mm}
	\begin{subfigure}{0.22\textwidth}
	\includegraphics[width=\textwidth]{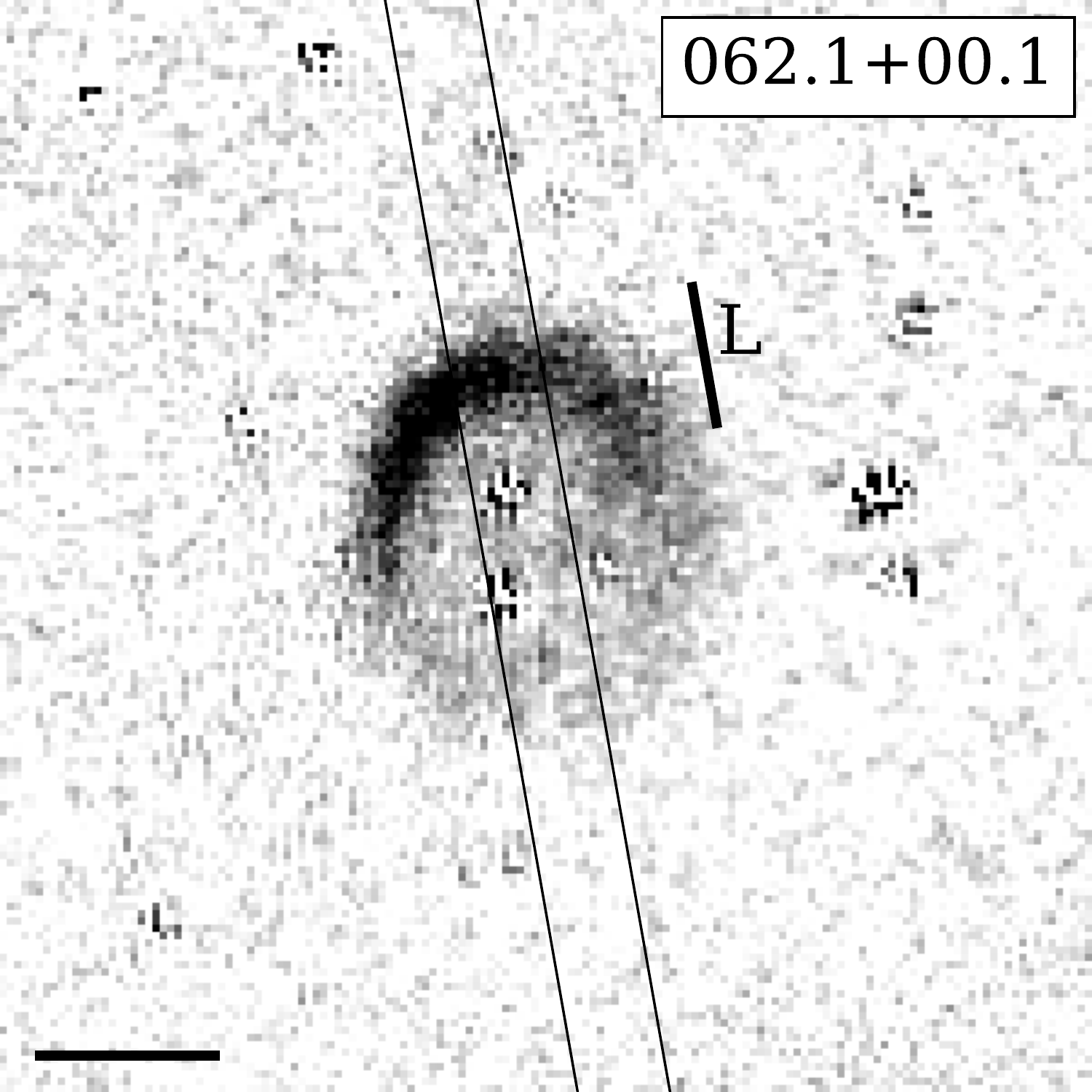}
	\end{subfigure}
	\hspace{2mm}
	\begin{subfigure}{0.22\textwidth}
	\includegraphics[width=\textwidth]{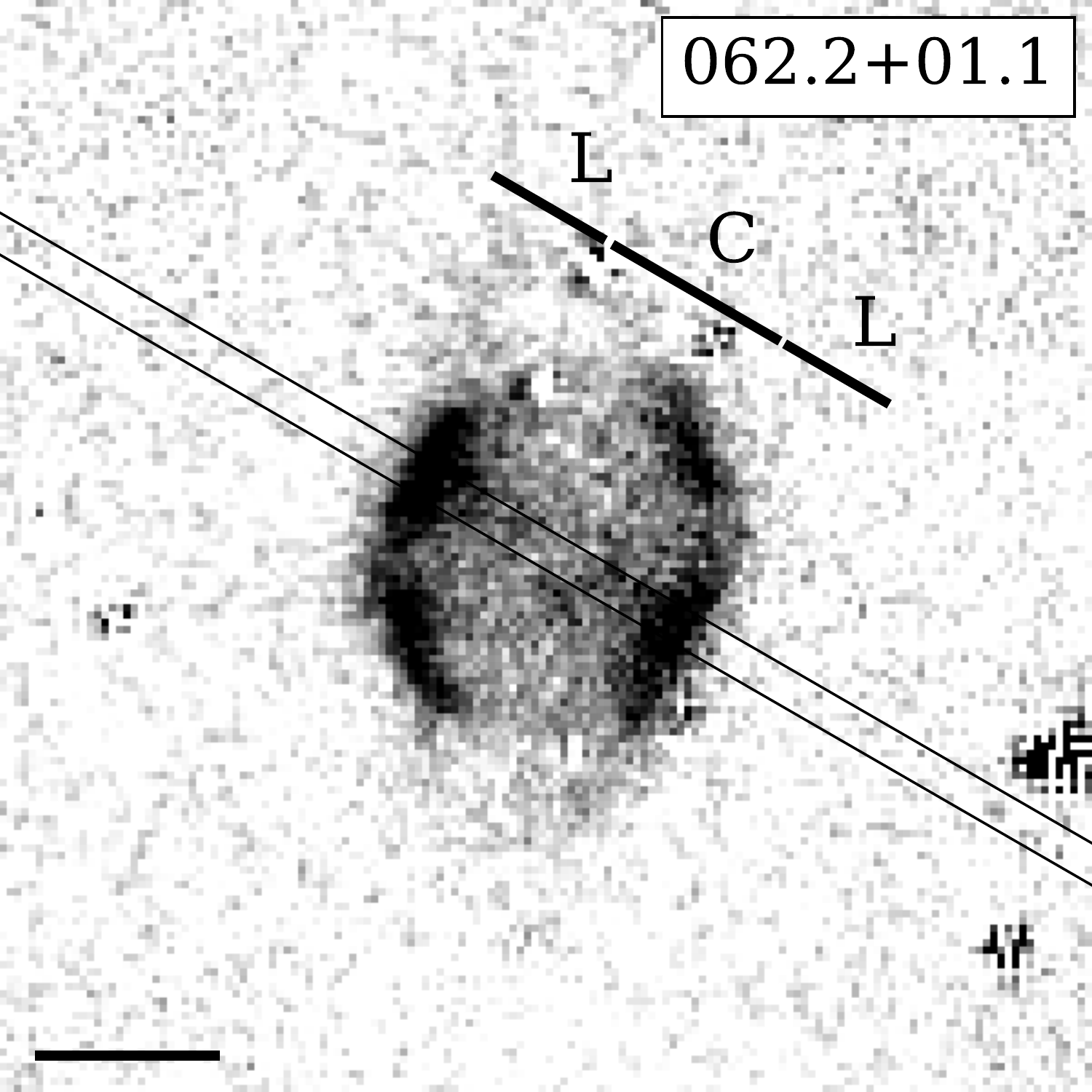}
	\end{subfigure}
	\begin{subfigure}{0.22\textwidth}
	\includegraphics[width=\textwidth]{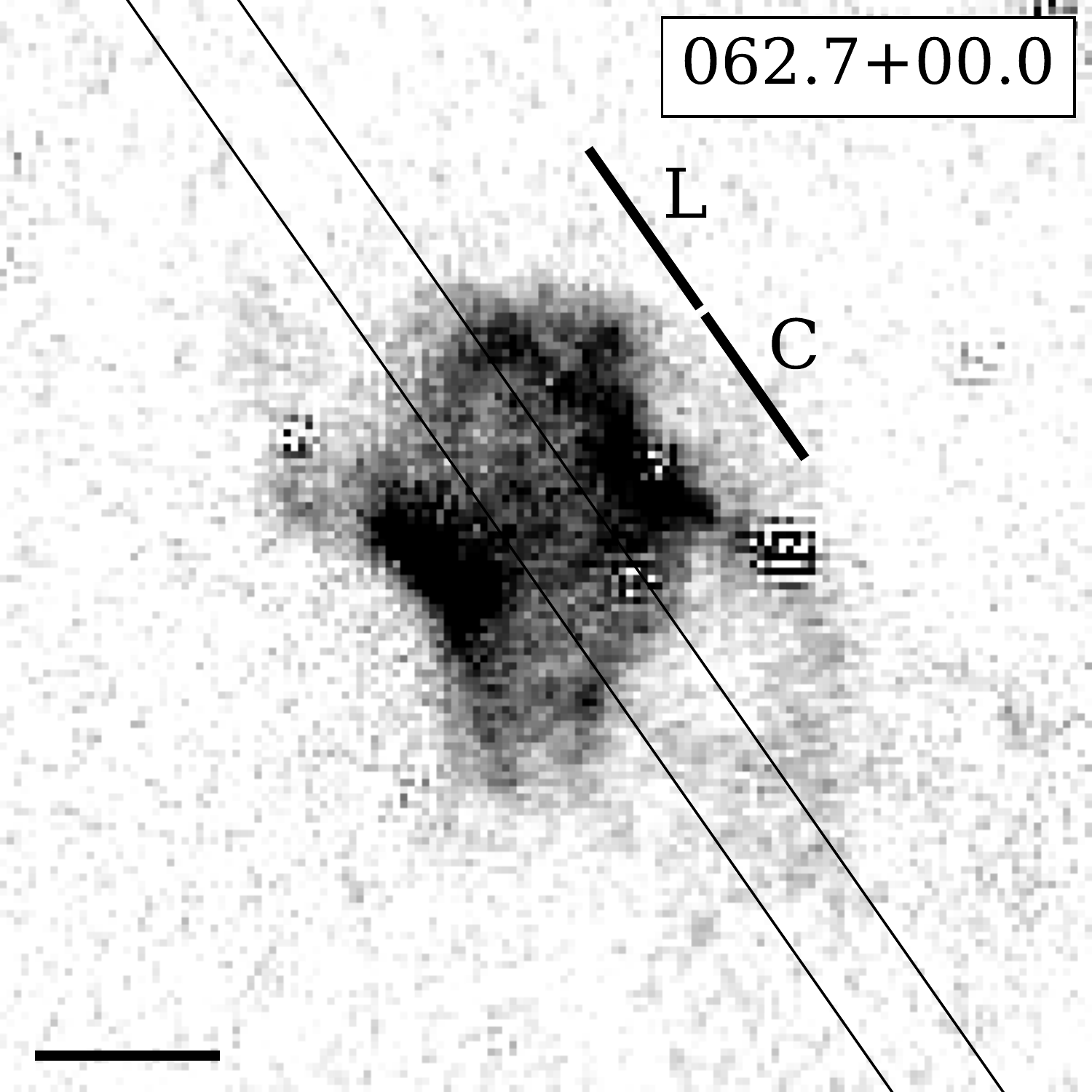}
	\end{subfigure}
	\vspace{4mm}
	\hspace{2mm}
	\begin{subfigure}{0.22\textwidth}
	\includegraphics[width=\textwidth]{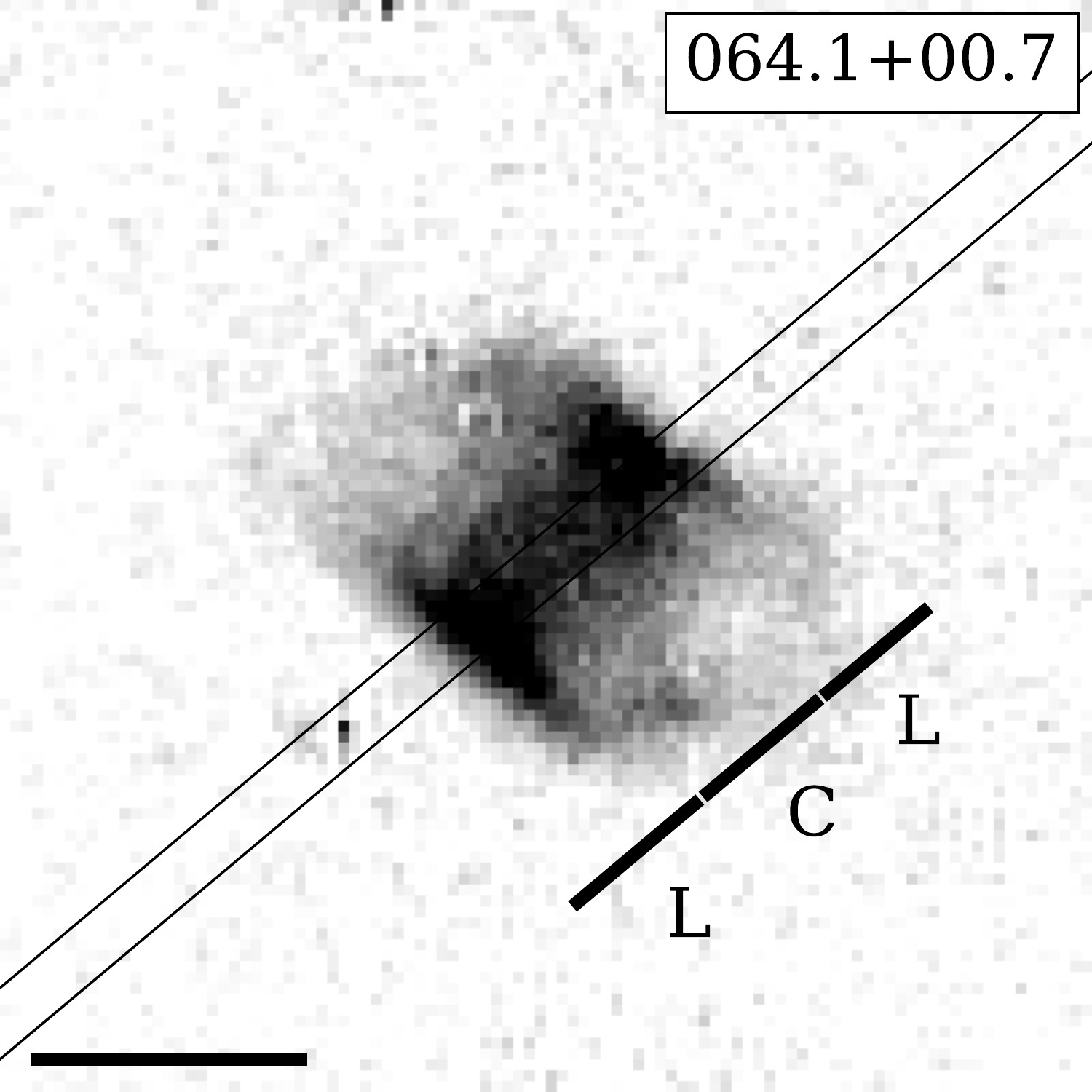}
	\end{subfigure}
	\hspace{2mm}
	\begin{subfigure}{0.22\textwidth}
	\includegraphics[width=\textwidth]{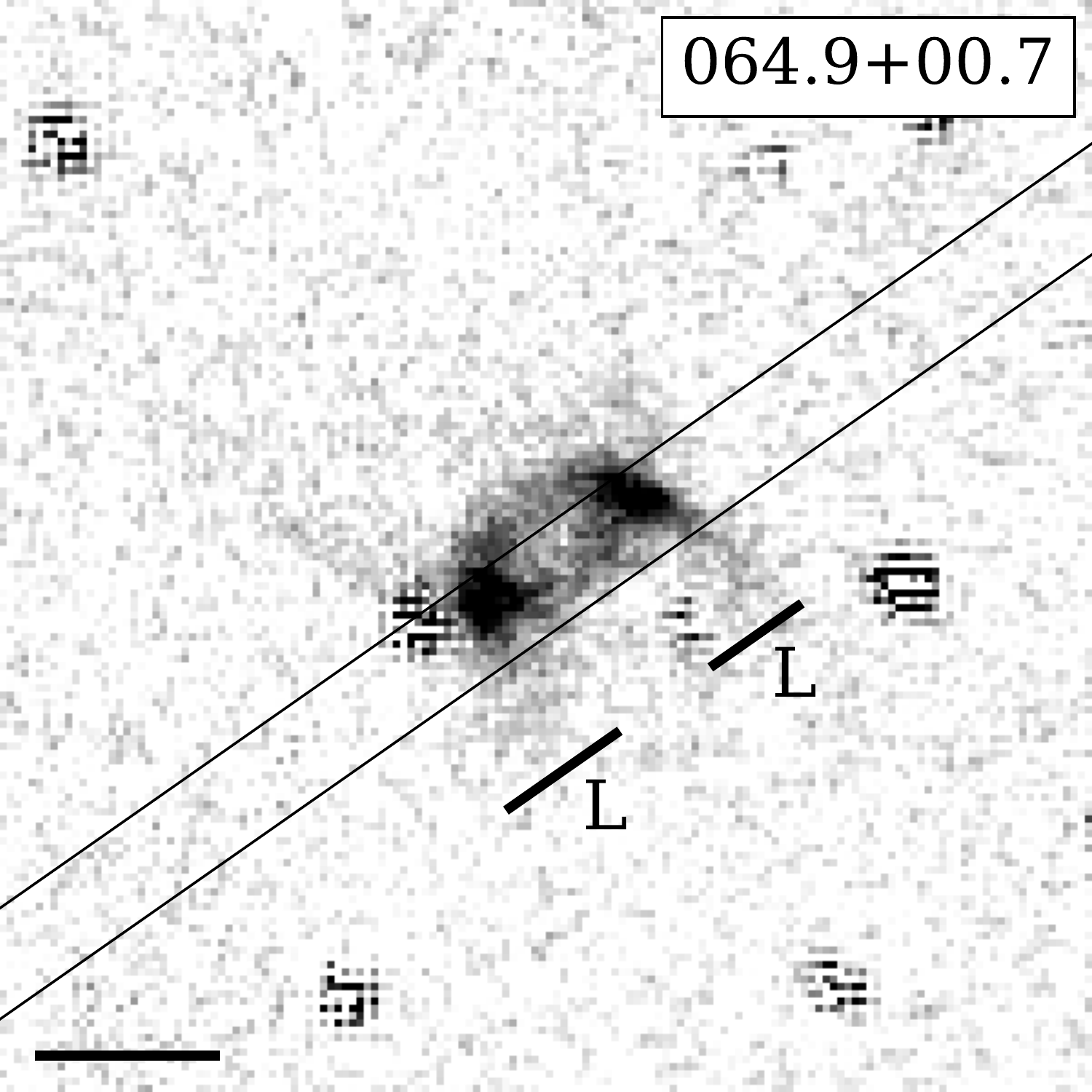}
	\end{subfigure}
\caption{\textbf{(cont.)} H$_2$ - K images for PN\,G048.2-00.4 to G064.9+00.7, excluding PN\,G050.5+00.0 and G059.7-00.8.}
\end{figure*}

\section{K-band spectra}

\begin{figure*}
\centering
\includegraphics[trim = 1.5cm 1.5cm 2.25cm 2.5cm, clip=true, width=1.0\linewidth]{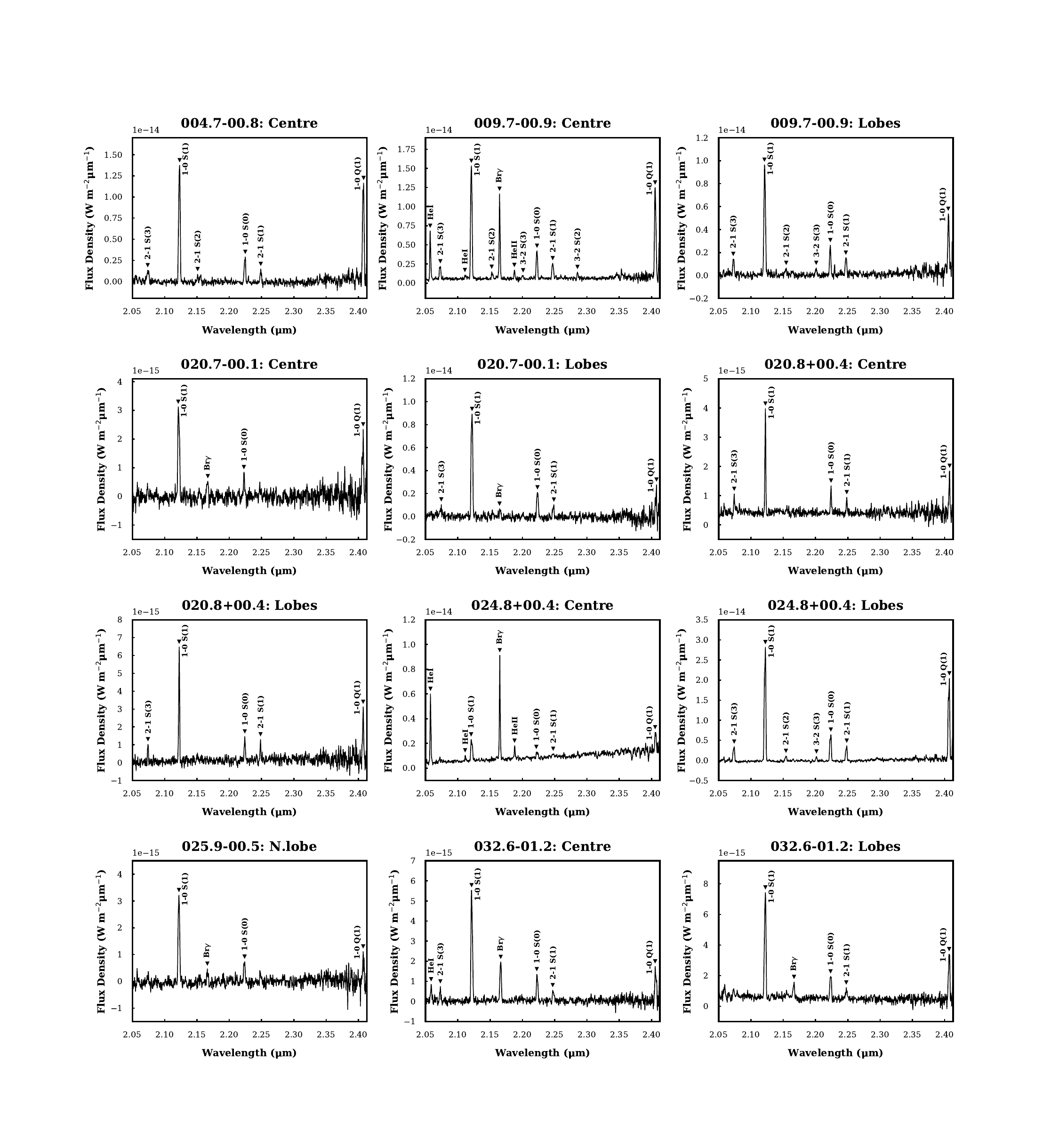}
\caption{K-band spectra for targets PN\,G004.7-00.8 to G032.6-01.2, in order of increasing Galactic longitude. This includes spectra extracted from different regions of the same target, which are given in the title. We label any emission lines present.}
\label{fig:spectra}
\end{figure*}

\begin{figure*}\ContinuedFloat
\centering
\includegraphics[trim = 1.5cm 1.5cm 2.25cm 2.5cm, clip=true, width=1.0\linewidth]{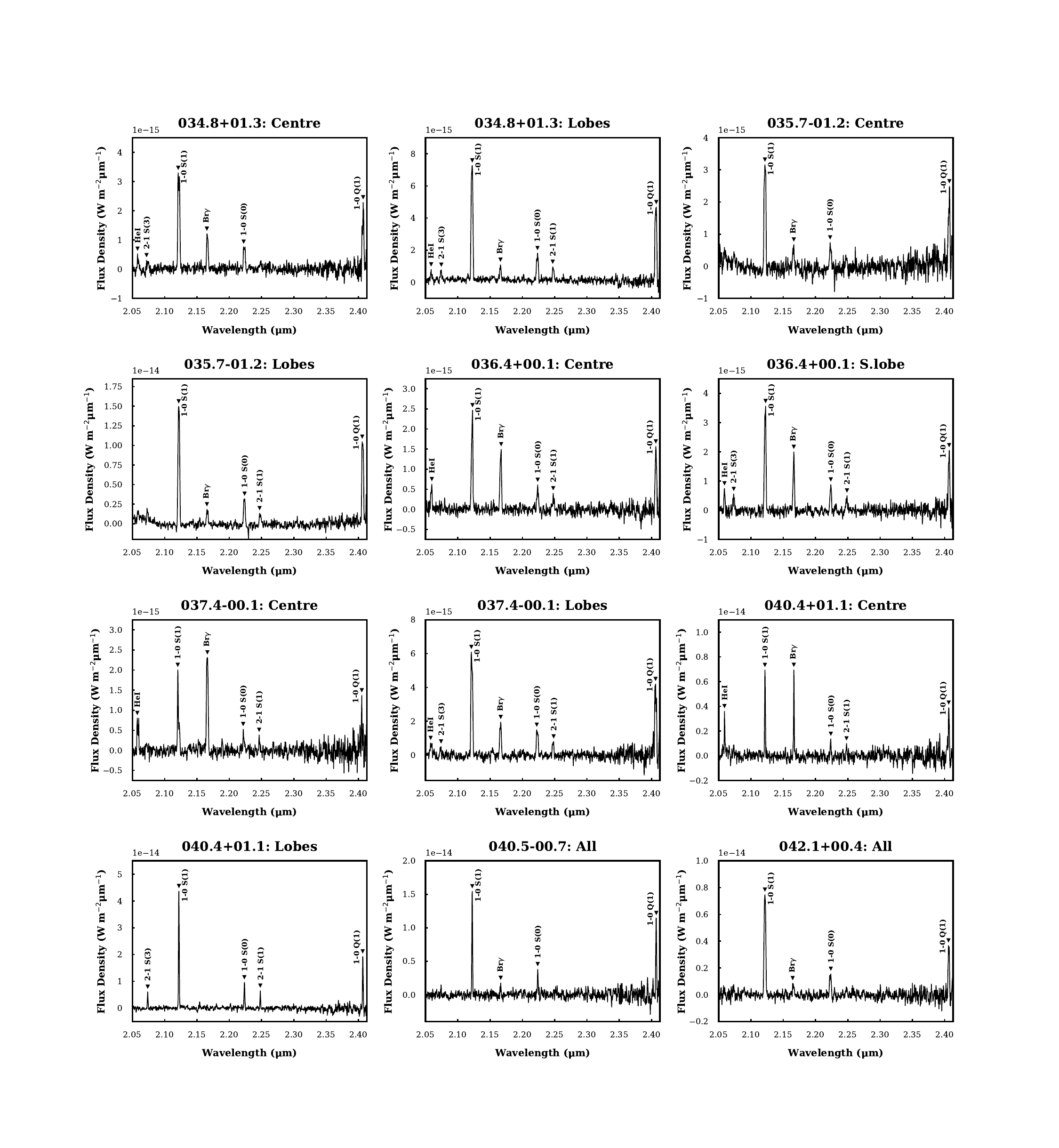}
\caption{\textbf{(cont.)} Spectra for PN\,G034.8+01.3 to G042.1+00.4.}
\end{figure*}

\begin{figure*}\ContinuedFloat
\centering
\includegraphics[trim = 1.5cm 1.5cm 2.25cm 2.5cm, clip=true, width=1.0\linewidth]{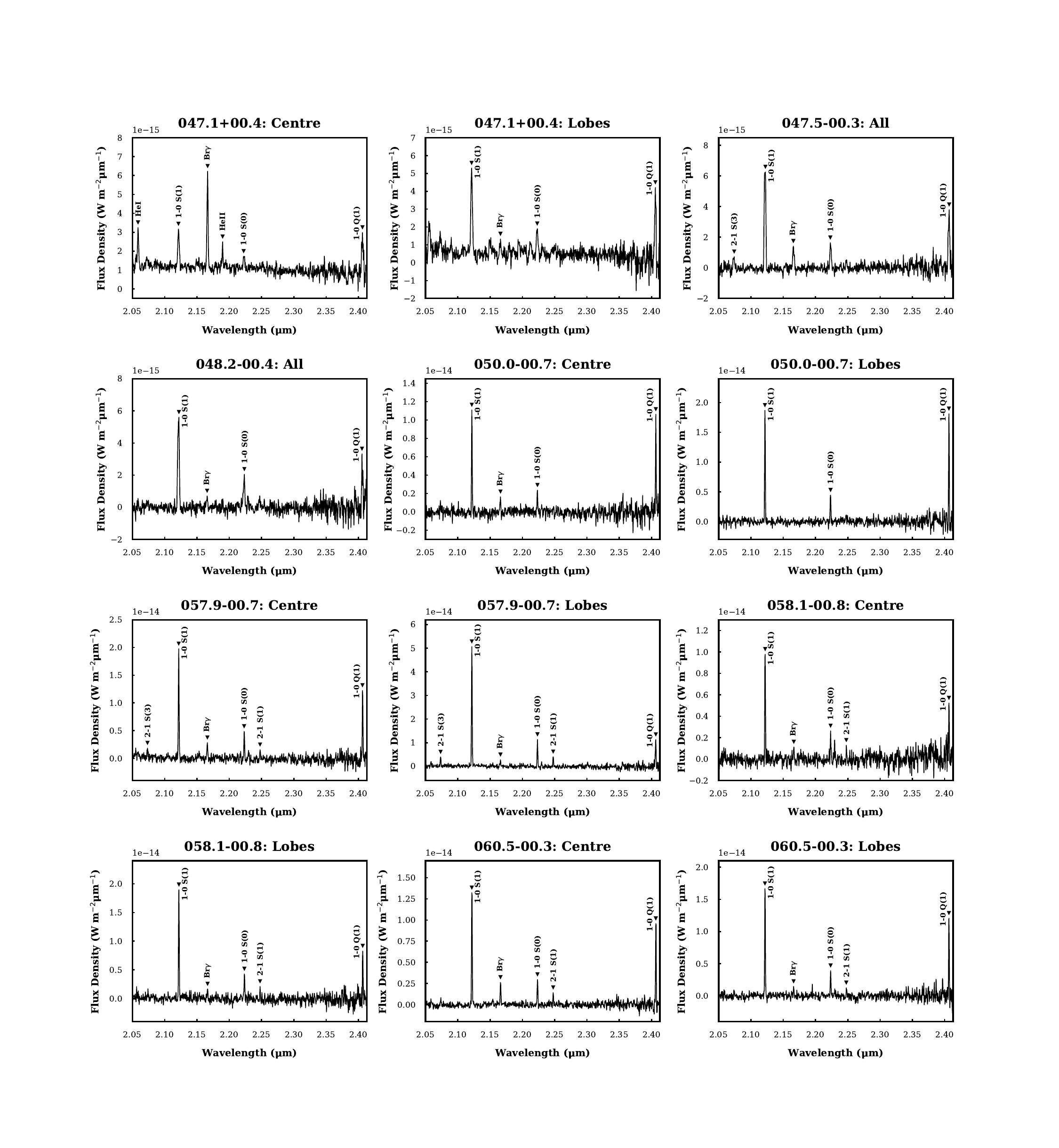}
\caption{\textbf{(cont.)} Spectra for PN\,G047.1+00.4 to G060.5-00.3, excluding PN\,G050.5+00.0 and G059.7-00.8.}
\end{figure*}	

\begin{figure*}\ContinuedFloat
\centering
\includegraphics[trim = 1.5cm 1.5cm 2.25cm 2.5cm, clip=true, width=1.0\linewidth]{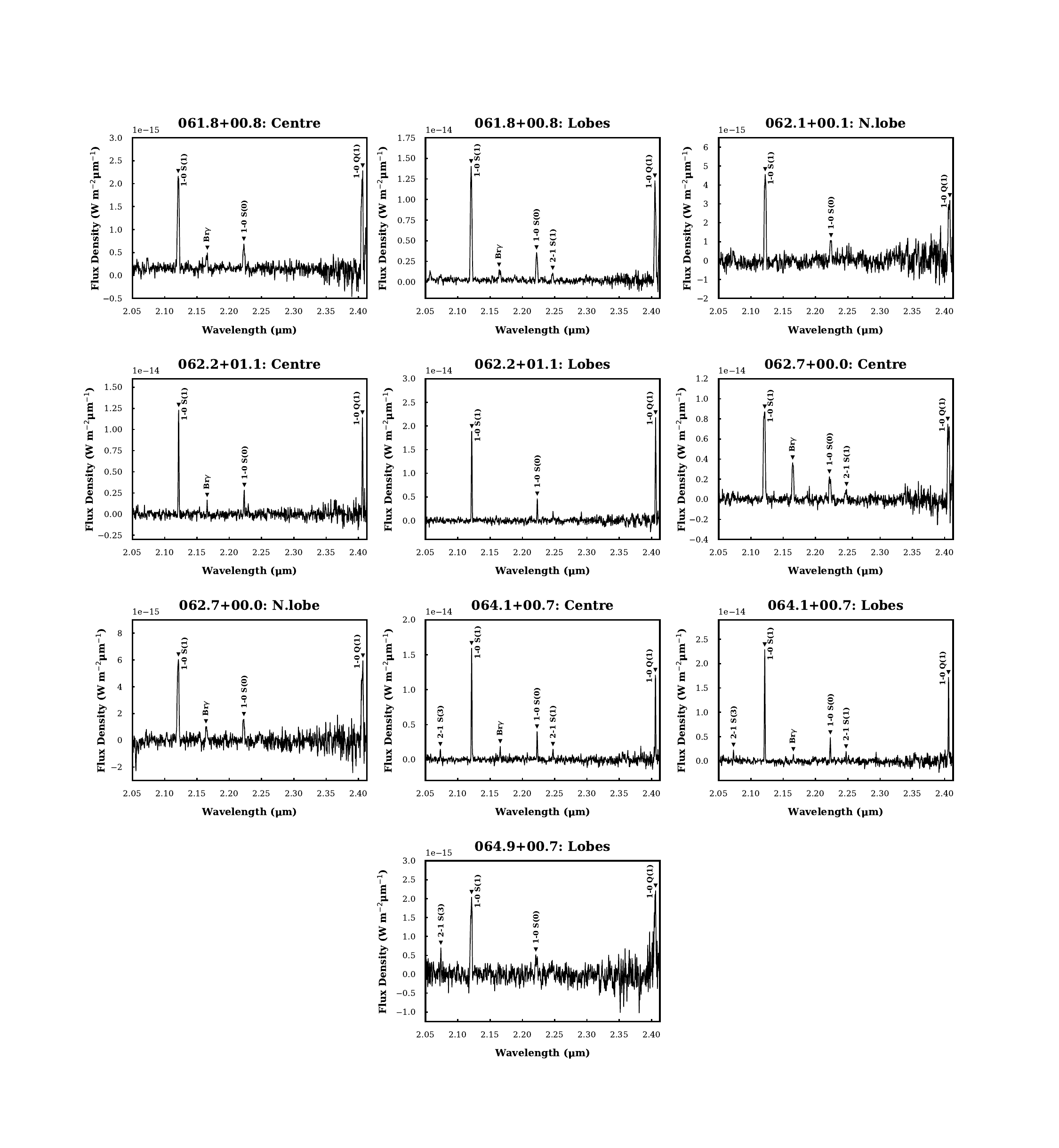}
\caption{\textbf{(cont.)} Spectra for PN\,G061.8+00.8 to G064.9+00.7.}
\end{figure*}

\section{Emission line fluxes} 

\begin{landscape}
		\begin{table}
		\centering
		\caption{Continuum-subtracted line fluxes and errors ($\times \, 10^{-19}$ Wm$^{-2}$) for K-band emission lines. Rest wavelengths of the emission lines are given in $\muup$m. Note these values are not corrected for extinction.}
		\fontsize{8.0}{9.6}\selectfont
		\begin{tabular}{llcccccccccccc}
		\hline
		ID & Region & \ion{He}{i} & 2-1 S(3) & \ion{He}{i} & 1-0 S(1) & 2-1 S(2) & Br$\gamma$ & \ion{He}{ii} & 3-2 S(3) & 1-0 S(0) & 2-1 S(1) & 3-2 S(2) & 1-0 Q(1) \\ 
		& & 2.0587 & 2.0735 & 2.1127 & 2.1218 & 2.1542 & 2.1661 & 2.1891 & 2.2014 & 2.2233 & 2.2477 & 2.2870 & 2.4066 \\
		\hline
		
		004.7-00.8 & Centre & --- & 47 $\pm$ 7 & --- & 398 $\pm$ 22 & 17 & --- & --- & --- & 78 $\pm$ 7 & 32 $\pm$ 5 & --- & 343 $\pm$ 22 \\
		
		009.7-00.9 & Centre & 99 $\pm$ 6 & 41 $\pm$ 5 & 9 & 412 $\pm$ 16 & 16 $\pm$ 3 & 169 $\pm$ 6 & 17 $\pm$ 2 & 7.6 $\pm$ 2.6 & 98 $\pm$ 5 & 56 $\pm$ 4 & 15 $\pm$ 3 & 332 $\pm$ 14 \\
		 & Lobes & --- & 33 $\pm$ 4 & --- & 280 $\pm$ 10 & 11 $\pm$ 4 & --- & --- & 11 $\pm$ 3 & 59 $\pm$ 3 & 39 $\pm$ 4 & --- & 147 $\pm$ 9 \\
		
		020.7-00.1 & Centre & --- & --- & --- & 99 $\pm$ 7 & --- & 16 $\pm$ 4 & --- & --- & 17 $\pm$ 2 & --- & --- & 59 $\pm$ 8 \\
		 & Lobes & --- & 29 $\pm$ 6 & --- & 282 $\pm$ 17 & --- & 15 $\pm$ 5 & --- & --- & 64 $\pm$ 5 & 25 $\pm$ 5 & --- & 49 $\pm$ 14 \\
		 
		020.8+00.4 & Centre & --- & 12 $\pm$ 2 & --- & 56 $\pm$ 2 & --- & --- & --- & --- & 15 $\pm$ 1 & 7.9 $\pm$ 1.4 & --- & 23 $\pm$ 3 \\
		 & Lobes & --- & 14 $\pm$ 4 & --- & 104 $\pm$ 4 & --- & --- & --- & --- & 25 $\pm$ 2 & 17 $\pm$ 3 & --- & 57 $\pm$ 6 \\
		
		024.8+00.4 & Centre & 78 $\pm$ 3 & --- & 5.4 $\pm$ 0.9 & 57 $\pm$ 8 & --- & 115 $\pm$ 3 & 13 $\pm$ 1 & --- & 23 $\pm$ 6 & 10 $\pm$ 3 & --- & 46 $\pm$ 7 \\
		 & Lobes & --- & 91 $\pm$ 8 & --- & 805 $\pm$ 65 & 27 $\pm$ 5 & --- & --- & 19 $\pm$ 4 & 185 $\pm$ 15 & 94 $\pm$ 9 & --- & 597 $\pm$ 51 \\
		
		025.9-00.5                  & North lobe & ---           & ---           & ---    & 95 $\pm$ 5   & ---      & 8.6 $\pm$ 2.9 & ---    & ---      & 22 $\pm$ 2 & ---        & ---      & 23 $\pm$ 7   \\
		
		032.6-01.2 & Centre     & 16 $\pm$ 3    & 11 $\pm$ 2    & ---    & 155 $\pm$ 9  & ---      & 53 $\pm$ 2    & ---    & ---      & 34 $\pm$ 3 & 13 $\pm$ 2 & ---      & 44 $\pm$ 7   \\
		                            & Lobes      & ---           & ---           & ---    & 211 $\pm$ 12 & ---      & 28 $\pm$ 3    & ---    & ---      & 43 $\pm$ 4 & 15 $\pm$ 2 & ---      & 77 $\pm$ 8   \\
		
		034.8+01.3 & Centre     & 8.7 $\pm$ 2.5 & 6.2 $\pm$ 2.8 & ---    & 117 $\pm$ 12 & ---      & 34 $\pm$ 3    & ---    & ---      & 25 $\pm$ 3 & ---        & ---      & 61 $\pm$ 8   \\
		                            & Lobes      & 11 $\pm$ 3    & 14 $\pm$ 3    & ---    & 234 $\pm$ 15 & ---      & 22 $\pm$ 3    & ---    & ---      & 50 $\pm$ 5 & 21 $\pm$ 2 & ---      & 140 $\pm$ 14  \\     
		                            
		035.7-01.2 & Centre     & ---        & ---           & ---    & 108 $\pm$ 10 & ---      & 17 $\pm$ 3  & ---        & ---      & 20 $\pm$ 2  & ---        & ---      & 69 $\pm$ 10  \\
		                            & Lobes      & ---        & ---           & ---    & 480 $\pm$ 28 & ---      & 53 $\pm$ 6  & ---        & ---      & 95 $\pm$ 8  & 40 $\pm$ 7 & ---      & 359 $\pm$ 27 \\
		
		036.4+00.1 & Centre     & 14 $\pm$ 2 & ---           & ---    & 67 $\pm$ 5   & ---      & 42 $\pm$ 2  & ---        & ---      & 15 $\pm$ 2  & 10 $\pm$ 2 & ---      & 44 $\pm$ 4   \\
		                            & South lobe & 15 $\pm$ 2 & 13 $\pm$ 3    & ---    & 102 $\pm$ 7  & ---      & 47 $\pm$ 3  & ---        & ---      & 24 $\pm$ 3  & 12 $\pm$ 2 & ---      & 55 $\pm$ 6   \\
		
		037.4-00.1 & Centre     & 21 $\pm$ 5 & ---           & ---    & 33 $\pm$ 5   & ---      & 75 $\pm$ 4  & ---        & ---      & 10 $\pm$ 2  & 5 $\pm$ 1  & ---      & 13 $\pm$ 4   \\
		                            & Lobes      & 15         & 8.9 $\pm$ 3.5 & ---    & 199 $\pm$ 16 & ---      & 54 $\pm$ 4  & ---        & ---      & 49 $\pm$ 5  & 22 $\pm$ 4 & ---      & 130 $\pm$ 20 \\
		
		040.4+01.1 & Centre     & 36 $\pm$ 7 & ---           & ---    & 99 $\pm$ 4   & ---      & 72 $\pm$ 4  & ---        & ---      & 20 $\pm$ 4  & 13 $\pm$ 4 & ---      & 67 $\pm$ 12  \\
		                            & Lobes      & ---        & 75 $\pm$ 6    & ---    & 579 $\pm$ 12 & ---      & ---         & ---        & ---      & 135 $\pm$ 9 & 75 $\pm$ 6 & ---      & 228 $\pm$ 14 \\
		
		040.5-00.7                  & All        & ---        & ---           & ---    & 204 $\pm$ 5  & ---      & 21 $\pm$ 8  & ---        & ---      & 49 $\pm$ 9  & ---        & ---      & 171 $\pm$ 11 \\
		
		042.1+00.4                  & All        & ---        & ---           & ---    & 244 $\pm$ 12 & ---      & 21 $\pm$ 4  & ---        & ---      & 46 $\pm$ 7  & ---        & ---      & 92 $\pm$ 11  \\
		
		047.1+00.4 & Centre     & 48 $\pm$ 6 & ---           & ---    & 61 $\pm$ 3   & ---      & 104 $\pm$ 3 & 21 $\pm$ 3 & ---      & 18 $\pm$ 3  & ---        & ---      & 43 $\pm$ 11  \\
		                            & Lobes      & ---        & ---           & ---    & 143 $\pm$ 5  & ---      & 16 $\pm$ 4  & ---        & ---      & 40 $\pm$ 5  & ---        & ---      & 75 $\pm$ 14  \\
		
		047.5-00.3                  & All        & ---        & 18 $\pm$ 4    & ---    & 202 $\pm$ 13 & ---      & 37 $\pm$ 4  & ---        & ---      & 40 $\pm$ 3  & ---        & ---      & 100 $\pm$ 11  \\
		
		048.2-00.4                  & All    & ---           & ---         & ---          & 180 $\pm$ 14 & ---         & 15 $\pm$ 3    & ---        & ---        & 51 $\pm$ 5   & ---        & ---         & 60 $\pm$ 14  \\
		
		050.0-00.7 & Centre & ---           & ---         & ---          & 140 $\pm$ 6  & ---         & 23 $\pm$ 6    & ---        & ---        & 30 $\pm$ 4   & ---        & ---         & 123 $\pm$ 12 \\
		                            & Lobes  & ---           & ---         & ---          & 234 $\pm$ 6  & ---         & ---           & ---        & ---        & 62 $\pm$ 7   & ---        & ---         & 222 $\pm$ 18 \\
		
		050.5+00.0 & Centre & 1920 $\pm$ 30 & ---         & 212 $\pm$ 19 & 209 $\pm$ 9  & ---         & 3890 $\pm$ 80 & 93 $\pm$ 4 & ---        & 67 $\pm$ 8   & 46 $\pm$ 4 & 182 $\pm$ 7 & 175 $\pm$ 20 \\
		                            & Lobes  & 109 $\pm$ 10  & 45 $\pm$ 12 & 15 $\pm$ 6   & 518 $\pm$ 15 & 28 $\pm$ 10 & 256 $\pm$ 9   & ---        & 32 $\pm$ 6 & 136 $\pm$ 10 & 89 $\pm$ 5 & 21 $\pm$ 7  & 364 $\pm$ 21 \\
		
		057.9-00.7 & Centre & ---           & 35 $\pm$ 10 & ---          & 277 $\pm$ 8  & ---         & 43 $\pm$ 5    & ---        & ---        & 61 $\pm$ 6   & 20 $\pm$ 6 & ---         & 158 $\pm$ 13 \\
		                            & Lobes  & ---           & 57 $\pm$ 6  & ---          & 669 $\pm$ 10 & ---         & 29 $\pm$ 9    & ---        & ---        & 141 $\pm$ 8  & 51 $\pm$ 6 & ---         & 148 $\pm$ 19 \\
		
		058.1-00.8 & Centre & ---           & ---         & ---          & 127 $\pm$ 7  & ---         & 16 $\pm$ 7    & ---        & ---        & 32 $\pm$ 6   & ---        & ---         & 64 $\pm$ 15  \\
		                            & Lobes  & ---           & ---         & ---          & 244 $\pm$ 6  & ---         & 24 $\pm$ 6    & ---        & ---        & 58 $\pm$ 7   & 18 $\pm$ 6 & ---         & 95 $\pm$ 15  \\
		
		059.7-00.8                  & Lobes  & ---           & 52 $\pm$ 10 & ---          & 503 $\pm$ 8  & ---         & 59 $\pm$ 7    & ---        & ---        & 119 $\pm$ 7  & 38 $\pm$ 6 & ---         & 378 $\pm$ 14 \\
		
		060.5-00.3 & Centre & ---           & ---         & ---          & 177 $\pm$ 5  & ---         & 39 $\pm$ 2    & ---        & ---        & 40 $\pm$ 3   & 15 $\pm$ 3 & ---         & 121 $\pm$ 10 \\
		                            & Lobes  & ---           & ---         & ---          & 215 $\pm$ 7  & ---         & 18 $\pm$ 5    & ---        & ---        & 49 $\pm$ 4   & 16 $\pm$ 5 & ---         & 137 $\pm$ 13 \\
		
		061.8+00.8 & Centre & ---           & ---         & ---          & 61 $\pm$ 3   & ---         & 8.8 $\pm$ 1.5 & ---        & ---        & 15 $\pm$ 2   & ---        & ---         & 61 $\pm$ 8   \\
		                            & Lobes  & ---           & ---         & ---          & 409 $\pm$ 17 & ---         & 36 $\pm$ 8    & ---        & ---        & 107 $\pm$ 7   & 31 $\pm$ 5 & ---         & 327 $\pm$ 25 \\
		
		062.1+00.1                  & North lobe & ---    & ---              & ---    & 145 $\pm$ 13 & ---      & ---         & ---    & ---      & 35 $\pm$ 4  & ---        & ---      & 103 $\pm$ 12 \\
		
		062.2+01.1 & Centre     & ---    & ---           & ---    & 166 $\pm$ 4  & ---      & 14 $\pm$ 3  & ---    & ---      & 41 $\pm$ 5  & ---        & ---      & 148 $\pm$ 13 \\
		                            & Lobes      & ---    & ---              & ---    & 255 $\pm$ 4  & ---      & ---         & ---    & ---      & 55 $\pm$ 4  & ---        & ---      & 269 $\pm$ 15 \\
		
		062.7+00.0 & Centre     & ---    & ---              & ---    & 309 $\pm$ 22 & ---      & 113 $\pm$ 8 & ---    & ---      & 67 $\pm$ 11 & 25 $\pm$ 6 & ---      & 244 $\pm$ 42 \\
		                            & North lobe & ---    & ---              & ---    & 191 $\pm$ 11 & ---      & 30 $\pm$ 5  & ---    & ---      & 43 $\pm$ 5  & ---        & ---      & 168 $\pm$ 26 \\
		
		064.1+00.7 & Centre     & ---    & 20 $\pm$ 5    & ---    & 206 $\pm$ 6  & ---      & 23 $\pm$ 4  & ---    & ---      & 52 $\pm$ 4  & 22 $\pm$ 4 & ---      & 146 $\pm$ 10 \\
		                            & Lobes      & ---    & 24 $\pm$ 6    & ---    & 282 $\pm$ 4  & ---      & 14 $\pm$ 4  & ---    & ---      & 64 $\pm$ 4  & 21 $\pm$ 6 & ---      & 204 $\pm$ 15 \\
		
		064.9+00.7                  & Lobes      & ---    & 7.4 $\pm$ 2.3 & ---    & 58 $\pm$ 6   & ---      & ---         & ---    & ---      & 18 $\pm$ 4  & ---        & ---      & 65 $\pm$ 9 \\
		
		\hline
		\end{tabular}
		\label{tab:linefluxes}
		\end{table}
\end{landscape}


\bsp	
\label{lastpage}
\end{document}